\documentclass[11pt]{article}

\usepackage[margin=1in]{geometry}
\usepackage{amsmath, amssymb, amsthm}
\usepackage{booktabs}
\usepackage{multirow}
\usepackage{longtable}
\usepackage{graphicx}
\usepackage{hyperref}
\usepackage{xcolor}
\usepackage{enumitem}
\usepackage{float}
\usepackage{caption}
\usepackage{subcaption}
\usepackage[numbers]{natbib}
\usepackage{url}
\usepackage{xurl}
\usepackage[T1]{fontenc}
\usepackage{pifont}
\newcommand{\cmark}{\ding{51}}
\newcommand{\xmark}{\ding{55}}
\usepackage{tikz}
\usetikzlibrary{arrows.meta, positioning, fit, backgrounds, calc, matrix}
\usepackage{pgfplots}
\pgfplotsset{compat=1.18}
\usepgfplotslibrary{statistics}

\definecolor{trustcolor}{RGB}{31, 119, 180}
\definecolor{reputcolor}{RGB}{255, 127, 14}
\definecolor{coopcolor}{RGB}{44, 160, 44}
\definecolor{defectcolor}{RGB}{214, 39, 40}
\definecolor{neutralcolor}{RGB}{127, 127, 127}

\hypersetup{
    colorlinks=true,
    linkcolor=blue!60!black,
    citecolor=green!60!black,
    urlcolor=blue!60!black,
    breaklinks=true,
}
\makeatletter
\g@addto@macro\UrlBreaks{\do\/\do\.\do\_\do\-\do\,\do\;\do\:}
\makeatother

\newcommand{\dij}{D_{ij}}

\title{Coopetition-Gym v1: \\ A Formally Grounded Platform for Mixed-Motive
Multi-Agent Reinforcement Learning under Strategic Coopetition}

\author{
Vik Pant \\
Faculty of Information \\
University of Toronto \\
\texttt{vik.pant@mail.utoronto.ca}
\and
Eric Yu \\
Faculty of Information and \\
Department of Computer Science \\
University of Toronto \\
\texttt{eric.yu@utoronto.ca}
}

\date{\vspace{-0.6em}\today\vspace{-1.2em}}

\begin{document}

\maketitle
\vspace{-2.0em}

\begin{abstract}
We present \textsc{Coopetition-Gym v1}, a formally grounded benchmark
platform for mixed-motive multi-agent reinforcement learning derived
from four prepublished technical reports on strategic coopetition:
\textbf{TR-1} on interdependence and complementarity
(\citealp{pant2025interdependence}, arXiv:2510.18802), \textbf{TR-2} on
trust and reputation dynamics (\citealp{pant2025trust},
arXiv:2510.24909), \textbf{TR-3} on collective action and loyalty
(\citealp{pant2026collective}, arXiv:2601.16237), and \textbf{TR-4} on
sequential interaction and reciprocity
(\citealp{pant2026reciprocity}, arXiv:2604.01240). The package
comprises twenty environments organized into four mechanism classes
corresponding to these four reports; each environment inherits a
closed-form payoff structure and a calibrated interdependence matrix
$\dij$ from the relevant report. All twenty environments use a uniform
scalar action space and expose a parameterized reward layer that the
user may configure across three modes (\emph{private},
\emph{integrated}, and \emph{cooperative}); this two-layer separation
of payoff from reward is what enables \emph{reward-type ablation}, the
package's principal methodological apparatus. Four environments are
calibrated to historically documented coopetitive relationships
(Samsung--Sony LCD at $98.3\%$ validation, Renault--Nissan Alliance at
$81.7\%$, Apache HTTP Server at $86.7\%$, and Apple iOS App Store at
$87.3\%$), establishing that the package's formal mechanisms capture
empirically observable coopetitive dynamics. The package exposes
three application programming interfaces (Gymnasium, PettingZoo
Parallel, and PettingZoo AEC) and a suite of $126$ reference
algorithms: $16$ training algorithms spanning independent learning and
centralized-training-with-decentralized-execution paradigms, $7$
game-theoretic oracles providing analytically grounded reference
policies, $2$ heuristic baselines, and $101$ constant-action policies
that span the cooperation continuum. A reference experimental study
trained the $16$ algorithms on every environment under every reward
configuration with seven random seeds, producing a $25{,}708$-file
training dataset and a $1{,}116$-file behavioral audit dataset, both
released under CC-BY-4.0 with Croissant 1.0 metadata. Four
methodological apparatuses accompany the package: a
\textbf{statistical-gate discipline} for cheap-test-before-claim
inferential commitment; a \textbf{controlled critic-learning-rate
ablation} isolating early-stage divergence in deterministic-policy
methods on high-agent-count environments; a \textbf{matrix-coverage
coverage-verification audit} certifying that every algorithm-environment pair
instantiates and trains without runtime exceptions; and
\textbf{continuous reliability diagnostics} replacing binary
success/failure with the per-cell finite-fraction
$f_{\mathrm{fin}}$. \textbf{Code:}
\url{https://github.com/vikpant/strategic-coopetition} (MIT).
\textbf{Datasets:}
\url{https://huggingface.co/datasets/vikpant/coopetition-gym-logs}
(CC-BY-4.0; Croissant 1.0 manifest at \texttt{croissant.json}).
\textbf{Documentation:}
\url{https://vikpant.github.io/strategic-coopetition/}. Total compute
cost of the reference experimental study is reported in
Appendix~\ref{app:compute} ($\$10{,}500$ USD on a fleet of cloud-hosted
NVIDIA RTX~5090 GPU instances).
\end{abstract}

\tableofcontents
\newpage

\section*{Foreword}
\label{sec:foreword}
\addcontentsline{toc}{section}{Foreword}

This technical report is the package reference for
\textsc{Coopetition-Gym v1}. It documents the package's mathematical
foundations, environment suite, programming interfaces, algorithm and
oracle pools, evaluation methodologies, validation record, and
reproducibility apparatus. The document is intended to serve as the
canonical reference source for users of the package: researchers
applying the package to their own questions, reviewers assessing
empirical claims that depend on the package, and downstream authors
who build on the package in their own work.

The package is a computational realization of four prepublished
technical reports on strategic coopetition. \textbf{TR-1}
(\citealp{pant2025interdependence}, arXiv:2510.18802) formalizes
\emph{interdependence and complementarity}: the static interdependence
matrix $\dij$, the synergy multiplier $\gamma$, and the integrated
utility expression that the package reuses as its canonical reward
layer. \textbf{TR-2} (\citealp{pant2025trust}, arXiv:2510.24909)
formalizes \emph{trust and reputation dynamics}: the two-layer model
of immediate trust $T_{ij}$ and exponentially smoothed reputation
$R_{ij}$ with the three-to-one negativity bias. \textbf{TR-3}
(\citealp{pant2026collective}, arXiv:2601.16237) formalizes
\emph{collective action and loyalty}: team-production payoffs, the
loyalty modifier $\theta_i$, and the closed-form free-riding Nash
equilibrium. \textbf{TR-4} (\citealp{pant2026reciprocity},
arXiv:2604.01240) formalizes \emph{sequential interaction and
reciprocity}: memory-bounded reciprocity signals, the bounded response
$\varphi(x) = \tanh(\kappa x)$, and the trust-gated reciprocity
modifier. The present document treats these four reports as the
mathematical substrate for the package's environment design and
reward parameterization. This package can be used and the equations
restated in this document can be read independently of the four
reports, with self-contained explanations provided where needed.

The empirical findings reported in Part~II are \emph{illustrative} of
the package's analytical utility rather than the document's principal
contribution. They demonstrate the kinds of claims the package
supports and the kinds of questions it exposes; extended treatment of
any specific finding lies outside the scope of a platform reference.
The package is positioned in the established tradition of platform
technical reports such as PettingZoo
(\citealp{terry2021pettingzoo}, arXiv:2009.14471, NeurIPS~$2021$) and
OpenSpiel (\citealp{lanctot2019openspiel}, arXiv:1908.09453).

\part{The Platform}

\section{Strategic Coopetition in Management Science}
\label{sec:coopetition_background}

Many readers of this technical report come to it through reinforcement
learning, multi-agent systems, or computational game theory rather
than through the management-science literature in which the term
\emph{coopetition} originated. Because the package's environments,
parameters, and validation cases are grounded in that literature, the
present section briefly situates strategic coopetition as an
established field of research inquiry within management science. The
exposition is intentionally short; readers comfortable with the
management-science framing of cooperation and competition as
\emph{distinguishable but co-occurring} strategic dimensions may skip
to Section~\ref{sec:intro}.

\subsection{Origin and conceptual core}

The term \emph{coopetition} is a portmanteau of \emph{cooperation}
and \emph{competition} introduced by~\citet{brandenburger1996coopetition}
in the management-strategy literature to describe relationships in
which two or more economic actors simultaneously cooperate to enlarge
a shared value pool and compete to capture shares of that pool. The
canonical maxim, ``cooperate to grow the pie, compete to split the
pie,'' captures the dual character that distinguishes coopetition
from pure competition (where the pie is fixed) and from pure
cooperation (where the split is pre-agreed). Coopetition is the
\emph{syncretic} phenomenon that emerges when actors reconcile
opposing strategic imperatives within a single ongoing relationship,
holding the cooperative and competitive sides of that relationship in
productive tension rather than collapsing one into the other.

The conceptual core of the field is that cooperation and competition
are \emph{esemplastic} dimensions of strategic action: they fuse
into a single coherent stance toward the partner rather than being
chosen alternately or alternated in time. An empirical alliance is
neither a series of cooperative episodes punctuated by competitive
ones, nor the reverse; it is a sustained orientation in which both
strategic logics operate concurrently and shape every decision the
actor takes. The shared value created through cooperation is the
substrate from which competitive value-capture flows, and the
distributional friction created by competition is the discipline that
keeps cooperation from collapsing into one-sided exploitation. The
research program that grew out of \citet{brandenburger1996coopetition}
has spent three decades elaborating the conditions, mechanisms, and
limits of this integrative combination across alliance studies, supply
chain analysis, platform economics, and inter-organizational
governance.

\subsection{Active taxonomies in coopetition theory}

Coopetition research is organized around several rich taxonomies that
remain under active investigation. We list five of the most-studied
distinctions because the package's environment design draws on each.

\paragraph{Uniaxial versus biaxial.} The uniaxial treatment models
cooperation and competition as two ends of a single continuum
\citep{bengtsson2000coopetition, padula2007coopetition}; an actor's
strategic choice is its position along that continuum. The biaxial
treatment, due to \citet{brandenburger1996coopetition} and developed
by \citet{luo2007coopetition} and \citet{gnyawali2011coopetition},
treats cooperation and competition as orthogonal axes; an actor can
be high or low on each independently. The two treatments yield
different formal structures and different implications for empirical
measurement; the package's v1 environments adopt the uniaxial
treatment because the underlying mathematical reports
(\citealp{pant2025interdependence}, \citealp{pant2025trust},
\citealp{pant2026collective}, \citealp{pant2026reciprocity}) are
formally bound to it.

\paragraph{Processual versus structural
(\citealp{bengtsson2014paradox}; \citealp{dahl2014processual}).}
Processual studies focus on the temporal unfolding of coopetitive
relationships, including phase transitions, trust accumulation, and
crisis recovery. Structural studies focus on the architectural
features that make a relationship coopetitive in the first place,
including dependency networks, bargaining-power asymmetries, and
resource interdependencies. The package realizes both: TR-2 trust
dynamics and TR-4 reciprocity are processual mechanisms, while the
TR-1 interdependence matrix and TR-3 collective-action structure are
structural.

\paragraph{Dyadic versus network
(\citealp{bengtsson2000coopetition}; \citealp{ritala2014coopetition}).}
Dyadic coopetition concerns relationships between two actors.
Network coopetition concerns ecosystems of three or more actors in
which any pair may be cooperating, competing, or coopeting at a given
moment. The package includes both: dyadic environments such as
\texttt{TrustDilemma-v0} and \texttt{SLCD-v0}, and network
environments such as \texttt{PlatformEcosystem-v0},
\texttt{ApacheProject-v0}, and \texttt{GraduatedSanction-v0}.

\paragraph{Direct versus indirect
(\citealp{nowak2005evolution}; \citealp{czakon2016architecture}).}
Direct coopetition is between two actors who interact bilaterally.
Indirect coopetition is mediated by third parties, by reputation
systems, or by shared institutional infrastructure. The package's
reputation-mediated environments (\texttt{ReputationMarket-v0},
\texttt{IndirectReciprocity-v0}) and its image-scoring formalism in
TR-4 instantiate the indirect regime.

\paragraph{Simultaneous versus sequential
(\citealp{padula2007coopetition}; \citealp{lado1997syncretic}).}
The simultaneous--sequential taxonomy is a methodological distinction
about how each \emph{decision period} is structured: under the
simultaneous treatment agents make a single joint move per period
without observing the other's choice first; under the sequential
treatment agents move in a defined order, and the second mover
conditions on the first mover's revealed action. The PettingZoo
Parallel API exposed by the package implements simultaneous-move
semantics; the PettingZoo AEC API implements sequential-move
semantics. The TR-4 reciprocity formalism is sequential in this
methodological sense because the cooperation signal at step $t$
conditions on a memory of partner actions over steps $t-k$ through
$t-1$.

The simultaneous--sequential distinction operates at the level of
\emph{move structure within a decision period}; it is logically
distinct from the integrative, dual-logic character of coopetition
as a sustained orientation, which operates at the level of an actor's
overall stance toward a relationship across periods. A relationship
in which both cooperation and competition are present concurrently as
strategic logics (the unifying frame above) can be operationalized
as either simultaneous-move or sequential-move at the per-step level
without contradiction; the unifying frame says \emph{both logics are
co-present at every period}, while the simultaneous--sequential frame
says \emph{within each period, the agents either reveal moves at once
or in turn}. Both characterizations apply to the same relationship
simultaneously and address different analytical layers.

\subsection{Why the package's mechanism classes follow this
literature}

The four mechanism classes around which the package is organized
(interdependence and complementarity; trust and reputation dynamics;
collective action and loyalty; sequential interaction and reciprocity)
are not arbitrary engineering choices. Each class corresponds to a
distinct mechanism that the management-science literature has
identified as central to how coopetitive relationships function in
practice. Interdependence (TR-1) operationalizes the structural
substrate; trust dynamics (TR-2) operationalize the processual
relational layer; collective action and loyalty (TR-3) extend the
treatment from dyadic to network settings; reciprocity (TR-4) carries
the simultaneous and sequential temporal logic. The four reports
together cover the structural-versus-processual and dyadic-versus-network
taxonomic axes, and the package's environment suite spans all five
distinctions reviewed above. Section~\ref{sec:intro} now turns to the
multi-agent reinforcement learning side of this work.

\section{Introduction}
\label{sec:intro}

\subsection{Motivation: the coopetition gap in MARL}

Multi-agent reinforcement learning (MARL) benchmarks have historically
emphasized either fully cooperative or strictly competitive settings.
Cooperative benchmarks, including the StarCraft Multi-Agent
Challenge~\citep{samvelyan2019smac}, Hanabi~\citep{bard2020hanabi},
Overcooked~\citep{carroll2019overcooked}, and the Multi-Particle
Environments~\citep{lowe2017maddpg}, test coordination under a shared
team objective in which the social optimum coincides with each agent's
optimal policy given truthful coordination. Competitive benchmarks,
including adversarial gaming suites, zero-sum poker variants, and the
competitive substrates of OpenSpiel~\citep{lanctot2019openspiel}, test
strategic play against opponents whose utility is directly opposed to
the learner's. Both regimes admit clean theoretical characterization:
cooperative settings reduce to joint-policy optimization; competitive
settings reduce to minimax analysis.

Mixed-motive settings occupy the space between. Agents have partially
overlapping and partially opposing objectives. Each agent's optimal
policy depends on partner policies, but the best-response mapping is
neither a coordination consensus (as in cooperative settings) nor a
minimax struggle (as in zero-sum settings). The analytical canonical
form is the general-sum stochastic game~\citep{shoham2008multiagent};
the behavioral canonical forms include the iterated Prisoner's
Dilemma, the public goods game, the Stag Hunt, and the Commons, each
of which has generated decades of theoretical~\citep{axelrod1984evolution,
ostrom1990governing, nowak2006five} and empirical~\citep{rand2013human,
fehr2000cooperation} study. Yet MARL benchmarks for mixed-motive
settings have received comparatively less systematic coverage than
their cooperative and competitive counterparts, despite being the
dominant structure of economically and socially important multi-agent
interactions: supply chains, platform ecosystems, research consortia,
joint ventures, licensing arrangements, and open-source coalitions.

The mixed-motive regime is theoretically rich because the relative
weight of cooperation and competition is a continuous parameter rather
than a binary choice, and because equilibrium behavior shifts
qualitatively with this parameter. A fully cooperative reward
collapses to team RL; a fully competitive reward produces zero-sum
play and collapses to minimax; the behavior of interest lies between.
\textsc{Coopetition-Gym v1} is designed to surface this interior and
to support methodologies, particularly reward-type ablation
(\S\ref{sec:eval}), that expose how algorithm rankings depend on
position along the cooperation-competition continuum.

\subsection{Positioning among MARL benchmarks}

Several MARL benchmarks include mixed-motive environments or are
adjacent to the design space this suite occupies. Clarifying the
relationship:

\paragraph{Melting Pot~\citep{leibo2021meltingpot, agapiou2022meltingpot}}
is the closest benchmark in spirit. It provides dozens of substrates
combining cooperation, competition, and social norms, and it supports
population-based evaluation. Melting Pot substrates are grid-world
pixel-based; Coopetition-Gym environments are continuous-action
abstract-state. The two benchmarks are complementary: Melting Pot
emphasizes visual and population-level emergence; Coopetition-Gym
emphasizes analytical transparency (closed-form equilibria, calibrated
interdependence coefficients, validated case studies) and supports
reward-type ablation as a first-class evaluation protocol.

\paragraph{The SSD line of research~\citep{leibo2017ssd, hughes2018inequity}}
extends matrix-game social dilemmas to temporally extended
2-dimensional pixel environments. The original SSD work introduced
\emph{Gathering} (a temporal extension of the Prisoner's Dilemma) and
\emph{Wolfpack} (a temporal extension of the Stag Hunt); subsequent
work added \emph{Cleanup} (public-good provision) and \emph{Harvest}
(commons tragedy) to the line of research. Coopetition-Gym v1 shares
the social-dilemma motivation but restricts to continuous-action
abstract-state formulations and provides formal equilibrium oracles
as reference policies.

\paragraph{PettingZoo~\citep{terry2021pettingzoo}} provides both a
multi-agent API specification (the Agent-Environment-Cycle and
Parallel interfaces) and a substantial library of reference
environments, including classical Atari multiplayer titles, MPE
(multi-agent particle environments), Butterfly cooperative-coordination
environments, classic board and card games (including Chess, Go,
Hanabi, and Texas Hold'em), and SISL multi-agent environments. Coopetition-Gym v1 implements the
PettingZoo Parallel and AEC APIs (\S\ref{sec:api}), making every
environment in our suite compatible with PettingZoo-native training
frameworks; relative to PettingZoo's own environment library,
Coopetition-Gym v1 contributes a tier of mixed-motive environments
with calibrated interdependence parameters and game-theoretic oracle
baselines that PettingZoo's library does not target.

\paragraph{OpenSpiel~\citep{lanctot2019openspiel}} is a broad framework
for reinforcement learning and search in $n$-player zero-sum,
cooperative, and general-sum games, with implementations of
algorithmic-game-theory tools (counterfactual regret minimization,
exploitability, best-response computation, fictitious play) alongside
deep RL algorithms (deep CFR, AlphaZero, MCTS, neural fictitious
self-play). Most OpenSpiel environments are discrete-action;
Coopetition-Gym complements this coverage by providing continuous-action
mixed-motive environments with calibrated payoff parameterization.

\paragraph{SMAC~\citep{samvelyan2019smac} and SMACv2~\citep{ellis2023smacv2}}
are fully cooperative (team-vs-script) and therefore do not cover
mixed-motive dynamics. Coopetition-Gym v1 occupies the adjacent niche
of mixed-motive evaluation, complementing rather than replacing
cooperative benchmarks; a thorough MARL evaluation portfolio can
exercise both regimes by drawing cooperative environments from SMAC
or SMACv2 and mixed-motive environments from Coopetition-Gym.

In short, Coopetition-Gym v1's distinguishing contribution relative to
existing benchmarks is the combination of (i) continuous-action abstract
environments, (ii) formally parameterized reward mutuality with
calibrated interdependence coefficients, (iii) game-theoretic oracle
baselines that provide theoretically grounded reference points, and (iv)
four empirically validated case study environments, a combination
not available in any prior MARL benchmark. Table~\ref{tab:positioning}
makes this comparison explicit by recording, for each of the four
distinguishing properties, whether each benchmark above provides it
as a first-class feature of its standard library.

\begin{table}[t]
\centering
\caption{Positioning of Coopetition-Gym v1 relative to existing MARL
benchmarks on four distinguishing properties: continuous-action
environments (CA), parameterized reward mutuality (PM),
game-theoretic oracle baselines (OB), and empirically validated case
studies (VCS). \cmark{} indicates that the property is provided as a
first-class feature of the benchmark's standard library; \xmark{}
indicates that it is not. The combination of all four properties is
unique to Coopetition-Gym v1 among the benchmarks reviewed in this
section.}
\label{tab:positioning}
\small
\begin{tabular}{lcccc}
\toprule
\textbf{Benchmark} & \textbf{CA} & \textbf{PM} & \textbf{OB} & \textbf{VCS} \\
\midrule
Melting Pot~\citep{leibo2021meltingpot, agapiou2022meltingpot} & \xmark & \xmark & \xmark & \xmark \\
SSD line of research~\citep{leibo2017ssd, hughes2018inequity} & \xmark & \xmark & \xmark & \xmark \\
PettingZoo~\citep{terry2021pettingzoo} & partial & \xmark & \xmark & \xmark \\
OpenSpiel~\citep{lanctot2019openspiel} & partial & \xmark & \cmark & \xmark \\
SMAC and SMACv2~\citep{samvelyan2019smac, ellis2023smacv2} & \xmark & \xmark & \xmark & \xmark \\
\midrule
\textbf{Coopetition-Gym v1} (this work) & \cmark & \cmark & \cmark & \cmark \\
\bottomrule
\end{tabular}
\end{table}

\textsc{Coopetition-Gym v1} fills this gap. The suite comprises twenty
Gymnasium- and PettingZoo-compatible environments drawn from four formal
mechanism classes, each with published derivation and validation:
interdependence and complementarity, trust and reputation dynamics,
collective action and loyalty, and sequential interaction and
reciprocity. In the reference experimental study configuration,
environments span $n = 2$ to $n = 6$ agents and episode horizons from
$40$ to $200$ steps; mechanism classes implement static payoff
structures (TR-1 interdependence), step-wise trust updates with
$3{:}1$ negativity bias (TR-2), loyalty multipliers that accumulate
over a memory window (TR-3), and bounded-response reciprocity over a
finite memory window (TR-4). The agent-count and horizon ranges
above describe the study configuration; source defaults coincide with
these values for $19$ of $20$ environments, and the
\texttt{ApacheProject-v0} source defaults expose phase-dependent
agent counts (8, 15, 25 agents under different phase modes) which
the reference study configures down to $5$ agents to align with the
$60$-step quarter-month horizon used in the validation rubric. Four
of the twenty environments are calibrated against historically
documented coopetitive relationships and are formally validated.

\subsection{Distinguishing features}

Three design choices distinguish \textsc{Coopetition-Gym} from existing
MARL benchmarks:

\paragraph{Uniform scalar action space.} Every environment uses the
continuous action space $[0, e_i]$ where $e_i$ is agent $i$'s per-step
endowment. The scalar action represents the agent's cooperation level.
A uniform action space across environments eliminates confounds that
arise when algorithm performance differences could be attributed to
action-space complexity interacting with environment mechanics. Scalar
actions also support direct comparison of learned policies across
environments via their cooperation-level distributions.

\paragraph{Parameterized reward mutuality.} Each environment supports
three reward configurations sharing the same payoff layer
(environment-generated rewards $\pi_i$) but differing in how rewards
are aggregated for learning: private reward ($U_i = \pi_i$), integrated
reward ($U_i = \pi_i + \sum_{j \ne i} \dij \pi_j$ with calibrated
$\dij$), and cooperative reward ($U_i = \frac{1}{n}\sum_j \pi_j$).
This parameterization enables \emph{reward-type ablation}: varying
reward mutuality while holding environment rules fixed, which reveals
whether algorithm rankings depend on the reward function or on the
environment itself. Non-ablation evaluation under a single reward
configuration (typically integrated) is also fully supported.

\paragraph{Game-theoretic oracle baselines.} The benchmark includes
seven oracle algorithms that compute analytically motivated reference
policies for each mechanism class: TR-1 Coopetitive Equilibrium, TR-2
Trust-Aware Equilibrium, and TR-3/TR-4 Nash and social-optimum bounds.
Oracle performance provides a theoretically grounded reference against
which learning algorithms can be evaluated without requiring a separate
``ground truth'' run.

\subsection{What this document specifies}

Part~I defines the benchmark. Section~\ref{sec:design} presents the
design principles. Section~\ref{sec:environments} specifies the
twenty environments. Section~\ref{sec:foundations} summarizes the
mathematical foundations from the four technical reports.
Section~\ref{sec:api} describes the three APIs.
Section~\ref{sec:algorithms} enumerates the algorithm suite and oracle
baselines. Section~\ref{sec:eval} presents the evaluation methodologies.
Section~\ref{sec:validation} reports the case study calibration and
validation. Section~\ref{sec:audit} describes the behavioral audit.
Section~\ref{sec:reproduce} describes the reproducibility package.

Part~II (Section~\ref{part:findings}) reports illustrative findings from
the reference evaluation. These findings are presented as demonstrations
of the benchmark's analytical utility, not as the benchmark's purpose;
extended treatments of individual findings appear in separate companion
papers.

Part~III (appendices) provides reference material: algorithm rankings,
network sensitivity analysis, case study discrimination, computational
cost breakdown, and dataset schemas.

\subsection{Intended audience and use cases}

\textsc{Coopetition-Gym v1} is designed to be useful to several
distinct research and practitioner communities whose methods,
priorities, and vocabularies only partially overlap. We list the
principal intended audiences below and describe, for each, how the
benchmark is expected to be used.

\paragraph{Machine learning and MARL researchers.} The benchmark
supports algorithm comparison and evaluation-robustness studies in
mixed-motive settings that existing cooperative-only or
competitive-only benchmarks do not cover. The reward-type ablation
methodology produces three algorithm rankings per environment, which
enables studies of whether published algorithm advantages are
robust to reward-function structure. The seven game-theoretic oracles
provide theoretically grounded reference points for claims about
cooperation, whereas single-algorithm or algorithm-vs-algorithm
comparisons do not. Researchers developing new algorithms in the
MADDPG, QMIX, MAPPO, SAC, or LOLA families will find the benchmark a
suitable diagnostic for mechanism-class-dependent performance.

\paragraph{Multi-agent systems and game theory researchers.}
The benchmark provides twenty general-sum stochastic games with
closed-form or numerically tractable equilibria. Researchers
studying coordination, coalition formation, equilibrium selection,
mechanism design, or computational game theory will find the
interdependence matrix $\dij$ a configurable parameter for varying relational
structure while holding game mechanics fixed. The benchmark's
game-theoretic oracles (\texttt{Oracle\_Nash},
\texttt{Oracle\_Loyalty}, \texttt{Oracle\_BoundedReciprocity}, and
others) can be used as reference solvers without engaging the RL
algorithm suite at all.

\paragraph{Behavioral economics and cognitive science researchers.}
The benchmark's formalisms implement the three-to-one negativity
bias in trust dynamics~\citep{rand2013human}, the public-goods
punishment mechanic~\citep{fehr2000cooperation}, the direct and
indirect reciprocity regimes~\citep{axelrod1984evolution,
nowak2006five}, and the commons governance design
principles~\citep{ostrom1990governing}. Researchers who wish to
ground their computational models in published behavioral-science
parameter values can adopt the benchmark's calibrated constants as
a starting point.

\paragraph{Strategic management and organizational research.} Four
of the twenty environments are calibrated to historically documented
coopetitive relationships: Samsung-Sony, Renault-Nissan, Apache HTTP
Server, and Apple iOS App Store. Researchers studying coopetition,
alliance management, platform strategy, or open-source governance
will find in these environments executable simulation models of
canonical cases from their literature.

\paragraph{Practitioners and policymakers.} The behavioral audit
methodology (\S\ref{sec:audit}) characterizes how a trained policy
responds to counterfactual cooperation levels and to temporal
deviation strategies. Practitioners deploying multi-agent systems
(platform designers, autonomous-vehicle coordination, algorithmic
trading, supply-chain optimization) can use the audit to evaluate
robustness to partner failure or adversarial manipulation before
deployment. Policymakers and AI-safety researchers can use the
same methodology to inspect how a candidate algorithm would
behave under exploitation pressures.

\paragraph{Educators and students.} The environments are pedagogically
tractable because they encode canonical game-theoretic and
coopetition problems (iterated Prisoner's Dilemma, public goods,
platform hold-up, gift exchange) in a uniform API. Course instructors
can use the benchmark to illustrate the classical dilemmas
computationally, to run small-scale student projects comparing
independent and centralized learning, or to build assignments around
the calibrated case studies. The reproducibility package provides a
single command-line entry point that a student can invoke to
regenerate every figure in this document.

\subsection{Content validation}

Every mathematical constant and equation in this document has been
verified against the source code. TR-1 and TR-2 formalisms live in
\texttt{coopetition\_gym/core/} (full implementations of
\citep{pant2025interdependence, pant2025trust}). TR-3 and TR-4
formalisms live in \texttt{coopetition\_gym/envs/}, specifically in
\texttt{envs/collective\_action\_envs.py} for TR-3 (the
\texttt{TR3Parameters} dataclass and associated team-production,
loyalty, and equilibrium functions) and \texttt{envs/reciprocity\_envs.py}
for TR-4 (the \texttt{TR4Parameters} dataclass, cooperation-signal,
memory-average, bounded-response, reciprocity-sensitivity, and
trust-gated-reciprocity functions). Both environment modules reference
the paper equation numbers directly in their source docstrings (TR-3
Eqs.~1--3; TR-4 Eqs.~19--25 and 44--45). The \texttt{core/} package
contains helper modules (\texttt{core/collective\_action.py},
\texttt{core/reciprocity.py}) with simplified support utilities; they
are not the authoritative implementations. Per-environment
specifications are extracted directly from \texttt{envs/} source files
and cross-checked against \texttt{experiments/config.py} for
study-configured overrides (for example, \texttt{SLCD-v0} with
horizon 40 in study vs source default 100). Case study validation
scores are drawn from the authoritative \texttt{TR\_validation/}
README files.

\subsection{Availability}

\begin{itemize}[leftmargin=*, itemsep=2pt]
\item \textbf{Code}: \url{https://github.com/vikpant/strategic-coopetition} (MIT license).
\item \textbf{Training dataset} (25{,}708 files):
  \url{https://huggingface.co/datasets/vikpant/coopetition-gym-v1} (CC-BY-4.0).
\item \textbf{Behavioral audit dataset} (1{,}116 files):
  \url{https://huggingface.co/datasets/vikpant/coopetition-gym-audit} (CC-BY-4.0).
\item \textbf{Croissant metadata} for both datasets:
  \url{https://github.com/vikpant/strategic-coopetition/tree/master/experiments/croissant}.
\end{itemize}

\section{Design Principles}
\label{sec:design}

Four design principles govern the benchmark: formal grounding, empirical
validation, reproducibility, and open artifacts. Each principle is a
response to a documented gap in the existing MARL benchmark literature
and is implemented at the level of the environment code, the dataset
schema, and the accompanying tooling. We articulate each principle,
motivate it by contrast with prevailing practice, and describe how the
benchmark operationalizes it.

\subsection{Formal grounding}

Every environment in \textsc{Coopetition-Gym v1} is derived from a
published mathematical formalism. The four source formalisms are the
technical reports~\citep{pant2025interdependence, pant2025trust,
pant2026collective, pant2026reciprocity} that formalize, respectively,
interdependence and complementarity, trust and reputation dynamics,
collective action and loyalty, and sequential interaction and
reciprocity. These formalisms are distinct but share a common
structural vocabulary: all define a payoff function
$\boldsymbol{\pi}(\mathbf{a})$ over cooperation actions and an
interdependence coefficient $\dij$ that governs cross-agent utility
coupling. They do not compete for explanatory territory; they address
complementary dimensions of coopetitive interaction.

The contrast with prevailing MARL benchmark practice is instructive.
Several widely used benchmarks specify environments by procedural
generation of tasks~\citep{leibo2021meltingpot, agapiou2022meltingpot}
or by inheritance from an external game
engine~\citep{samvelyan2019smac, ellis2023smacv2}. Both approaches
yield realistic environments but make the game-theoretic status of
each scenario (the Nash equilibrium correspondence, the Pareto
frontier, the presence or absence of stable cooperation) difficult
to characterize analytically. \textsc{Coopetition-Gym v1} inherits the
opposite trade-off: every environment has a closed-form or
numerically tractable equilibrium characterization derived in its
source technical report, at the cost of scenarios being more stylized
than those in engine-driven benchmarks. For a benchmark whose
scientific purpose is to study how algorithms handle coopetitive
mechanisms in isolation and in combination, we judge this trade-off
appropriate: controlled games with known solutions are what the
reward-type ablation methodology (\S\ref{sec:eval}) requires.

The shared structural vocabulary across the four TRs is not incidental.
It follows from the coopetition research
program~\citep{brandenburger1996coopetition, bengtsson2000coopetition,
gnyawali2011coopetition, bouncken2015coopetition} which treats
cooperation and competition as independent axes of relationship
structure rather than as endpoints of a one-dimensional spectrum.
Interdependence matrix $\dij$ encodes the cooperation axis; the payoff
function $\boldsymbol{\pi}(\mathbf{a})$ encodes the competition axis
and resource-allocation structure. A benchmark whose environments
share this vocabulary admits a cross-mechanism comparison that is
otherwise unavailable: one may ask whether an algorithm that learns
cooperation via trust (TR-2) also learns it via reciprocity (TR-4),
and the question has a precise answer because the algorithm is
evaluated against a payoff structure whose parameters have identical
meaning in both cases.

\subsection{Empirical validation}

Four of the twenty environments are calibrated to historically
documented coopetitive relationships. Calibration consists of
extracting $\dij$ values and other formalism parameters from
qualitative coding of archival sources, then verifying that the
calibrated environment produces simulation trajectories qualitatively
consistent with documented historical outcomes. The protocol follows
the Behavioral Correspondence schema defined in each TR's validation
suite: a sixty-item rubric coded by two independent analysts against
historical strategic-interaction records.

The four validated case studies are \texttt{SLCD-v0} (Samsung-Sony LCD
joint venture, $59/60 = 98.3\%$\footnote{The Samsung-Sony SLCD score
of $59/60 = 98.3\%$ reflects the CAiSE 2026 peer-reviewed version
of~\citep{pant2025interdependence}, which incorporated a one-point
refinement to the cooperation-dynamics criterion during peer review.
The earlier arXiv preprint reports $59/60 = 98.3\%$. We use the
peer-reviewed score throughout this technical report.}),
\texttt{RenaultNissan-v0}
(Renault-Nissan Alliance, $49/60 = 81.7\%$), \texttt{ApacheProject-v0}
(Apache HTTP Server community, $52/60 = 86.7\%$), and
\texttt{AppleAppStore-v0} (Apple iOS App Store ecosystem,
$48/55 = 87.3\%$). Validation scores and the
Calibration/Discrimination protocol are reported in
Section~\ref{sec:validation} and Appendix~\ref{app:case_studies}.

Calibration matters beyond face validity. In the experimental
economics tradition~\citep{fehr2000cooperation, rand2013human}, the
external-validity question for any stylized cooperation game is
whether its qualitative dynamics survive instantiation with real
parameter values. The four validated environments provide that
instantiation, at the cost of specificity: each case study is one
trajectory through parameter space, not a distribution. Users who
require population-level external validity should read the case
studies as existence proofs rather than as representative samples.

\subsection{Reproducibility}

All randomness in \textsc{Coopetition-Gym v1} is seeded through the
standard NumPy/PyTorch seeding protocol. Given a seed, algorithm
hyperparameters, and hardware within floating-point tolerance,
trajectories are deterministic; independent executions on
\texttt{ubuntu-latest} GitHub-Actions runners reproduce any released
trajectory to within relative error $1.7 \times 10^{-7}$ on the
canonical regression test (\texttt{Constant\_50} on
\texttt{TrustDilemma-v0}, seed $99$). The reproducibility package
(\S\ref{sec:reproduce}) provides a single command-line entry point
\texttt{experiments/campaign.py} that regenerates every row and
figure in Part II of this document from the released datasets.

Reproducibility is specified at three levels:
\begin{enumerate}[leftmargin=*, itemsep=2pt]
\item \emph{Byte-exact replay} of pre-computed trajectories from the
      released dataset, requiring only the MIT-licensed Python
      package and the CC-BY-4.0 dataset. Ten-minute runtime; no GPU
      required.
\item \emph{Single-environment training regression}: training a
      specified algorithm on a specified environment with released
      hyperparameters and a specified seed should produce a learning
      curve within one standard error of the released one.
      Hour-scale runtime; GPU recommended but not required.
\item \emph{Full-study regeneration}: reproducing the entire
      $25{,}708$-file training dataset and the $1{,}116$-file
      behavioral audit dataset. Multi-day runtime on a multi-GPU
      cluster; total compute cost is reported in
      Appendix~\ref{app:compute}.
\end{enumerate}
Provenance is tracked by git tag \texttt{v1.0.0}, which
captures the exact code state that produced the released datasets.
Documentation-only clarifications in later patch releases ($1.0.1+$)
do not change any computational behavior.

\subsection{Open artifacts}

The benchmark code is released under MIT license; datasets are
released under CC-BY-4.0. No proprietary dependencies, no gated
content, no behind-a-paywall artifacts. Researchers may extend the
benchmark, redistribute the datasets, and publish derivative work
without restriction beyond attribution. The code and datasets are
hosted on GitHub and HuggingFace respectively, both FAIR-aligned
repositories with stable identifiers and public download counts.
The release schedule is staggered: a baseline subset is released at
the time of the first conference submission; the full ablation
dataset follows after the associated arXiv timestamp, protecting
independent evaluation while ensuring open access within months
rather than years. The release schedule, the licensing terms, and the
reproducibility tiers above together define the open-artifacts surface
that downstream users interact with: anyone with an internet
connection can reproduce the headline result with a ten-minute
byte-exact replay, and anyone with a multi-GPU instance can reproduce
the entire reference evaluation end-to-end. No component of the
artifact set is gated by registration, paywall, or institutional
affiliation.

\section{The Twenty Environments}
\label{sec:environments}

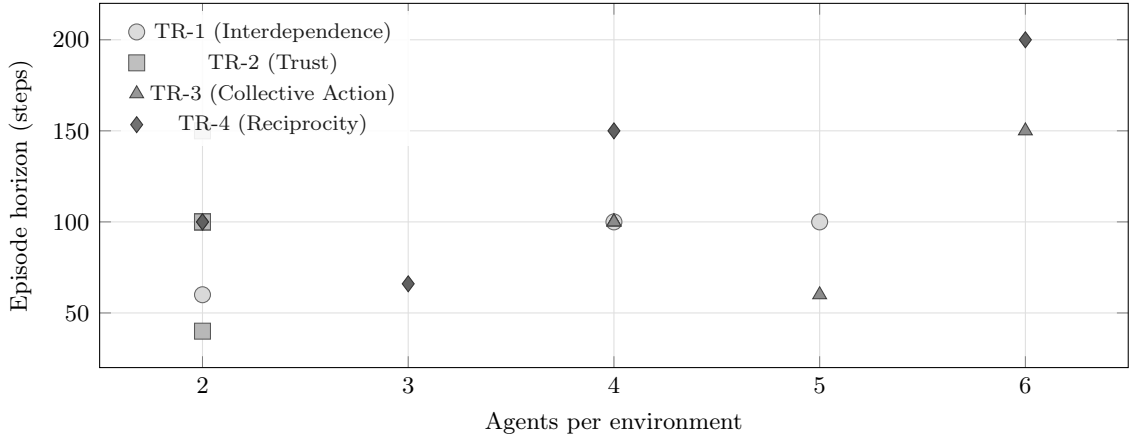
\begin{figure}[H]
\centering
\begin{tikzpicture}
\begin{axis}[
    width=0.92\textwidth,
    height=6.4cm,
    xlabel={Agents per environment},
    ylabel={Episode horizon (steps)},
    xlabel style={font=\footnotesize},
    ylabel style={font=\footnotesize},
    xticklabel style={font=\footnotesize},
    yticklabel style={font=\footnotesize},
    legend style={font=\scriptsize, at={(0.02,0.98)}, anchor=north west, draw=none, fill=white, fill opacity=0.9},
    grid=major, grid style={gray!25},
    xmin=1.5, xmax=6.5, ymin=20, ymax=220,
    xtick={2,3,4,5,6},
    only marks,
    mark size=3pt,
]
\addplot[mark=*, mark options={fill=black!15, draw=black!60}] coordinates {(2,100) (5,100) (4,100) (2,100) (2,60)};
\addplot[mark=square*, mark options={fill=black!28, draw=black!65}] coordinates {(2,100) (2,150) (2,40) (2,100) (2,100)};
\addplot[mark=triangle*, mark options={fill=black!45, draw=black!75}] coordinates {(4,100) (4,100) (6,150) (5,60) (4,100)};
\addplot[mark=diamond*, mark options={fill=black!62, draw=black!80}] coordinates {(2,100) (2,100) (4,150) (6,200) (3,66)};
\legend{TR-1 (Interdependence), TR-2 (Trust), TR-3 (Collective Action), TR-4 (Reciprocity)}
\end{axis}
\end{tikzpicture}
\caption{Coverage of the twenty environments along the
agent-count and episode-horizon axes. Each marker is one
environment; marker shape encodes the mechanism-class tier and
marker shading encodes the same. The suite spans $2$-agent dyadic
games to $6$-agent commons, and short ($n=40$) to long ($n=200$)
episode horizons. TR-1 and TR-2 are dyadic-heavy; TR-3 and TR-4
populate the multi-agent and longer-horizon regions. Several
markers overlap at $(2, 100)$ (the canonical dyadic, century-step
configuration); these are the canonical-form environments around
which the package's other environments vary.}
\label{fig:env_coverage}
\end{figure}

\textsc{Coopetition-Gym v1} comprises twenty environments organized into
four mechanism-class tiers (TR-1 through TR-4) of five environments each.
The tiers are introduced one at a time in
\S\ref{subsec:tr1}--\S\ref{subsec:tr4}; we state the formal constructs
of each tier when it is reached. Table~\ref{tab:envs} summarizes the
environments and Figure~\ref{fig:env_coverage} plots their coverage along
the agent-count and episode-horizon axes.

\begin{table}[H]
\centering
\small
\caption{The twenty environments of \textsc{Coopetition-Gym v1}.}
\label{tab:envs}
\begin{tabular}{llrrl}
\toprule
Tier & Environment & Agents & Horizon & Notes \\
\midrule
\multirow{5}{*}{TR-1} & \texttt{PartnerHoldUp-v0}           & 2 & 100 & Asymmetric vertical relationship \\
                      & \texttt{PlatformEcosystem-v0}       & 5 & 100 & Platform + N developers \\
                      & \texttt{DynamicPartnerSelection-v0} & 4 & 100 & Reputation-based matching \\
                      & \texttt{SynergySearch-v0}           & 2 & 100 & Hidden complementarity \\
                      & \texttt{RenaultNissan-v0}           & 2 & 60  & Validated: 49/60 (81.7\%) \\
\midrule
\multirow{5}{*}{TR-2} & \texttt{TrustDilemma-v0}            & 2 & 100 & Continuous iterated PD with trust \\
                      & \texttt{RecoveryRace-v0}            & 2 & 150 & Post-crisis trust recovery \\
                      & \texttt{SLCD-v0}                    & 2 & 40  & Validated: 58/60 (98.3\%) \\
                      & \texttt{CooperativeNegotiation-v0}  & 2 & 100 & Multi-round negotiation \\
                      & \texttt{ReputationMarket-v0}        & 2 & 100 & Public reputation market \\
\midrule
\multirow{5}{*}{TR-3} & \texttt{TeamProduction-v0}          & 4 & 100 & Free-rider baseline \\
                      & \texttt{LoyaltyTeam-v0}             & 4 & 100 & Above-Nash via loyalty \\
                      & \texttt{CoalitionFormation-v0}      & 6 & 150 & Entry/exit dynamics \\
                      & \texttt{ApacheProject-v0}           & 5 & 60  & Validated: 52/60 (86.7\%) \\
                      & \texttt{PublicGoods-v0}             & 4 & 100 & Classic public goods with punishment \\
\midrule
\multirow{5}{*}{TR-4} & \texttt{ReciprocalDilemma-v0}       & 2 & 100 & Direct reciprocity with memory \\
                      & \texttt{GiftExchange-v0}            & 2 & 100 & Asymmetric employer-worker \\
                      & \texttt{IndirectReciprocity-v0}     & 4 & 150 & Image-scoring reputation \\
                      & \texttt{GraduatedSanction-v0}       & 6 & 200 & Commons with graduated punishment \\
                      & \texttt{AppleAppStore-v0}           & 3 & 66  & Validated: 48/55 (87.3\%) \\
\bottomrule
\end{tabular}
\end{table}

\subsection{Tier TR-1: Interdependence and Complementarity}
\label{subsec:tr1}

Environments in TR-1 implement the formalism
of~\citet{pant2025interdependence} (arXiv:2510.18802). Two
constructs from that report carry the formal weight of the tier and
of every environment in the package: the \emph{interdependence
matrix} $\dij$ and the \emph{synergy multiplier} $\gamma$. We
introduce both in detail because they are central to every reward
expression downstream.

\paragraph{The interdependence matrix $\dij$.} The matrix
$\mathbf{D} \in [0,1]^{n \times n}$ has entries $\dij \in [0, 1]$
(read: ``$D$-sub-$i$-$j$'') indexed by ordered agent pairs. Each
entry $\dij$ specifies the weight that agent $i$ places on agent
$j$'s payoff in its own utility: roughly, ``how much of agent $j$'s
gain or loss flows into agent $i$'s reward.'' At one extreme,
$\dij = 0$ means agent $i$ is indifferent to agent $j$'s payoff
(purely self-interested setting). At the other extreme, $\dij = 1$
means agent $i$ values agent $j$'s payoff as much as its own
(perfect alignment). Mixed-motive coopetition occupies the interior
$0 < \dij < 1$.

The matrix is in general \emph{asymmetric}: $\dij \ne D_{ji}$ is
permitted and is the typical case in real coopetitive
relationships. The Samsung-Sony LCD joint venture, for example, is
calibrated as $D_{\text{Sony}, \text{Samsung}} = 0.86$,
$D_{\text{Samsung}, \text{Sony}} = 0.64$ in
\citet{pant2025interdependence}; the asymmetry reflects Sony's
heavier dependence on Samsung's fabrication capacity than the
reverse. The diagonal entries $D_{ii}$ are conventionally undefined
or set to $1$ (an agent always values its own payoff in full).
$\dij$ enters every utility expression in the package through the
weighted partner-payoff term $\sum_{j \ne i} \dij \cdot \pi_j(\mathbf{a})$
of the integrated utility (Eq.~\ref{eq:integrated_utility}); the
\emph{reward-type ablation methodology} (\S\ref{sec:eval}) of the
package varies precisely whether and how this matrix is allowed to
modulate the reward signal.

\paragraph{The synergy multiplier $\gamma$.} The scalar
$\gamma \in [0, 1]$ is the weight placed on the joint-coordination
bonus term $g(\mathbf{a})$ in the total-value expression
(Eq.~\ref{eq:total_value}). Whereas $\dij$ encodes the
\emph{bilateral} (pairwise) coupling between agents, $\gamma$
encodes a \emph{multilateral} coordination premium: when every
agent contributes a positive cooperation level the geometric-mean
synergy $g(\mathbf{a}) = (\prod_i a_i)^{1/n}$ is nontrivial and
adds to every agent's utility, whereas when any single agent
free-rides ($a_i = 0$) the geometric mean collapses to zero and the
entire $\gamma g(\mathbf{a})$ term vanishes. This weakest-link
property is what distinguishes the synergy multiplier from a
naive additive cooperation bonus: a single defector destroys the
multilateral premium for all agents, not merely her own share.
The value $\gamma = 0.65$ is calibrated from the Samsung-Sony LCD
case study and is held fixed across the v1 environments unless an
environment explicitly overrides it.

\paragraph{Tier-specific specialization.} TR-1 environments
represent coopetitive relationships in which the payoff structure
encoded by $\dij$ and $\gamma$ is stable over time, so the learning
challenge lies entirely in the agents' action choices rather than
in any evolving trust or loyalty state. The remaining tiers (TR-2
through TR-4) inherit the same $\dij$ and $\gamma$ constructs but
introduce additional state (trust $T_{ij}$, loyalty $\theta_i$,
reciprocity $\rho_{ij}$) that evolves over the course of an episode.

The tier covers three canonical coopetitive relationship archetypes:
asymmetric vertical coupling (\texttt{PartnerHoldUp-v0}), multi-sided
platform coordination (\texttt{PlatformEcosystem-v0},
\texttt{DynamicPartnerSelection-v0}), and discovery-of-complementarity
under uncertainty (\texttt{SynergySearch-v0}). The validated case
study is the Samsung-Sony LCD joint venture (\texttt{RenaultNissan-v0}
in TR-2 is a second calibrated dyad but grounded in trust dynamics).
The tier is well-suited to users who want to study how algorithms
handle static coopetitive coupling before introducing the
confounding effect of dynamic relational state.

\subsection{Tier TR-2: Trust and Reputation Dynamics}
\label{subsec:tr2}

Environments in TR-2 implement the two-layer trust model of
\citet{pant2025trust} (arXiv:2510.24909). Immediate trust $T_{ij}$
between agent $i$ and agent $j$ is a scalar on $[0, 1]$ that updates
each step based on whether agent $j$'s cooperation action deviated
upward or downward from a baseline. Updates are asymmetric with a
$3{:}1$ negativity bias ($\lambda^- = 0.30$, $\lambda^+ = 0.10$),
meaning that trust erodes three times faster in response to
defection than it builds in response to cooperation. The
empirically-observed ``three good for one bad'' ratio in human
relational trust~\citep{rand2013human} is the motivation for this
asymmetry. A second layer, reputation $R_{ij}$, accumulates the
immediate trust trajectory under exponential smoothing and represents
the long-memory component of the trust relationship.

Taken together these dynamics expose agents to trajectory-dependent
payoffs in which early behavior shapes later payoff possibilities.
The TR-2 tier is the most mature of the four tiers in v1 because the
full formalism is implemented in the \texttt{core/trust\_dynamics.py}
module, and it is the tier on which the benchmark's behavioral audit
produces the sharpest discrimination between algorithms. Applications
include research on trust-aware algorithm design, reputation-market
analysis, post-crisis recovery (the \texttt{RecoveryRace-v0}
environment), and the Samsung-Sony liquid-crystal-display
calibration that anchors the tier's empirical validity.

\subsection{Tier TR-3: Collective Action and Loyalty}
\label{subsec:tr3}

Environments in TR-3 implement the collective action formalism of
\citet{pant2026collective} (arXiv:2601.16237). The central construct
is team production with loyalty accumulation: a group of agents
each choose a contribution level, team output rises super-linearly
with total contribution via a productivity factor $\omega$ and a
returns-to-scale parameter $\beta$, and each agent's share of team
output is modulated by a loyalty score $\theta_i$ that grows when
the agent sustains cooperation over a memory window of $k = 3$--$10$
steps. Loyalty creates a path to above-Nash cooperation that
fixed-action policies cannot exploit, because loyalty must be
accumulated through sustained contribution and is reset by
defection.

The tier covers the canonical collective action regimes studied in
institutional economics since Ostrom~\citep{ostrom1990governing}:
the standard free-rider baseline (\texttt{TeamProduction-v0}),
loyalty-driven above-Nash cooperation (\texttt{LoyaltyTeam-v0}),
entry-exit coalition dynamics (\texttt{CoalitionFormation-v0}), the
classical public goods game with punishment option
(\texttt{PublicGoods-v0}; \citealp{fehr2000cooperation}), and an
empirically validated open-source software commons calibrated to
the Apache HTTP Server community (\texttt{ApacheProject-v0}).
TR-3 environments have the highest agent counts ($n = 4$--$6$) in
the suite and exercise the free-rider-dynamics regime most
extensively. They are appropriate for research on collective action
problems, institutional design, commons governance, and algorithms
whose inductive biases target sustained cooperation.

\subsection{Tier TR-4: Sequential Interaction and Reciprocity}
\label{subsec:tr4}

Environments in TR-4 implement the reciprocity formalism of
\citet{pant2026reciprocity} (arXiv:2604.01240). The core construct is
memory-bounded reciprocity with graduated sanctions: each agent
maintains a finite-window memory of its partners' recent actions,
computes a reciprocity signal from that memory, and adjusts its own
cooperation accordingly. Sanctions escalate in response to continued
defection and de-escalate in response to renewed cooperation, so
the tier formalizes the direct reciprocity principles dating to
Axelrod's tit-for-tat work~\citep{axelrod1984evolution} and the
subsequent generalizations to graduated sanctions and indirect
reciprocity~\citep{nowak2006five}.

The tier covers the canonical reciprocity regimes: direct
reciprocity between two agents (\texttt{ReciprocalDilemma-v0}),
asymmetric exchange where one side holds structural bargaining power
(\texttt{GiftExchange-v0} formalizing the employer-worker gift
exchange), indirect reciprocity mediated by image scoring across
$n = 4$ agents (\texttt{IndirectReciprocity-v0}), graduated
sanctions over a commons with $n = 6$ agents
(\texttt{GraduatedSanction-v0}, the Ostrom design-principle
instantiation), and the empirically calibrated Apple iOS App Store
platform ecosystem (\texttt{AppleAppStore-v0}, the largest
validated n-agent case study in the suite). TR-4 environments are
the most temporally structured in the suite because the reciprocity
state is a function of a window of prior actions rather than a
single instantaneous update, and they are therefore especially
suited to research on policy architectures with explicit memory or
recurrence.

\section{Mathematical Foundations}
\label{sec:foundations}

The four formalisms share a common structural vocabulary. This section
summarizes the equations that directly govern environment dynamics. The
full derivations appear in the four source technical
reports~\citep{pant2025interdependence, pant2025trust, pant2026collective,
pant2026reciprocity}.

\subsection{Theoretical framework}

The four technical reports collectively specify a family of stochastic
games parameterized by an interdependence matrix $\dij$ and mechanism-
specific dynamics. In the general form, each agent $i \in \{1, \ldots, n\}$
at step $t$ chooses a cooperation action $a_i(t) \in [0, e_i]$. The
environment generates a base payoff vector $\boldsymbol{\pi}(\mathbf{a})$
whose form depends on the mechanism class; the integrated utility for
agent $i$ is:
\begin{equation}
U_i(\mathbf{a}) = \pi_i(\mathbf{a}) + \sum_{j \ne i} \dij \cdot \pi_j(\mathbf{a}) + M_i(\mathbf{a}, \mathbf{h}_t)
\label{eq:general_utility}
\end{equation}
\vspace{-0.4em}
\hfill\textit{\small [Generalized form composing
\cite{pant2025interdependence}, Eq.~1, with the mechanism-specific
modifiers introduced in \cite{pant2025trust},
\cite{pant2026collective}, and \cite{pant2026reciprocity}]}
\vspace{0.5em}

where $\mathbf{h}_t$ is the history of prior actions and $M_i$ is a
mechanism-specific modifier capturing TR-2 trust state, TR-3 loyalty
state, or TR-4 reciprocity state. For TR-1 environments $M_i = 0$;
for TR-2, $M_i$ encodes trust-gated payoff adjustment; for TR-3, $M_i$
encodes the loyalty modifier; for TR-4, $M_i$ encodes the reciprocity
modifier (Equation~\ref{eq:tr4_utility_recip} below).

The reward signal used for policy learning is a configurable aggregation
of $\boldsymbol{\pi}$:
\begin{equation}
R_i^{\text{private}}(\mathbf{a}) = \pi_i(\mathbf{a}), \qquad
R_i^{\text{integrated}}(\mathbf{a}) = U_i(\mathbf{a}), \qquad
R_i^{\text{cooperative}}(\mathbf{a}) = \frac{1}{n} \sum_j \pi_j(\mathbf{a})
\end{equation}
The reward-type ablation methodology (\S\ref{sec:eval}) varies this
aggregation while holding $\boldsymbol{\pi}$ fixed. This separation of
the payoff layer from the reward layer is novel in the MARL benchmark
literature; it enables systematic investigation of how learning
outcomes depend on the reward function's mutuality structure separately
from the environment's transition structure.

The game-theoretic status of this family is that of a general-sum
stochastic game~\citep{shapley1953stochastic, shoham2008multiagent}.
Symmetric Nash equilibria exist for TR-1 and TR-3 environments in
closed form (derived below for TR-3); equilibria for TR-2 and TR-4
require numerical computation because the state dynamics (trust,
reciprocity) create non-stationary best-response mappings. The
benchmark provides oracle algorithms for both closed-form and
numerically computed equilibria (\S\ref{sec:algorithms}).

\subsection{Payoff functions (TR-1)}

The \texttt{value\_functions.py} module implements two individual-value
specifications. The logarithmic form (default) is used for all
environments in v1:

\paragraph{Logarithmic individual value (default).}
\begin{equation}
f_i(a_i) = \theta \ln(1 + a_i), \quad \theta = 20.0
\label{eq:individual_value_log}
\end{equation}
\vspace{-0.4em}
\hfill\textit{\small [From \cite{pant2025interdependence}, Eq.~13]}
\vspace{0.5em}

\paragraph{Power individual value (alternative).}
\begin{equation}
f_i(a_i) = a_i^\beta, \quad \beta = 0.75
\label{eq:individual_value_power}
\end{equation}
\vspace{-0.4em}
\hfill\textit{\small [From \cite{pant2025interdependence}, Eq.~10, \S7.4]}
\vspace{0.5em}

Selected via \texttt{ValueFunctionParameters.specification}
$\in$ \{\texttt{LOGARITHMIC}, \texttt{POWER}\}. Both specifications
exhibit diminishing marginal returns: $f_i'(a_i) > 0$ and
$f_i''(a_i) < 0$ for $a_i > 0$, and $\lim_{a_i \to 0^+} f_i'(a_i)
= +\infty$ for the logarithmic form and $+\infty$ for the power form
when $\beta < 1$. The Inada-type boundary condition on the derivative
guarantees that each agent's individually optimal cooperation level
is strictly positive in the absence of extraction cost, which is the
classical precondition for cooperation to be an interior equilibrium
phenomenon rather than a corner case~\citep{shoham2008multiagent}.
The logarithmic form was preferred in the S-LCD calibration on
qualitative fit grounds; the power form is retained as the TR-1
\S7.4 alternative for users who want to vary curvature. We discuss
the functional-form choice further below.

The choice between logarithmic and power specifications is familiar
from consumer and production theory. Logarithmic utility implies
constant relative risk aversion equal to $1$ and is the limiting case
of CRRA utility as $\sigma \to 1$; power utility $a_i^\beta$ with
$\beta < 1$ corresponds to a Cobb--Douglas production technology with
a single input and constant returns to scale only as $\beta \to 1$.
Both forms appear in the classical cooperation-games
literature~\citep{fehr2000cooperation, nowak2006five} because both
preserve the qualitative equilibrium structure of the underlying
dilemma; neither dominates the other on theoretical grounds. The
benchmark's choice to expose both as a configurable parameter
reflects the TR-1 author's view that the mechanism of interest --
diminishing marginal returns to cooperation -- is robust to the
specific functional form, and that a benchmark should not privilege
one parameterization of the mechanism over another.

\paragraph{Synergy (geometric mean).}
\begin{equation}
g(\mathbf{a}) = \left(\prod_{i=1}^{n} a_i\right)^{1/n}
\label{eq:synergy}
\end{equation}
\vspace{-0.4em}
\hfill\textit{\small [From \cite{pant2025interdependence}, Eq.~14]}
\vspace{0.5em}

\paragraph{Total value.}
\begin{equation}
V(\mathbf{a} \mid \gamma) = \sum_i f_i(a_i) + \gamma \cdot g(\mathbf{a})
\label{eq:total_value}
\end{equation}
\vspace{-0.4em}
\hfill\textit{\small [From \cite{pant2025interdependence}, Eq.~16]}
\vspace{0.5em}

with $\gamma = 0.65$ calibrated for the S-LCD case study. The
geometric-mean synergy is strongly super-additive (zero cooperation
by any agent zeroes the synergy term), and in this sense enforces
a weakest-link complementarity~\citep{brandenburger1996coopetition,
gnyawali2011coopetition}. Unlike an additive-synergy specification
$\gamma \sum_i a_i$, the geometric mean cannot be dominated by a
single agent's contribution and therefore preserves the benchmark's
character as a coopetitive (rather than additive-value) environment
regardless of the degree of agent specialization. The calibrated
$\gamma = 0.65$ places the synergy weight below the sum of individual
values at interior equilibria (ensuring $f_i$ retains economic
meaning) but high enough that full cooperation Pareto-dominates the
Nash equilibrium by a wide margin (ensuring the environment is not
trivial). The \texttt{ValueFunctionParameters} dataclass validates
$\gamma \in [0, 1]$, $\theta > 0$, and $\beta \in (0, 1]$ at
initialization (\texttt{coopetition\_gym/core/value\_functions.py},
lines 87--94).

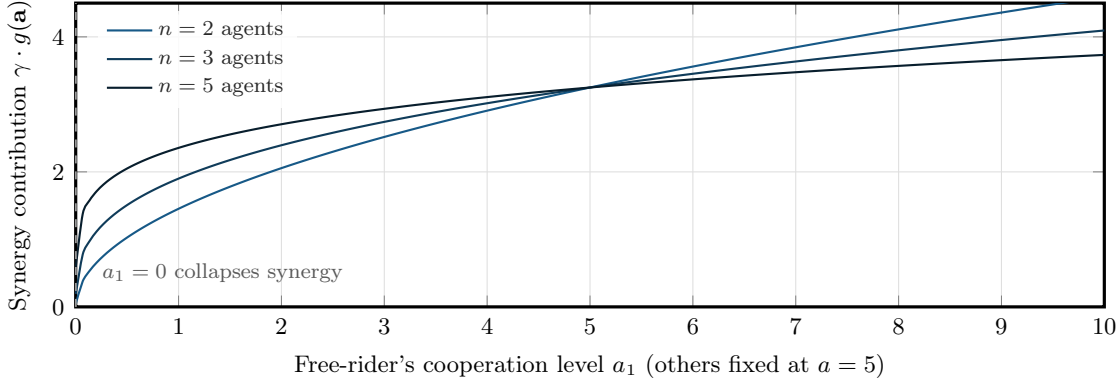
\begin{figure}[ht]
\centering
\begin{tikzpicture}
\begin{axis}[
    width=0.92\textwidth,
    height=5.6cm,
    xlabel={Free-rider's cooperation level $a_1$ (others fixed at $a = 5$)},
    ylabel={Synergy contribution $\gamma \cdot g(\mathbf{a})$},
    xlabel style={font=\footnotesize},
    ylabel style={font=\footnotesize},
    xticklabel style={font=\footnotesize},
    yticklabel style={font=\footnotesize},
    legend style={font=\scriptsize, at={(0.02,0.98)}, anchor=north west, draw=none, fill=white, fill opacity=0.85},
    grid=major, grid style={gray!25},
    xmin=0, xmax=10, ymin=0, ymax=4.5,
    enlargelimits=false,
    no markers, smooth, very thick,
    domain=0.001:10, samples=150,
]
\addplot[trustcolor!75!black, thick] {0.65 * sqrt(x*5)};
\addplot[trustcolor!50!black, thick] {0.65 * (x*25)^(1/3)};
\addplot[trustcolor!25!black, thick] {0.65 * (x*625)^(1/5)};
\addplot[black!40, dashed, thick] coordinates {(0,0) (0,4.5)};
\node[font=\scriptsize, anchor=west, black!65] at (axis cs:0.15, 0.50) {$a_1 = 0$ collapses synergy};
\legend{$n = 2$ agents, $n = 3$ agents, $n = 5$ agents}
\end{axis}
\end{tikzpicture}
\caption{Weakest-link property of the geometric-mean synergy.
Holding all other agents at cooperation level $a = 5$ and varying
the focal agent's contribution $a_1 \in [0, 10]$, the synergy
contribution $\gamma \cdot g(\mathbf{a})$ is shown for three
agent-count configurations. As $a_1 \to 0$ the synergy term
collapses to zero regardless of how many other agents are
cooperating, illustrating the weakest-link property: a single
defector destroys the multilateral premium for all agents. The
function is concave in $a_1$, so the marginal synergy benefit of
additional cooperation diminishes as $a_1$ grows; this concavity is
what makes the cooperative equilibrium interior rather than corner.}
\label{fig:synergy_weakest_link}
\end{figure}

\subsection{Integrated utility (TR-1)}

The integrated utility composes the per-agent private payoff
$\pi_i(\mathbf{a})$ with a $\dij$-weighted contribution from every
other agent's payoff, yielding the central reward expression that
all four mechanism classes inherit:
\begin{equation}
U_i(\mathbf{a}) = \pi_i(\mathbf{a}) + \sum_{j \ne i} \dij \, \pi_j(\mathbf{a})
\label{eq:integrated_utility}
\end{equation}
\vspace{-0.4em}
\hfill\textit{\small [From \cite{pant2025interdependence}, Eq.~1]}
\vspace{0.5em}

The package reuses this expression as the canonical reward layer
over which the reward-type ablation methodology is parameterized.
The reward-type ablation methodology (\S\ref{sec:eval}) varies
\emph{whether} this $\dij$-weighted summation is applied, replacing
the integrated form with the private form ($\dij \equiv 0$, recovering
$U_i = \pi_i$) or with the cooperative form ($U_i = \frac{1}{n}
\sum_j \pi_j$, the team-mean reward) while leaving the underlying
payoff vector $\boldsymbol{\pi}(\mathbf{a})$ unchanged. Holding
$\boldsymbol{\pi}$ fixed across the three reward configurations is
what isolates reward-structure effects from environment-mechanism
effects in subsequent empirical analyses.

\begin{figure}[ht]
\centering
\begin{tikzpicture}
\begin{axis}[
    width=6.8cm, height=6.8cm,
    enlargelimits=false,
    axis on top,
    xtick={0,1}, ytick={0,1},
    xticklabels={Samsung, Sony},
    yticklabels={Samsung, Sony},
    xlabel={Agent $j$ (whose payoff is weighted)}, xlabel style={font=\footnotesize},
    ylabel={Agent $i$ (placing the weight)}, ylabel style={font=\footnotesize},
    xticklabel style={font=\footnotesize}, yticklabel style={font=\footnotesize},
    colormap={softblues}{rgb255=(247,251,255) rgb255=(222,235,247) rgb255=(198,219,239) rgb255=(158,202,225) rgb255=(107,174,214) rgb255=(66,146,198) rgb255=(33,113,181)},
    colorbar,
    colorbar style={
      width=0.30cm,
      yticklabel style={font=\scriptsize},
      title={$\dij$}, title style={font=\scriptsize}
    },
    point meta min=0, point meta max=1,
    nodes near coords*={\pgfmathprintnumber\pgfplotspointmeta},
    every node near coord/.append style={font=\small, color=black, anchor=center, /pgf/number format/fixed, /pgf/number format/precision=2},
]
\addplot [matrix plot, mesh/cols=2, point meta=explicit] coordinates {
  (0,0) [1.00] (1,0) [0.64]
  (0,1) [0.86] (1,1) [1.00]
};
\end{axis}
\end{tikzpicture}
\caption{Interdependence matrix $\mathbf{D}$ for the Samsung-Sony LCD
joint venture (\texttt{SLCD-v0}), calibrated from documented
strategic dependencies in \citet{pant2025interdependence}. Row $i$
indexes the agent placing the weight; column $j$ indexes the agent
whose payoff is weighted. The diagonal $\dij = 1$ entries reflect
that each agent fully values its own payoff. The off-diagonal
entries are asymmetric: Sony places weight $0.86$ on Samsung's
payoff, while Samsung places weight $0.64$ on Sony's payoff. The
asymmetry encodes Sony's heavier dependence on Samsung's
fabrication capacity. The integrated reward function (Eq.~\ref{eq:integrated_utility})
weights partner payoffs by these off-diagonal entries.}
\label{fig:dij_heatmap}
\end{figure}
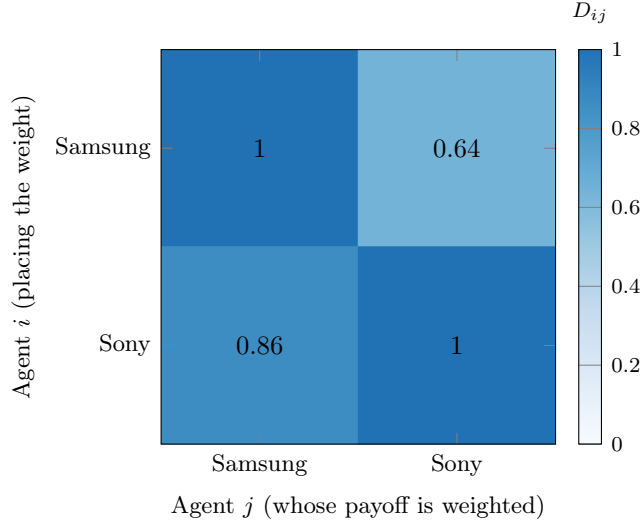

The coefficient $\dij \in [0,1]$ quantifies the weight agent $i$
places on agent $j$'s payoff. This formulation is not an
assumption about agent preferences in the sense of
\citet{harsanyi1967games}, where utility functions describe
sovereign preference structure; rather it encodes the institutional
or relational structure of the strategic context. Two agents whose
operations are tightly coupled (e.g., platform and developer
ecosystem) have high $\dij$ because each agent's welfare depends
substantially on the other's payoff through shared production,
cross-subsidization, or joint exposure to exogenous risk. Two
agents whose operations are nearly independent have low $\dij$ even
if they are in principle benevolent toward each other.

Symmetric relationships satisfy $\dij = D_{ji}$; asymmetric
relationships (e.g., platform--developer) satisfy $\dij \ne D_{ji}$.
The S-LCD calibration is $D_{\text{Samsung,Sony}} = 0.64$,
$D_{\text{Sony,Samsung}} = 0.86$. The asymmetry reflects that Sony's
LCD supply was dependent on Samsung's fabrication capacity to a
greater degree than the reverse, a structural fact independently
documented in the business-history record. Asymmetric $\dij$ is a
departure from the symmetric two-player stochastic-game templates
that dominate the MARL benchmark
literature~\citep{leibo2017ssd, samvelyan2019smac} and is what
enables the AI-3 paper's D$_{ij}$-scaling findings
(\S\ref{sec:dij}) to characterize how an algorithm's learned
policy behaves as a function of structural-coupling strength.

The default observation configuration includes agent $i$'s row
$D_{i,:}$ of the interdependence matrix as part of its observation
vector (\texttt{interdependence\_visible=True} in
\texttt{ObservationConfig}). An alternative configuration that
hides $D$ from the observation supports a separate line of
investigation in which the interdependence structure is a hidden
prior to be inferred from trajectories rather than a visible
environment attribute; that configuration is the design target for
the AI-8 coopetition-aware algorithm paper~\citep{repro}.

\subsection{Trust dynamics (TR-2, full implementation in v1)}

\paragraph{Implementation status.} The \texttt{trust\_dynamics.py}
module is a full implementation of the TR-2
formalism~\citep{pant2025trust}, not a skeleton. All core equations
below appear in the source code verbatim; their presence in v1
makes TR-2 environments the most mature tier of the benchmark.

\paragraph{Cooperation signal.}
\begin{equation}
s_{ij} = \tanh(\kappa \cdot (a_j - \bar{a}_j)), \quad \kappa = 1.0
\label{eq:coop_signal}
\end{equation}
\vspace{-0.4em}
\hfill\textit{\small [From \cite{pant2025trust}, Eq.~7]}
\vspace{0.5em}

where $\bar{a}_j$ is agent $j$'s baseline cooperation level (computed
by the environment from endowments and prior actions) and
$\kappa > 0$ is the \emph{response-sensitivity parameter} (sometimes
called the trust-update gain). With $\kappa = 1.0$ the cooperation
signal saturates at $\pm 1$ when the partner's deviation reaches
roughly $\pm 1.5$ units; larger $\kappa$ produces faster saturation
(the tanh becomes step-like and trust updates discretize), smaller
$\kappa$ produces slower saturation (trust updates approach a linear
response). The hyperbolic tangent bounds $s_{ij} \in (-1, 1)$ and
provides a smooth sign-preserving compression of the raw deviation
$a_j - \bar{a}_j$.
The concave shape of $\tanh$ on the positive axis implies that the
trust-building signal diminishes as the deviation grows, a saturation
property consistent with the psychometric literature on
trust~\citep{rand2013human}: a partner who cooperates at twice the
baseline does not increase trust by twice as much as a partner who
cooperates at $1.5\times$ the baseline. The \texttt{TrustParameters}
dataclass enforces $\kappa > 0$ (\texttt{trust\_dynamics.py}
line 109).

\paragraph{Asymmetric update with negativity bias.}
\begin{equation}
\begin{aligned}
T_{ij}(t{+}1) =\ & T_{ij}(t) + \lambda^+ \max(0, s) \cdot (1 - T_{ij}(t)) \\
& - \lambda^- \max(0, -s) \cdot T_{ij}(t)
\end{aligned}
\label{eq:trust_update}
\end{equation}
\vspace{-0.4em}
\hfill\textit{\small [From \cite{pant2025trust}, Eq.~12]}
\vspace{0.5em}

with $\lambda^+ = 0.10$ (the \emph{trust-building rate}: the
fraction by which the gap $1 - T_{ij}$ closes when the partner
behaves cooperatively at a unit signal) and $\lambda^- = 0.30$ (the
\emph{trust-erosion rate}: the fraction by which the current trust
level $T_{ij}$ is reduced when the partner defects at a unit signal).
The $3{:}1$ ratio $\lambda^- / \lambda^+ = 3.0$ encodes
\emph{negativity bias}: trust erodes three times faster than it
builds. This asymmetry is
empirically supported across multiple behavioral domains, including
observational studies of relational trust in marriages, reputation
dynamics in repeated games, and consumer-trust effects of
product failures~\citep{rand2013human}; the $3{:}1$ magnitude is
within the range reported in that literature. The consequence for
learning is that an agent whose policy occasionally defects from an
otherwise cooperative trajectory incurs a disproportionate penalty
that persists for many steps; this is a different dynamic from the
symmetric-update model that underlies most MARL social-dilemma
work~\citep{leibo2017ssd} and is one of the reasons TR-2
environments discriminate so sharply between algorithm paradigms
(\S\ref{sec:ctde}).

The multiplicative $(1 - T_{ij})$ factor in the building term ensures
$T_{ij} \le 1$; the multiplicative $T_{ij}$ factor in the erosion
term ensures $T_{ij} \ge 0$. Taken together these multiplicative
bounds make Eq.~\ref{eq:trust_update} a bounded-range dynamic system
whose fixed points at $s_{ij} = 0$ are the entire interval
$[0, 1]$ (any constant trust level is stationary under a zero
cooperation signal). Transient dynamics therefore encode the entire
informational content of the trust layer; an algorithm that cannot
condition its policy on these transients will fail to exploit the
coopetitive structure of TR-2 environments. The dataclass validates
$\lambda^+, \lambda^- \in (0, 1]$ at initialization
(\texttt{trust\_dynamics.py} lines 99--102).

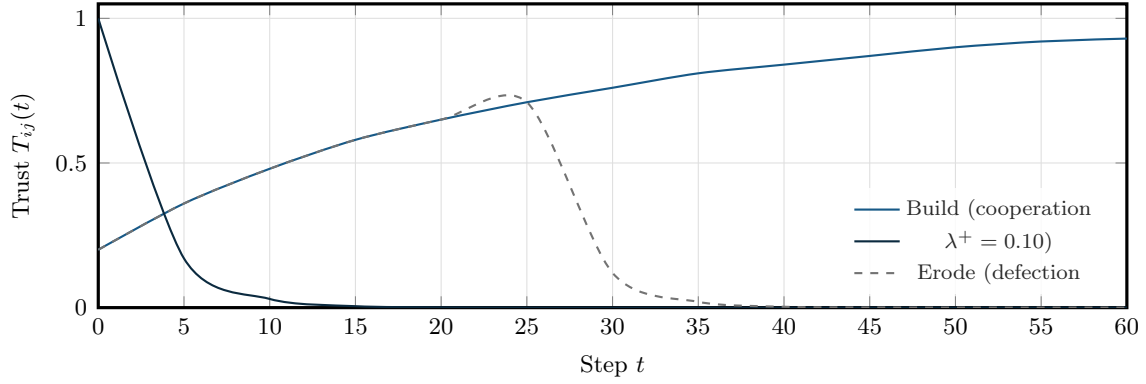
\begin{figure}[ht]
\centering
\begin{tikzpicture}
\begin{axis}[
    width=0.92\textwidth,
    height=5.6cm,
    xlabel={Step $t$},
    ylabel={Trust $T_{ij}(t)$},
    xlabel style={font=\footnotesize},
    ylabel style={font=\footnotesize},
    xticklabel style={font=\footnotesize},
    yticklabel style={font=\footnotesize},
    legend style={font=\scriptsize, at={(0.98,0.04)}, anchor=south east, draw=none, fill=white, fill opacity=0.85},
    grid=major, grid style={gray!25},
    xmin=0, xmax=60, ymin=0, ymax=1.05,
    enlargelimits=false,
    no markers, smooth, very thick,
]
\addplot[trustcolor!75!black, thick] coordinates {
  (0,0.20) (5,0.36) (10,0.48) (15,0.58) (20,0.65) (25,0.71)
  (30,0.76) (35,0.81) (40,0.84) (45,0.87) (50,0.90) (55,0.92) (60,0.93)
};
\addplot[trustcolor!35!black, thick] coordinates {
  (0,1.00) (5,0.17) (10,0.03) (15,0.005) (20,0.001)
  (25,0.0002) (30,0.00003) (60,0.000001)
};
\addplot[black!55, thick, dashed] coordinates {
  (0,0.20) (5,0.36) (10,0.48) (15,0.58) (20,0.65) (25,0.71)
  (30,0.12) (35,0.02) (40,0.003) (45,0.0005) (50,0.00008) (60,0.000001)
};
\legend{Build (cooperation, $\lambda^+ = 0.10$),
        Erode (defection, $\lambda^- = 0.30$),
        Build then defect at step 25}
\end{axis}
\end{tikzpicture}
\caption{Trust dynamics under the asymmetric update of
Eq.~\ref{eq:trust_update}. Three trajectories illustrate the
$3{:}1$ negativity-bias regime. The dark blue trajectory shows
trust building from $T_{ij}(0) = 0.20$ under sustained cooperation
($s = +1$): trust approaches the upper bound asymptotically over
roughly $60$ steps. The medium blue trajectory shows trust eroding
from $T_{ij}(0) = 1.0$ under sustained defection ($s = -1$): trust
collapses to near-zero in approximately $15$ steps. The dashed
trajectory illustrates the consequence of a single defection in an
otherwise cooperative trajectory: $25$ steps of trust accumulation
are erased in $5$ steps, an asymmetric reset that the
trust-aware oracle and TR-2 environments exploit to discriminate
sharply between sustained-cooperation and intermittent-cooperation
policies.}
\label{fig:trust_dynamics}
\end{figure}

\paragraph{Reputation layer.} Reputation $R_{ij}$ accumulates from
the immediate trust trajectory with exponential smoothing;
parameters are specified in \texttt{TrustParameters}. The two-layer
model allows agents to respond to long-run reputation while trust
reacts quickly to recent behavior. The separation of layers admits
an information-processing interpretation~\citep{nowak2006five}: the
immediate layer mediates direct-reciprocity responses
(Axelrod-style tit-for-tat~\citep{axelrod1984evolution}), while the
reputation layer mediates indirect-reciprocity responses
(image scoring, social-norm enforcement). The TR-4 formalism
extends this distinction into full reciprocity dynamics with
memory; TR-2 and TR-4 together span the spectrum from
short-memory to long-memory trust-driven cooperation mechanisms.

\subsection{Implementation architecture: where each formalism lives}

A note on where the four TR formalisms are implemented in the source tree.
V03 incorrectly located TR-3 and TR-4 in the \texttt{core/} helper
modules; the authoritative locations are:

\begin{itemize}[leftmargin=*, itemsep=2pt]
\item \textbf{TR-1 (interdependence)}: \texttt{core/value\_functions.py},
      \texttt{core/interdependence.py}, \texttt{core/equilibrium.py}.
      Full formalism in \texttt{core/}.
\item \textbf{TR-2 (trust dynamics)}: \texttt{core/trust\_dynamics.py}.
      Full formalism in \texttt{core/}.
\item \textbf{TR-3 (collective action)}: full formalism in
      \texttt{envs/collective\_action\_envs.py} via the
      \texttt{TR3Parameters} dataclass and associated utility
      functions. Source module docstring enumerates Eqs.~1--3 of the
      TR-3 paper and the free-riding equilibrium. A simplified
      helper module \texttt{core/collective\_action.py} provides
      auxiliary utilities but is not the authoritative
      implementation.
\item \textbf{TR-4 (reciprocity)}: full formalism in
      \texttt{envs/reciprocity\_envs.py} via the
      \texttt{TR4Parameters} dataclass and associated utility
      functions. Source module docstring enumerates Eqs.~19--25 and
      44--45 of the TR-4 paper. A simplified helper module
      \texttt{core/reciprocity.py} provides auxiliary utilities but
      is not the authoritative implementation.
\end{itemize}

The \texttt{envs/base.py} \texttt{AbstractCoopetitionEnv} class
composes TR-1 and TR-2 mechanics through the payoff and trust models.
TR-3 and TR-4 environment classes subclass the base environment and
inject their respective formalisms via overrides (for example,
\texttt{BaseTR4Env.\_compute\_reciprocity\_modifier()} implementing
Eq.~44).

\paragraph{On the helper modules \texttt{core/collective\_action.py} and
\texttt{core/reciprocity.py}.} These two modules provide auxiliary
state-tracking containers and helper classes
(\texttt{CollectiveActionState}, \texttt{ReciprocityState}) imported
by the env files. They are \emph{not} the authoritative
implementations of the TR-3 and TR-4 paper formalisms; that role
belongs to \texttt{envs/collective\_action\_envs.py} and
\texttt{envs/reciprocity\_envs.py} respectively. The helper modules
are retained in the public API across v1.x because the env files
depend on their exported names. Architectural consolidation (e.g.,
relocating the helpers into the env files or renaming them to
\texttt{core/*\_support.py}) is reserved for v2.0.0, where a
SemVer-major break is acceptable.

\paragraph{Provenance note.} The code state of the \texttt{coopetition-gym}
package that produced the 25{,}708-file training dataset and the
1{,}116-file behavioral audit dataset is preserved at the git tag
\texttt{v1.0.0} on \texttt{master}. Documentation
clarifications to the helper modules (rewritten module docstrings;
architectural pointers in the env files) are released as
\texttt{coopetition-gym 1.0.1} without any computational, API, or
output changes. The \texttt{v1.0.0} tag provides exact
byte-level reproducibility of study-era package behavior for users
who require it.

\subsection{Collective action mechanics (TR-3)}

The TR-3 tier formalizes the following problem. A group of $n$ agents
each choose a contribution level $a_i \in [0, a_{\max}]$. The group
produces a joint output that rises super-linearly with total
contribution, so all agents collectively benefit from high group
cooperation. Each agent's individual utility is the sum of its share
of the group output, a private cost of its own contribution, and a
loyalty modifier that rewards sustained cooperation. The formalism's
central claim is that the loyalty modifier admits a path to
above-Nash cooperation that is not accessible to agents who play
fixed actions: loyalty must be accumulated over time, and free-riders
cannot capture the loyalty component by a one-shot deviation.

The full formalism is implemented in
\texttt{envs/collective\_action\_envs.py}. The \texttt{TR3Parameters}
dataclass defines the verified constants:

\begin{center}
\small
\begin{tabular}{lll}
\toprule
Parameter & Value & Role \\
\midrule
$\omega$    & $25.0$ & Productivity factor \\
$\beta$     & $0.7$  & Returns to scale (diminishing, $<1$) \\
$c$         & $1.0$  & Effort cost coefficient \\
$\phi_B$    & $0.8$  & Loyalty benefit strength \\
$\phi_C$    & $0.3$  & Cost tolerance strength \\
$a_{\max}$  & $50.0$ & Maximum effort bound \\
\bottomrule
\end{tabular}
\end{center}

\paragraph{Parameter semantics.} The \emph{productivity factor}
$\omega = 25.0$ is the multiplicative scale of the team-production
function; larger $\omega$ raises the absolute return to total
cooperation, and the calibrated value places interior equilibria in
a regime that is neither trivially solved nor numerically unstable
for the algorithms in the suite. The \emph{returns-to-scale} exponent
$\beta = 0.7$ controls concavity of the production function; with
$\beta < 1$ the production is strictly concave (diminishing returns),
guaranteeing an interior Nash equilibrium, and as $\beta \to 1$
returns become linear and free-riding dominates. The \emph{effort
cost coefficient} $c = 1.0$ is the marginal private cost of
contribution: each unit of effort $a_i$ that an agent contributes
costs $c \cdot a_i$ in private utility. The \emph{loyalty benefit
strength} $\phi_B = 0.8$ is the fraction of teammates' average payoff
that flows to a fully loyal agent ($\theta_i = 1$) through the
loyalty channel; the \emph{cost tolerance strength} $\phi_C = 0.3$ is
the fraction of effort cost that a fully loyal agent effectively
absorbs without utility loss, modeling the empirical observation that
committed members tolerate short-run sacrifice for long-run
collective benefit. The \emph{maximum effort bound} $a_{\max} = 50.0$
is the upper bound of the per-agent action space $[0, a_{\max}]$;
$a_{\max}$ together with $\omega$ and $\beta$ jointly determine the
magnitude of returns.

\paragraph{Team production.}
The group's total output is a
concave power function of the sum of individual contributions. The
productivity factor $\omega = 25.0$ sets the scale of the output; the
exponent $\beta = 0.7 < 1$ encodes diminishing returns to total
contribution so that the marginal product of an additional unit of
cooperation decreases as total contribution grows. The concavity
ensures that the environment has an interior (as opposed to
corner-solution) equilibrium.
\begin{equation}
Q(\mathbf{a}) = \omega \left( \sum_{i=1}^{n} a_i \right)^{\beta}
\label{eq:tr3_production}
\end{equation}
\vspace{-0.4em}
\hfill\textit{\small [From \cite{pant2026collective}, Eq.~1]}
\vspace{0.5em}

\paragraph{Base team payoff.}
Each agent receives a $1/n$
share of the group output and pays a private cost proportional to its
own contribution. The cost coefficient $c = 1.0$ and the share $1/n$
together produce the classical free-rider incentive: an agent's
marginal private cost $c$ is linear in $a_i$, while its marginal
share of group output is only $1/n$ of the marginal total product,
so each agent prefers to let others contribute.
\begin{equation}
\pi_i^{\text{team}} = \frac{1}{n} Q(\mathbf{a}) - c \cdot a_i
\label{eq:tr3_payoff}
\end{equation}
\vspace{-0.4em}
\hfill\textit{\small [From \cite{pant2026collective}, Eq.~2]}
\vspace{0.5em}

\paragraph{Teammate-average payoff.} Agent $i$'s teammates obtain
the following average payoff, which serves as the reference from
which agent $i$'s loyalty modifier is computed. When teammates are
doing well, an agent's loyalty earns a larger reward.
\begin{equation}
\bar{\pi}_{-i} = \frac{1}{n - 1} \sum_{j \ne i} \pi_j^{\text{team}}
\label{eq:tr3_teammates}
\end{equation}
\vspace{-0.4em}
\hfill\textit{\small [From \cite{pant2026collective}, intermediate definition supporting Eq.~3]}
\vspace{0.5em}

\paragraph{Loyalty modifier.}
The loyalty modifier $L_i$ is
the sum of two terms, both scaled by agent $i$'s loyalty score
$\theta_i \in [0, 1]$. The first term, $\phi_B \cdot \bar{\pi}_{-i}$,
is a \emph{loyalty benefit}: the agent receives a fraction of its
teammates' welfare, weighted by its own loyalty. This models the
intuition that loyal members of a successful team share in the
team's success. The second term, $\phi_C \cdot c \cdot a_i$, is a
\emph{cost tolerance}: a loyal agent effectively bears less of its
own cooperation cost (because $\phi_C > 0$), a mathematical rendering
of the claim that committed members tolerate short-run sacrifice for
long-run collective benefit. The values $\phi_B = 0.8$ and
$\phi_C = 0.3$ place the benefit term at 80\% of teammate welfare and
the cost tolerance at 30\% of the agent's private cost, conditional
on maximum loyalty $\theta_i = 1$.
\begin{equation}
L_i = \theta_i \cdot \left[ \phi_B \cdot \bar{\pi}_{-i} + \phi_C \cdot c \cdot a_i \right]
\label{eq:tr3_loyalty}
\end{equation}
\vspace{-0.4em}
\hfill\textit{\small [From \cite{pant2026collective}, Eq.~3]}
\vspace{0.5em}

The loyalty score $\theta_i \in [0, 1]$ is agent $i$'s normalized
cooperation rate over a memory window (specifically, the time-
averaged ratio $a_i / a_{\max}$ over the past \texttt{loyalty\_horizon}
steps, as implemented in the environment's state-tracking utilities).
Loyalty therefore cannot be accumulated instantaneously: it requires
sustained cooperation over multiple steps.

\paragraph{Loyalty-augmented utility.} Agent $i$'s full utility is
the sum of its team payoff and its loyalty modifier. A free-riding
agent (low $a_i$) will have a low $\theta_i$ over time and therefore
receive a small $L_i$, so the loyalty channel is closed to agents
who do not sustain cooperation.
\begin{equation}
U_i = \pi_i^{\text{team}} + L_i
\label{eq:tr3_utility}
\end{equation}
\vspace{-0.4em}
\hfill\textit{\small [From \cite{pant2026collective}, Eq.~4]}
\vspace{0.5em}

\paragraph{Free-riding equilibrium (closed form).} Under the private
reward configuration (no loyalty channel, so $L_i$ is absent from the
learning signal), the symmetric Nash equilibrium contribution level is
obtained by setting each agent's first-order condition on $\pi_i^{
\text{team}}$ to zero and solving:
\begin{equation}
a^{*}_{\text{Nash}} = \left( \frac{\omega \beta}{n \cdot c} \right)^{1 / (1 - \beta)}
\label{eq:tr3_nash}
\end{equation}
\vspace{-0.4em}
\hfill\textit{\small [From \cite{pant2026collective}, \S6.1]}
\vspace{0.5em}

This is the well-known Cobb-Douglas free-riding equilibrium: each
agent contributes at the level at which its private marginal cost
equals its share of the marginal team output. For the calibrated
constants ($\omega = 25$, $\beta = 0.7$, $c = 1$, $n = 4$) the
equilibrium is $a^{*}_{\text{Nash}} \approx 3.53$, an order of
magnitude below the socially optimal contribution. The loyalty
modifier $L_i$ lifts cooperative behavior above $a^{*}_{\text{Nash}}$
when agents maintain high $\theta_i$, producing the above-Nash
cooperation characteristic of TR-3 environments. The gap between the
free-riding equilibrium and the loyalty-augmented optimum is the
quantity that TR-3 algorithms compete to close.

\paragraph{Team cohesion.} An environment-level summary statistic
measuring the extent to which the team's most-dependent agents are
also its most loyal agents. High cohesion ($\mathcal{C}$ close to 1)
indicates that loyalty is concentrated on the agents whose welfare
depends most on the team; low cohesion indicates a fragile team in
which loyalty is misallocated.
\begin{equation}
\mathcal{C} = \frac{\sum_i D_{T, i} \cdot \theta_i}{\sum_i D_{T, i}}
\label{eq:tr3_cohesion}
\end{equation}
\vspace{-0.4em}
\hfill\textit{\small [From \cite{pant2026collective}, Eq.~5]}
\vspace{0.5em}

where $D_{T, i}$ is agent $i$'s team-dependency weight.

\subsection{Reciprocity mechanics (TR-4)}

The TR-4 tier formalizes memory-bounded reciprocity with graduated
sanctions. The formalism composes three building blocks in sequence.
First, each agent computes a \emph{cooperation signal} from its
partner's recent actions: has the partner been cooperating more or
less than the recent average? Second, the signal is mapped through a
\emph{bounded response function} (the hyperbolic tangent) so that
neither extreme cooperation nor extreme defection can produce an
unbounded reciprocity reaction. Third, the bounded response is
weighted by two modulating factors: a \emph{reciprocity sensitivity}
term that depends on the agent's structural interdependence with its
partner, and a \emph{trust gate} that attenuates reciprocity when
trust is low. The end result is a single reciprocity modifier that
adds to each agent's utility at each step, representing the
psychologically and institutionally motivated cooperative adjustment
that the agent makes in response to its partner's recent behavior.

Full formalism is implemented in \texttt{envs/reciprocity\_envs.py}.
The \texttt{TR4Parameters} dataclass defines the verified constants:

\begin{center}
\small
\begin{tabular}{lll}
\toprule
Parameter & Value & Role \\
\midrule
$\rho_0$    & $1.0$ & Base reciprocity strength \\
$\eta$      & $1.0$ & Dependency elasticity \\
$\kappa$    & $1.0$ & Response sensitivity \\
$k$         & $5$   & Memory window length \\
$\lambda_R$ & $1.0$ & Reciprocity weight \\
$\omega$    & $0.6$ & Dependency amplification \\
\bottomrule
\end{tabular}
\end{center}

\paragraph{Parameter semantics.} The \emph{base reciprocity strength}
$\rho_0 = 1.0$ is the multiplicative coefficient on the reciprocity
sensitivity $\rho_{ij}$; larger values amplify the reciprocal
response. The \emph{dependency elasticity} $\eta = 1.0$ is the
exponent on $\dij$ in $\rho_{ij} = \rho_0 \, \dij^{\eta}$; $\eta > 1$
would super-linearly amplify reciprocity in high-dependency dyads,
$\eta < 1$ would compress it. The \emph{response sensitivity}
$\kappa = 1.0$ is the slope of the bounded-response function
$\varphi(x) = \tanh(\kappa x)$ at $x = 0$; larger $\kappa$ produces
sharper, step-like responses to small deviations. The
\emph{memory window length} $k = 5$ is the number of past steps over
which the partner's recent baseline $\bar{a}_j$ is averaged; the
calibrated value models short-window conditional cooperation
consistent with the experimental-economics literature. The
\emph{reciprocity weight} $\lambda_R = 1.0$ is a multiplicative weight
on the entire reciprocity modifier, placing the reciprocity channel
on equal footing with the trust channel. The \emph{dependency
amplification} $\omega = 0.6$ enters as the $(1 + \omega \dij)$
factor in the trust-gated reciprocity modifier and is distinct from
the TR-3 productivity factor that uses the same Greek letter.

\paragraph{Cooperation signal.} The signal $s_{ij}(t)$
measures how much agent $j$'s action at step $t$ deviates from its
own recent baseline. Positive signal means $j$ cooperated above the
recent norm (a pleasant surprise from $i$'s perspective); negative
signal means $j$ defected below the recent norm (an unpleasant one).
The signal is the elementary quantity from which all downstream
reciprocity computations derive.
\begin{equation}
s_{ij}(t) = a_j(t) - \bar{a}_j(t)
\label{eq:tr4_signal}
\end{equation}
\vspace{-0.4em}
\hfill\textit{\small [From \cite{pant2026reciprocity}, Eq.~19]}
\vspace{0.5em}

\paragraph{Memory-windowed baseline.} The recent norm
$\bar{a}_j(t)$ is the simple average of agent $j$'s actions over
the last $k = 5$ steps. The finite-window baseline encodes bounded
memory: an agent reciprocates against recent behavior, not
against behavior from arbitrary time in the past.
\begin{equation}
\bar{a}_j(t) = \frac{1}{\min(k, t - 1)} \sum_{\tau = \max(1, t - k)}^{t - 1} a_j(\tau)
\label{eq:tr4_memory}
\end{equation}
\vspace{-0.4em}
\hfill\textit{\small [From \cite{pant2026reciprocity}, Eq.~20]}
\vspace{0.5em}

The memory window $k = 5$ is distinct from TR-3's
\texttt{loyalty\_horizon} (which defaults to $10$ in the helper
module). The two windows correspond to different cognitive primitives:
loyalty is a long-memory accumulator, whereas reciprocity is a
short-memory responsiveness.

\paragraph{Bounded response.} The unbounded cooperation
signal is passed through the hyperbolic tangent to produce a bounded
reciprocity response in $(-1, 1)$. Two properties of $\tanh$ matter
for the formalism. First, it is saturating: very large positive or
negative signals do not produce proportionally large responses, so
the reciprocity channel cannot be overwhelmed by a single extreme
observation. Second, it is smooth and sign-preserving, so small
signals produce small responses whose direction matches the signal.
\begin{equation}
\varphi(x) = \tanh(\kappa \cdot x), \qquad \kappa = 1.0
\label{eq:tr4_response}
\end{equation}
\vspace{-0.4em}
\hfill\textit{\small [From \cite{pant2026reciprocity}, Eq.~21]}
\vspace{0.5em}

\begin{figure}[ht]
\centering
\begin{tikzpicture}
\begin{axis}[
    width=0.92\textwidth,
    height=5.6cm,
    xlabel={Cooperation deviation $x$},
    ylabel={Bounded response $\varphi(x) = \tanh(\kappa x)$},
    xlabel style={font=\footnotesize},
    ylabel style={font=\footnotesize},
    xticklabel style={font=\footnotesize},
    yticklabel style={font=\footnotesize},
    legend style={font=\scriptsize, at={(0.02,0.98)}, anchor=north west, draw=none, fill=white, fill opacity=0.85},
    grid=major, grid style={gray!25},
    xmin=-3, xmax=3, ymin=-1.1, ymax=1.1,
    enlargelimits=false,
    no markers, smooth, very thick,
    domain=-3:3, samples=120,
]
\addplot[trustcolor!75!black, thick] {tanh(0.5*x)};
\addplot[trustcolor!50!black, thick] {tanh(1.0*x)};
\addplot[trustcolor!25!black, thick] {tanh(2.0*x)};
\addplot[black!40, dashed, thick] coordinates {(-3,-1) (3,-1)};
\addplot[black!40, dashed, thick] coordinates {(-3,1) (3,1)};
\legend{$\kappa = 0.5$ (gradual), $\kappa = 1.0$ (calibrated default), $\kappa = 2.0$ (sharp)}
\end{axis}
\end{tikzpicture}
\caption{Bounded reciprocity response $\varphi(x) = \tanh(\kappa x)$
for three values of the response-sensitivity parameter $\kappa$. All
three curves saturate at $\pm 1$ (dashed horizontal asymptotes), so
no extreme cooperation deviation can produce an unbounded reciprocity
reaction. Larger $\kappa$ produces sharper, step-like responses that
discretize the reciprocal reaction; smaller $\kappa$ produces a more
gradual response that approaches a linear regime in a neighborhood
of zero. The calibrated default $\kappa = 1.0$ saturates near
$x = \pm 1.5$, placing the saturation boundary at roughly the
typical magnitude of a single-step deviation in calibrated TR-4
environments.}
\label{fig:tr4_bounded_response}
\end{figure}
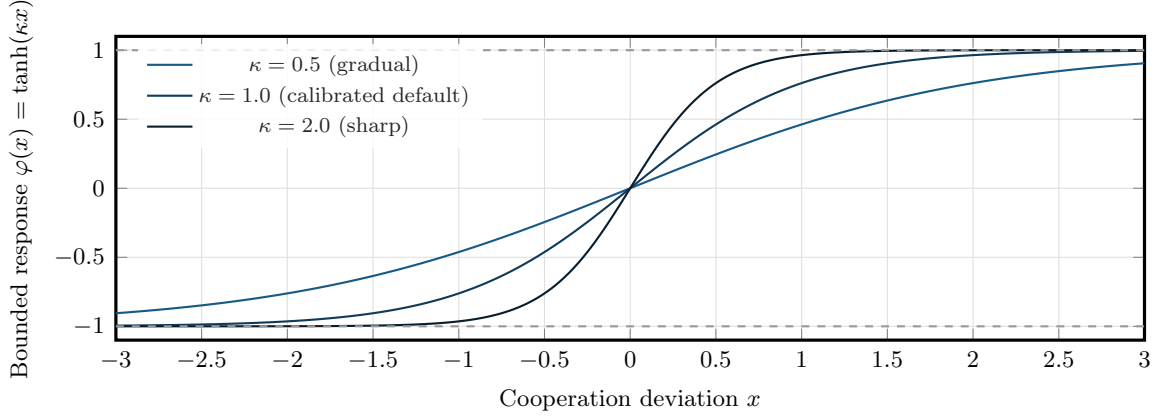

\paragraph{Reciprocity sensitivity.} The base reciprocity
strength is scaled by the structural dependency $\dij$ raised to the
elasticity $\eta$. With $\rho_0 = \eta = 1$ (the experimental study defaults)
the sensitivity is simply $\rho_{ij} = \dij$: an agent reciprocates
more strongly against partners on whom it is more structurally
dependent. Economically, this formalizes the observation that people
are more sensitive to the kindness or unkindness of people who
matter more to their welfare.
\begin{equation}
\rho_{ij} = \rho_0 \cdot \dij^{\eta}
\label{eq:tr4_rho}
\end{equation}
\vspace{-0.4em}
\hfill\textit{\small [From \cite{pant2026reciprocity}, Eq.~23]}
\vspace{0.5em}

\paragraph{Trust-gated reciprocity modifier.} The full
reciprocity modifier $U_i^{\text{recip}}$ is a sum over partners $j$
of four factors: the current trust $T_{ij}$ (which gates the response
so that low trust attenuates reciprocity regardless of signal
magnitude), a dependency-amplification term $(1 + \omega \dij)$ that
strengthens reciprocity further in high-dependency relationships
beyond what the baseline sensitivity $\rho_{ij}$ already encodes,
the sensitivity $\rho_{ij}$ itself, and the bounded response
$\varphi(s_{ij})$. The weight $\lambda_R = 1$ places the reciprocity
channel on equal footing with the trust channel.
\begin{equation}
\begin{aligned}
U_i^{\text{recip}} = &\ \lambda_R \sum_{j \ne i} T_{ij} \cdot (1 + \omega \cdot \dij) \\
& \cdot \rho_{ij} \cdot \varphi(s_{ij})
\end{aligned}
\label{eq:tr4_utility_recip}
\end{equation}
\vspace{-0.4em}
\hfill\textit{\small [From \cite{pant2026reciprocity}, Eq.~44]}
\vspace{0.5em}

The trust gate deserves comment: an agent who has low trust in its
partner will largely ignore the partner's recent cooperation or
defection, because both are processed through the $T_{ij}$
multiplier. This is a mathematical rendering of the intuition that
reciprocity requires a baseline level of trust to function; below
that baseline, the agent treats the partner's actions as
uninformative noise. The gate is one of the key mechanisms that
makes TR-4 environments resist exploitation by late-defection
strategies (see \S\ref{sec:oracle}).

\paragraph{Complete utility.} An agent's full utility in a
TR-4 environment is the sum of four components: the private payoff,
the TR-1 interdependence-integrated component, the TR-2 trust-
modulated component, and the TR-4 reciprocity modifier. The
composition is additive by design, so each component's contribution
to the utility gradient can be separately inspected.
\begin{equation}
U_i = \pi_i^{\text{base}} + U_i^{\text{interdep}} + U_i^{\text{trust}} + U_i^{\text{recip}}
\label{eq:tr4_complete}
\end{equation}
\vspace{-0.4em}
\hfill\textit{\small [From \cite{pant2026reciprocity}, Eq.~45]}
\vspace{0.5em}

TR-4 environments therefore compose all four TR-series mechanisms
within a single expression. This composition is the reason the
tier is positioned last: it exercises every other tier's machinery
simultaneously and requires the most mature environment
infrastructure.

\section{Application Programming Interfaces}
\label{sec:api}

\textsc{Coopetition-Gym} exposes three APIs to match different
user preferences for multi-agent environment interfaces.

\subsection{Gymnasium API (single-agent style)}

Agents are represented as elements of a single action vector. Standard
Gymnasium \texttt{reset}/\texttt{step} signature. Observation is a
flat array encoding each agent's state contribution plus the
interdependence row $D[i,:]$ when the default observation configuration
is used. Appropriate for users training with stable-baselines3 and
similar single-agent RL frameworks.

\begin{verbatim}
import coopetition_gym
env = coopetition_gym.make("TrustDilemma-v0")
obs, info = env.reset(seed=42)
obs, rewards, terminated, truncated, info = env.step([60.0, 55.0])
\end{verbatim}

\subsection{PettingZoo Parallel API (simultaneous moves)}

Agents act simultaneously. Standard PettingZoo Parallel signature.
Per-agent observation and reward dictionaries. Appropriate for users
training with PettingZoo-native MARL frameworks.

\begin{verbatim}
env = coopetition_gym.make_parallel("PlatformEcosystem-v0")
observations, infos = env.reset(seed=42)
actions = {agent: 50.0 for agent in env.agents}
observations, rewards, terminations, truncations, infos = env.step(actions)
\end{verbatim}

\subsection{PettingZoo AEC API (sequential moves)}

Agent-environment-cycle semantics. Standard PettingZoo AEC signature.
Appropriate for environments where simultaneous moves are a modeling
simplification and explicit turn order is preferred.

\begin{verbatim}
env = coopetition_gym.make_aec("TrustDilemma-v0")
env.reset(seed=42)
for agent in env.agent_iter():
    observation, reward, termination, truncation, info = env.last()
    if termination or truncation:
        action = None
    else:
        action = env.action_space(agent).sample()
    env.step(action)
\end{verbatim}

\subsection{Observation configuration}

\texttt{ObservationConfig} controls what each agent observes. The
default configuration includes each agent's interdependence row
$D[i,:]$ in the observation vector. Setting
\texttt{interdependence\_visible=False} creates an evaluation axis
that tests whether agents can learn cooperative policies without
observing the interdependence structure. The default models agents
who know their strategic context; the hidden configuration models
agents who must infer it from interaction history.

\paragraph{Hidden-$\dij$ regime as a research direction.} The
\texttt{interdependence\_visible=False} configuration is a
research-oriented evaluation regime that the package exposes as a
first-class configuration option through \texttt{ObservationConfig}
without requiring environment-side modifications: every environment
in the package supports both visible and hidden $\dij$ via the same
configuration switch. The reference evaluation uses the default
visible-$\dij$ configuration throughout. The hidden-$\dij$ regime is
the natural setting for downstream research on
coopetition-aware algorithm design, including algorithms that
maintain explicit posteriors over the unobserved $\dij$ structure and
update them through interaction, and on comparative analysis of
inferred-versus-observed cooperation policies. The configuration support
for the regime is documented here as a substrate fact, not as a
particular finding.

\section{Algorithm Suite and Oracle Baselines}
\label{sec:algorithms}

The reference algorithm suite comprises 126 algorithms organized into
four classes: $16$ training algorithms, $7$ game-theoretic oracles,
$2$ heuristic baselines, and $101$ constant-action policies. The suite
is designed to span the principal paradigm axes along which MARL
algorithms
differ~\citep{tampuu2017independent, lowe2017maddpg, papoudakis2021benchmarking}:
independent versus centralized training, actor-critic versus
value-based methods, on-policy versus off-policy data collection,
stochastic versus deterministic policy representations, and
parameter-sharing versus parameter-independent architectures.
Algorithm choice was guided by three criteria: (i) representative of
a paradigm that has been adopted in a published benchmark or
real-system study; (ii) implementable with a common interface that
supports continuous actions and variable agent counts without
exceeding the $4$-GB VRAM constraint of the reference training
hardware; and (iii) compatible with the reward-type ablation
methodology without paradigm-specific modifications.

\subsection{Training algorithms (16)}

The sixteen training algorithms are organized into two paradigm
classes. The classification is consequential: paradigm-class
differences account for the principal mechanism-dependent reversals
discussed in Part II.

\paragraph{Independent learners (7).} Each agent trains an independent
policy and critic using its own observations and rewards only. No
centralized information channel. IPPO (Independent PPO;
\citealp{schulman2017ppo, yu2022mappo}) runs the on-policy PPO
algorithm separately per agent. IA2C (Independent A2C) analogously
runs the advantage-actor-critic update. ISAC
(Independent Soft Actor-Critic; \citealp{haarnoja2018sac}) applies
per-agent SAC with twin critics and automatic entropy tuning; ISAC
is the overall strongest independent learner in our experimental study and the
paradigm-boundary counterpoint to the CTDE methods below. LOLA
(Learning with Opponent-Learning Awareness;
\citealp{foerster2018lola}) explicitly accounts for opponents'
learning dynamics through a multi-step lookahead term; our
implementation uses \texttt{torch.func.functional\_call} for the
inner gradient step. SelfPlay\_PPO trains one policy through
self-play. IndependentREINFORCE is a stochastic-policy baseline
that omits the value-function baseline. FCP (Fictitious Co-Play;
\citealp{strouse2021fcp}) constructs a population of checkpoints
and trains against sampled population members to improve robustness
against non-stationary partners.

\paragraph{CTDE methods (9).} Algorithms that use centralized
training with decentralized execution: each agent's critic at
training time has access to the joint observation, while at
execution time each agent acts from its own observation only. The
paradigm traces to \citet{lowe2017maddpg} (MADDPG); we include the
MADDPG family plus its twin-delayed variant MATD3, the minimax
variant M3DDPG, and the maximum-entropy variant MASAC. Value-based
CTDE methods include QMIX \citep{rashid2018qmix} with its
monotonic mixing network (implemented via \texttt{torch.abs}-gated
hypernetwork), VDN \citep{sunehag2018vdn} with additive factorization,
and COMA \citep{foerster2018coma} with its counterfactual baseline.
Policy-gradient CTDE methods are represented by MAPPO
\citep{yu2022mappo} and by MeanFieldAC, a mean-field approximation
that scales to large agent populations. MeanFieldAC is restricted
to environments with $n \ge 3$ agents because the mean-field
approximation is degenerate for $n = 2$ (the mean-field action
equals the single partner action). We documented an instability
in MASAC on a subset of TR-3 environments (onset near $83\%$ of
training progress; $14/140$ affected files) in
\S\ref{sec:failure}. This finding is enabled by the benchmark's
reward-type ablation protocol and is not visible in single-reward
evaluation.

\paragraph{Paradigm-class implications for coopetitive environments.}
A recurring empirical question across the MARL benchmark literature
is whether CTDE consistently dominates independent
learning~\citep{lowe2017maddpg, yu2022mappo, papoudakis2021benchmarking}.
The consensus in cooperative-only benchmarks is that CTDE provides
modest but reliable gains. The benchmark results of \S\ref{sec:ctde}
complicate this consensus: in mechanism classes that reward
sustained reciprocity (TR-2, TR-4), independent learning
consistently matches or exceeds CTDE; in mechanism classes that
reward coordination or free-riding control (TR-1, TR-3), CTDE
provides its expected advantage. The paradigm boundary is
mechanism-dependent, not universal, and the interpretation we
advance is that centralized training destabilizes exactly the
reciprocity-exploiting equilibria that independent learning
discovers incrementally from local reward signals.

Full hyperparameter specifications appear in
\texttt{experiments/config.py} in the reproducibility package.
Hyperparameters were selected to match the defaults of the source
papers where possible and are frozen across all environments and
all seeds within a training run; no per-environment tuning is
permitted in the reference evaluation protocol.

\subsection{Game-theoretic oracle baselines (7)}

Oracle algorithms compute analytically motivated reference policies
from environment parameters. They do not train; they evaluate.

\begin{itemize}[leftmargin=*, itemsep=2pt]
\item \texttt{Oracle\_Equilibrium} (TR-1 Nash reference)
\item \texttt{Oracle\_TrustAware} (TR-2 trust-aware reference)
\item \texttt{Oracle\_Nash} (TR-3 lower bound)
\item \texttt{Oracle\_Loyalty} (TR-3 upper bound via social optimum)
\item \texttt{Oracle\_SocialOptimum} (TR-3 upper bound, equivalent to Oracle\_Loyalty)
\item \texttt{Oracle\_ReciprocityEquilibrium} (TR-4 lower bound)
\item \texttt{Oracle\_BoundedReciprocity} (TR-4 upper bound)
\end{itemize}

Appendix~\ref{app:env_oracle_ref} specifies the reference oracle per
environment used for Gap-percentage computation.

\subsection{Heuristic baselines (2)}

\paragraph{Random.} Uniform random cooperation level sampled each
step. Tests whether an algorithm has learned anything nontrivial beyond
random policy.

\paragraph{TitForTat.} Conditional reciprocity: match partner's
previous cooperation level. Tests whether an algorithm outperforms
a well-known non-learning strategy that is optimal in many dyadic
settings.

\subsection{Constant-action policies (101)}

\texttt{Constant\_00} through \texttt{Constant\_100}: each plays the
same cooperation fraction ($0\%$ through $100\%$ in $1\%$ increments)
every step. This fine-grained sweep enables non-parametric
characterization of the payoff landscape as a function of uniform
cooperation level and supports the ``highest fixed-action return''
calculation used in oracle benchmarking.

\section{Evaluation Methodologies}
\label{sec:eval}

\textsc{Coopetition-Gym v1} supports four evaluation methodologies:
single-reward evaluation, reward-type ablation, oracle benchmarking,
and behavioral audit. The four methodologies are orthogonal: they
answer different questions about a trained policy, and a thorough
evaluation will use all four. Users may adopt any subset, but the
reward-type ablation and behavioral audit methodologies together
are the principal contribution of the benchmark and warrant
particular attention.

\begin{figure}[ht]
\centering
\begin{tikzpicture}[
  node distance=0.6cm,
  layer/.style={rectangle, rounded corners, draw, thick, minimum width=4.4cm, minimum height=0.9cm, align=center, font=\small},
  payoff/.style={layer, fill=trustcolor!15, draw=trustcolor!70!black},
  reward/.style={layer, fill=coopcolor!15, draw=coopcolor!70!black},
  modeb/.style={rectangle, rounded corners, draw, minimum width=2.6cm, minimum height=0.7cm, align=center, font=\footnotesize},
  arr/.style={-{Latex[length=2mm]}, thick}
]
\node[payoff] (env) {Environment transition rules\\(trust, loyalty, reciprocity)};
\node[payoff, below=of env] (pi) {Payoff vector $\boldsymbol{\pi}(\mathbf{a})$\\(per-agent environment payoff)};
\node[font=\footnotesize, right=2.0cm of pi] (note1) {payoff layer};
\node[modeb, fill=black!06, draw=black!50, below left=1.4cm and -0.5cm of pi] (priv) {private\\$R_i = \pi_i$};
\node[modeb, fill=black!10, draw=black!55, below=1.4cm of pi] (intg) {integrated\\$R_i = \pi_i + \sum_{j \ne i} \dij \pi_j$};
\node[modeb, fill=black!14, draw=black!60, below right=1.4cm and -0.5cm of pi] (coop) {cooperative\\$R_i = \frac{1}{n}\sum_j \pi_j$};
\node[reward, below=1.5cm of intg] (alg) {Reward signal $R_i$ delivered to learning algorithm};
\node[font=\footnotesize, right=2.0cm of alg] (note2) {reward layer};
\draw[arr] (env) -- (pi);
\draw[arr] (pi.south) -- ++(0,-0.3) -| (priv.north);
\draw[arr] (pi) -- (intg);
\draw[arr] (pi.south) -- ++(0,-0.3) -| (coop.north);
\draw[arr] (priv.south) -- ++(0,-0.4) -| (alg.north);
\draw[arr] (intg) -- (alg);
\draw[arr] (coop.south) -- ++(0,-0.4) -| (alg.north);
\end{tikzpicture}
\caption{Two-layer separation of payoff and reward in
\textsc{Coopetition-Gym v1}. This package's environments produce a
payoff vector $\boldsymbol{\pi}(\mathbf{a})$ deterministically given
the joint action $\mathbf{a}$ and the environment's mechanism state
(trust, loyalty, reciprocity). The reward signal delivered to the
learning algorithm is constructed from $\boldsymbol{\pi}$ by one of
three reward functions: \emph{private} (each agent receives only its
own payoff), \emph{integrated} (each agent receives a calibrated
$\dij$-weighted combination of partner payoffs), or
\emph{cooperative} (each agent receives the mean payoff). The
reward-type ablation methodology varies this construction while
holding the payoff layer fixed, isolating reward-structure effects
from environment-mechanism effects.}
\label{fig:two_layer}
\end{figure}
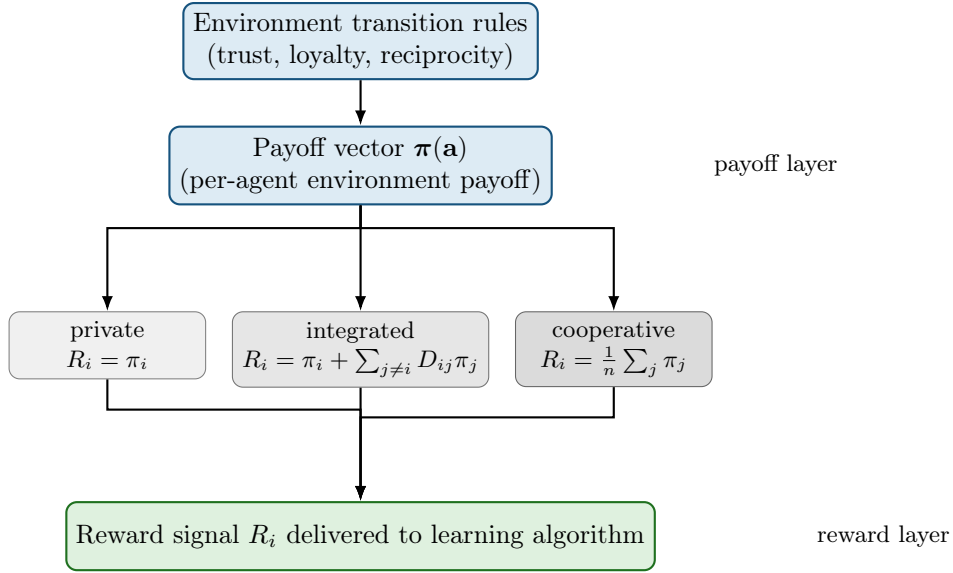

The motivation for providing multiple evaluation methodologies is
that single-scalar comparisons on a fixed reward function have a
long-documented failure mode in MARL: two algorithms with
indistinguishable returns on one reward specification may exhibit
qualitatively different behavior on another, and the ranking
published in a benchmark becomes an artifact of the benchmark's
reward choice rather than a property of the algorithms being
compared~\citep{papoudakis2021benchmarking}. The reward-type
ablation methodology below is our response to that failure mode.

\subsection{Single-reward evaluation}

The standard MARL evaluation protocol. Select the integrated reward
configuration, run training algorithms with default hyperparameters,
report mean episodic return after training. Produces one ranking of
algorithms per environment. Compatible with existing MARL evaluation
conventions, including the protocols used in
SMAC~\citep{samvelyan2019smac}, Melting Pot~\citep{leibo2021meltingpot},
and the SSD family~\citep{leibo2017ssd}. The reference evaluation's
main-results section ($16{,}835$ files in
\texttt{baseline\_integrated/}) provides the reference output of
this protocol. We retain single-reward evaluation as a supported
methodology because it is the modality in which cross-benchmark
comparison is most transparent: a reader familiar with Melting Pot or
SMAC returns can calibrate their expectations about what a given
return number implies.

\subsection{Reward-type ablation}

The reward-type ablation methodology varies reward mutuality across
three configurations while holding environment rules fixed:
\begin{align}
R_i^{\text{private}}(\mathbf{a})     &= \pi_i(\mathbf{a}) \\
R_i^{\text{integrated}}(\mathbf{a})  &= U_i(\mathbf{a}) = \pi_i(\mathbf{a}) + \sum_{j \ne i} \dij \pi_j(\mathbf{a}) + M_i \\
R_i^{\text{cooperative}}(\mathbf{a}) &= \frac{1}{n} \sum_j \pi_j(\mathbf{a})
\end{align}
Each configuration produces a separate ranking of algorithms per
environment. Differences between rankings diagnose whether algorithm-
ranking claims are robust to reward-function structure or are
artifacts of the particular reward function used. The theoretical
content of this methodology is that the reward function and the
environment transition structure are logically separable properties
of the MDP/stochastic-game specification, and each can be varied
independently. In practice many MARL benchmarks conflate the two:
an environment ships with one reward function, and ``the benchmark''
refers to that environment--reward pair. The reward-type ablation
protocol disentangles the two and allows the benchmark user to
attribute a performance result to the transition structure, to the
reward structure, or to their interaction.

The private configuration tests how algorithms perform when given
the minimal information of their own payoff only; the cooperative
configuration tests how they perform when given the maximum shared
signal (full average payoff); the integrated configuration tests
the TR-1 structural-coupling interpretation in which the
interdependence matrix $\dij$ encodes the relational structure of
the multi-agent system. Changes in algorithm ranking across
configurations are diagnostic of the sensitivity of the algorithm
to this structural property of the reward; in particular, the AI-4
implicit-cooperation findings (\S\ref{sec:implicit}) rely on
the observation that several algorithms achieve above-Nash
cooperation under $R^{\text{private}}$, where no shared reward
signal is available. This observation is meaningful only because
the methodology separates reward from transition.

\subsection{Oracle benchmarking}

Evaluation against analytically computed reference policies. For
each environment, the designated reference oracle
(Appendix~\ref{app:env_oracle_ref}) provides a non-learned baseline
whose value derives from a game-theoretic solution concept rather
than from trained policy behavior. The \emph{Gap percentage}
between an algorithm and its reference oracle is:
\begin{equation}
\mathrm{Gap\%}(A, e) = \frac{R(A, e) - R(\mathrm{Oracle}_e)}{|R(\mathrm{Oracle}_e)|} \times 100
\label{eq:gap_percent}
\end{equation}
Positive values indicate the algorithm exceeds the oracle reference;
negative values indicate the algorithm underperforms. Gap-percentage
evaluation provides a theoretically grounded yardstick that is
independent of empirical algorithm-vs-algorithm comparisons and
admits a direct interpretation: a positive Gap\% against
\texttt{Oracle\_Nash} means the algorithm has discovered a
cooperative equilibrium that dominates the non-cooperative Nash
equilibrium; a positive Gap\% against \texttt{Oracle\_Loyalty}
means the algorithm has exceeded the social optimum derivable from
the TR-3 loyalty formalism, which is a stronger claim warranting
inspection.

Oracle benchmarking is supported for all four TR tiers with
tier-appropriate solution concepts~\citep{nash1950equilibrium,
shapley1953stochastic, ostrom1990governing}. TR-1 environments use
\texttt{Oracle\_Equilibrium}, a Nash reference computed by
iterative best response on the closed-form TR-1 payoff structure.
TR-2 environments use \texttt{Oracle\_TrustAware}, which integrates
the trust dynamics of Eq.~\ref{eq:trust_update} into the best-
response computation. TR-3 environments use two oracles,
\texttt{Oracle\_Nash} as a lower bound (pure free-riding
equilibrium) and \texttt{Oracle\_Loyalty} as an upper bound (social
optimum accounting for the loyalty modifier). TR-4 environments
similarly pair \texttt{Oracle\_ReciprocityEquilibrium} as a lower
bound with \texttt{Oracle\_BoundedReciprocity} as an upper bound.
The lower/upper pairing on TR-3 and TR-4 admits a sharper
diagnostic than a single reference: algorithms that achieve
returns between the two oracles occupy the interior of the
cooperative region; algorithms below the lower bound fail to
discover even the non-cooperative equilibrium; algorithms above
the upper bound exceed the formalism's theoretically computable
cooperative ceiling, a diagnostic we use to motivate the
adaptive-sequence analysis of \S\ref{sec:oracle}.

\subsection{Behavioral audit}

A policy-behavior evaluation orthogonal to return-based ranking.
Characterizes how an algorithm's policy responds to counterfactual
cooperation levels (static response-surface audit) and to temporal
deviation strategies (temporal deviation audit). Enables safety- and
alignment-focused analysis of trained policies. The audit records
the \emph{policy} under structured perturbations rather than the
return, and therefore distinguishes algorithms that achieve similar
returns through qualitatively different strategies. This distinction
matters for the multi-agent-system deployment question: two
algorithms with indistinguishable mean episodic returns may
nevertheless differ sharply in their response to exploitation or to
partner failure, and the behavioral audit surfaces this difference
at evaluation time rather than at deployment time. The methodology
and reference results appear in Section~\ref{sec:audit}.

\section{Case Study Validation}
\label{sec:validation}

Four of the twenty environments are calibrated to historically
documented coopetitive relationships. Calibration consists of
extracting interdependence coefficients $\dij$ and other parameters
from qualitative coding of strategic dependencies documented in
archival sources, then verifying that the calibrated environment
produces simulation trajectories qualitatively consistent with the
documented historical outcomes.

\begin{table}[H]
\centering
\small
\caption{Validation scores for the four calibrated case study environments.
Scoring follows the Behavioral Correspondence protocol defined in the
TR-validation suites.}
\label{tab:validation}
\begin{tabular}{lllrl}
\toprule
Environment & Case study & TR & Score & Source \\
\midrule
\texttt{SLCD-v0}          & Samsung-Sony LCD JV (2004--11)  & TR-1 & 59/60 (98.3\%) & CAiSE 2026 \\
\texttt{RenaultNissan-v0} & Renault-Nissan Alliance         & TR-2 & 49/60 (81.7\%) & TR-validation \\
\texttt{ApacheProject-v0} & Apache HTTP Server              & TR-3 & 52/60 (86.7\%) & TR-validation \\
\texttt{AppleAppStore-v0} & Apple iOS App Store             & TR-4 & 48/55 (87.3\%) & TR-validation \\
\bottomrule
\end{tabular}
\end{table}

Validation methodology and full discrimination analyses (reward-function
configurations distinguishable at case study accuracy) appear in
Appendix~\ref{app:case_studies}.

\section{Behavioral Audit}
\label{sec:audit}

The benchmark ships with a two-part behavioral audit methodology.

\subsection{Static response-surface audit}

For each \texttt{(algorithm, environment, seed)} triple, sweep uniform
cooperation from $0\%$ to $100\%$ of endowment in $5\%$ increments
($21$ levels), recording per-agent returns at each level. Additionally,
at four test cooperation levels ($20\%$, $40\%$, $60\%$, $80\%$), test
whether unilateral deviation (agent 0 reduces cooperation by $50\%$
while other agents maintain) yields a gain for the deviator at the
expense of partners, classifying the outcome as \emph{exploitative}
when agent 0 gains and others lose.

The static audit characterizes the payoff landscape's exposed
exploitation gradient. Finding: the exploitation count is
\emph{algorithm-independent}. All tested policies produce identical
exploitation classifications per environment, because the audit uses
fixed actions applied to the environment, not trained-policy actions.
The gradient is a structural property of the environment.

\subsection{Temporal deviation audit}

For each \texttt{(environment, seed)} pair, test five temporal
strategies against a baseline of all agents cooperating at $50\%$ of
endowment: (i) full defection throughout; (ii) binary late-defection
at switchpoints spanning $50\%$ to $99\%$ of the episode (nine
switchpoints); (iii) early defection for $10\%$--$30\%$ of the episode;
(iv) gradual linear ramp-down over the final $20\%$ of the episode;
(v) single-step final-step defection. Classification: whether agent 0
gains and partners lose.

Finding: binary switchpoint strategies are universally blocked across
the benchmark ($0$ of $504$ tests exploitative). Gradual ramp-down
produces marginal exploitation on $6$ of $20$ environments
($+0.004\%$ to $+0.41\%$ of baseline return), a detection-threshold
effect in the per-step sanction mechanisms.

Full audit results appear in Appendix~\ref{app:audit_full}.

\section{Statistical-Gate Methodology}
\label{sec:statistical-gate}

Empirical multi-agent reinforcement-learning benchmark evaluation
routinely encounters distributional anomalies (surprising return
shifts, apparent bimodality, and failure-rate changes under sampling
extension) that invite mechanistic interpretation. A central
methodological question is how to discipline the inference from
observation to mechanism: when should a surprising distribution license
a mechanism claim in the paper text, and when should it not? This
section documents the inferential discipline applied throughout the
reference evaluation. The discipline is: \emph{run a cheap statistical
test before writing the mechanism claim, and write only the claim the
test supports.} The discipline is operationalized through three
gates (Hartigan's dip test for bimodality, an exploration-budget
diagnostic for hyperparameter-artifact objections, and a
pre-registered censoring rule for computationally intractable cells),
each of which
converts a visual or intuitive mechanism hypothesis into an empirical
commitment. The gates apply uniformly to the findings reported in
\S\ref{part:findings} and to any downstream paper derived from the
benchmark.

\subsection{Hartigan's dip test as a mechanism-claim gate}
\label{subsec:dip-test-gate}

Hartigan's dip test~\citep{hartigan1985dip} computes the maximum
distance between the empirical distribution function of a sample and
the closest unimodal distribution function. The test statistic, dip,
is combined with a bootstrap or tabulated critical value to produce a
p-value against the null hypothesis of unimodality. The test is
distribution-free and requires no assumption about the underlying
distribution's shape beyond whether it has one mode or more. It is
appropriate for per-seed returns because the seed is a natural
independent-observation unit in RL benchmark evaluation.

Our discipline:

\begin{enumerate}[leftmargin=*, itemsep=2pt]
\item \textbf{A distributional observation triggers the gate.} Whenever
  per-seed returns look bimodal, step-shaped, or visually anomalous in
  a way that a bimodal-convergence mechanism hypothesis would
  explain, and whenever the mechanism hypothesis is novel enough to
  warrant text in the paper, the gate fires.
\item \textbf{The cheap test runs first.} Hartigan's dip test on the
  per-seed return distribution (pooled across reward modes for maximum
  $n$, or per-mode if the cell-level distribution is the target).
  At $n \geq 10$ the test is usable; at $n \geq 16$ the test is
  comfortable.
\item \textbf{The test's outcome gates the paper-text claim.} If
  $p < 0.05$, the bimodal mechanism is empirically supported and the
  paper-text claim may enter with an explicit test reference. If
  $p \geq 0.05$, the bimodal claim does \emph{not} enter the paper:
  the text reflects what the test supports (``high-variance unimodal'')
  or what the test does not rule out (``consistent with unimodality,
  further investigation required'').
\item \textbf{No mechanism speculation without a supporting test.} The
  paper does not write ``bimodal convergence is a plausible
  mechanism'' as speculation; either the test rejects unimodality and
  the paper writes the mechanism, or the test does not reject and the
  paper withholds the claim.
\end{enumerate}

This discipline prevents a common failure mode in benchmark
evaluation: a surprising distribution triggers an intuitive mechanism
interpretation, the interpretation enters the paper as a ``plausible''
or ``candidate'' explanation, and downstream readers carry the
interpretation as a fact. The test-before-claim rule forces the
author to commit to the claim the data actually supports, rather than
to the mechanism that is visually plausible.

\subsection{Paradigm case study 1: $\beta = 0.90$ bounce-back on the
two-dimensional sensitivity grid}
\label{subsec:case-study-1}

The two-dimensional action-space extension (\S\ref{sec:2d-slcd})
conducts a sensitivity sweep on the $(c_i, p_i)$ action space of
\texttt{SLCD-v0} across an $\eta \times \beta$ grid with $30$ seeds
per cell. At the $(\eta=0.40, \beta=0.90)$ cell, the equilibrium
appropriation $p^*$ distribution exhibits a non-monotonic recovery
pattern relative to neighboring cells: $p^*$ is higher at this cell
than at the surrounding $\beta$ values. Visual inspection suggests two
candidate mechanisms:

\begin{itemize}[leftmargin=*, itemsep=2pt]
\item \textbf{Bimodal convergence.} The $p^*$ distribution at this cell
  is bimodal, with a subset of seeds landing at a low-$p^*$ attractor
  and another subset at a high-$p^*$ attractor. The observed mean
  recovery is a mixture of the two modes.
\item \textbf{High-variance unimodal.} The $p^*$ distribution is
  unimodal but has a larger variance than at neighboring cells. The
  observed recovery is a mean shift with increased tail weight, not a
  structural change in the attractor landscape.
\end{itemize}

The two candidates have different implications for the benchmark's
utility: a bimodal-convergence finding would suggest that the
two-dimensional extension admits pathological attractor structures in
some parameter regions, whereas a high-variance unimodal finding would
suggest only that the region is noisier than its neighbors.

We applied the dip test. The primary suspect cell
($\eta=0.40, \beta=0.90$, $n=30$ ISAC seeds) yielded $\text{dip}=0.055$,
$p=0.88$: the test \emph{fails to reject} unimodality at
$\alpha=0.05$ by a wide margin. Six control cells, each with $n \geq
22$ seeds, returned $p$-values in $[0.147, 0.993]$: none reject
unimodality. The paper-text finding is therefore ``the $\beta=0.90$
recovery is a high-variance unimodal phenomenon, not a bimodal
convergence to two attractors.'' The bimodal-convergence mechanism
is withdrawn from the paper text; the $p=0.88$ dip-test outcome is
referenced explicitly in the 2D-SLCD finding.

\subsection{Paradigm case study 2: PlatformEcosystem return drift for
deterministic-policy algorithms}
\label{subsec:case-study-2}

A second distributional anomaly emerged from the $10$-seed extension
study: on \texttt{PlatformEcosystem-v0} (a two-agent mixed-motive
environment with asymmetric endowments), four deterministic-policy
algorithms (MATD3, MADDPG, M3DDPG, LOLA) showed return shifts of
$-59\%$ to $-93\%$ between the $7$-seed baseline and the $10$-seed
extension, all directionally downward. Visual inspection of the pooled
per-seed returns at $n=16$ per algorithm suggests two clean clusters:
a low-return cluster at approximately $\{2.5\text{k}, 3.4\text{k}\}$
and a high-return cluster at approximately $\{42\text{k}, 117\text{k}\}$,
with an empty gap between them. Under the bimodal-convergence
hypothesis, the original $7$-seed sample drew predominantly from the
upper cluster; the $10$-seed extension added lower-cluster samples.

We applied the dip test per algorithm, pooling across reward modes to
maximize $n$. The outcomes were not uniform:

\begin{itemize}[leftmargin=*, itemsep=2pt]
\item \textbf{MATD3 ($n=16$):} $\text{dip}=0.142$, $p=0.0036$. Rejects
  unimodality at $p < 0.01$. The bimodal-convergence mechanism is
  empirically supported; the paper-text finding is ``MATD3 on
  PlatformEcosystem-v0 exhibits bimodal convergence.''
\item \textbf{MADDPG ($n=16$):} $\text{dip}=0.139$, $p=0.0050$.
  Rejects unimodality at $p<0.01$. Paper-text finding mirrors MATD3:
  ``MADDPG on PlatformEcosystem-v0 exhibits bimodal convergence.''
\item \textbf{M3DDPG ($n=16$):} $\text{dip}=0.088$, $p=0.349$.
  Fails to reject unimodality. The raw data are suggestive of
  bimodality on visual inspection, but the test does not support the
  claim at $n=16$. The paper-text finding is ``M3DDPG on
  PlatformEcosystem-v0 exhibits high-variance unimodal returns;
  bimodality is not empirically supported by the current sample.''
\end{itemize}

Three algorithms, two different verdicts. The discipline enforces
separate treatment: the bimodal claim enters the paper for MATD3 and
MADDPG with the dip-test reference; for M3DDPG the paper writes the
high-variance unimodal framing that the test supports, despite the
visual temptation. This is the test-before-claim discipline in its
exact operational form: the same visual pattern across three
algorithms produces three different inferential commitments because
the test outcomes differ. A less disciplined paper would write
``the DDPG family exhibits bimodal convergence on
PlatformEcosystem-v0'' as a three-algorithm claim; our paper does not.

\subsection{Exploration-budget diagnostic for hyperparameter-artifact
objections}
\label{subsec:exploration-budget}

A distinct but related gate applies to the IPPO low-entropy
exploration collapse observed in the $2$D-SLCD sensitivity sweep
(\S\ref{sec:2d-slcd}). Under certain $(\eta, \beta)$ configurations,
IPPO collapses to $p=0$ (zero appropriation), which could be
interpreted as (a) a genuine exploration failure of IPPO under the
specific reward configuration, or (b) a hyperparameter artifact of
IPPO's default entropy coefficient ($\text{ent\_coef}=0.01$).
Interpretation (a) would be a finding about IPPO's exploration
behavior on this task family; interpretation (b) would be a
configuration-setting criticism of the IPPO implementation rather
than a finding.

The diagnostic: sweep IPPO's entropy coefficient across
$\{0.005, 0.01, 0.05, 0.20\}$ at the collapse-prone cell
$(\eta=0.50, \beta=0.30)$ and observe whether the collapse persists.
At all four values, including the highest ($0.20$, which is 20$\times$
the default), IPPO collapses to $p=0$ on a majority of seeds. The
collapse is therefore not attributable to insufficient entropy
regularization; it is a structural behavior of IPPO under this reward
configuration. The paper-text finding ``IPPO exhibits low-entropy
exploration collapse on this class of mixed-motive 2D action spaces''
is gated by the entropy sweep: reviewer objections of the form
``increase the entropy coefficient'' are pre-empted by the diagnostic.

\subsection{Pre-registered censoring rule and dual-symbol table markup}
\label{subsec:censoring}

A third gate concerns empirical cells that cannot be resolved within
the experimental study's computational horizon. Two cell conditions require
distinct treatment:

\begin{itemize}[leftmargin=*, itemsep=2pt]
\item \textbf{NaN-producing cells.} Some algorithm-environment
  combinations produce \texttt{NaN} training returns with high
  reproducibility across seeds and reward modes (e.g., the
  deterministic-policy reward-mode-conditional divergence documented
  in \S\ref{subsec:ddpg-rmc-nan}). These cells \emph{have} data; the
  data is \texttt{NaN}. Under robust-statistics aggregation (including
  the interquartile mean) \texttt{NaN} cannot participate: it is a
  missing-value condition in a principled sense, not a
  below-baseline score.
\item \textbf{Computationally intractable cells.} A small number of
  algorithm-environment combinations (e.g., M3DDPG on the $6$-agent
  \texttt{ApacheProject-v0} at $2$--$3$ steps/second, $\sim 90$ hours
  per run at the $1$M-step cap) may not land within the submission
  horizon even with additional wall-clock. These
  cells would in principle produce valid data given more wall-clock;
  within the reporting horizon, however, they are indeterminate.
\end{itemize}

The pre-registered censoring rule is:

\begin{enumerate}[leftmargin=*, itemsep=2pt]
\item \textbf{Mark NaN cells with \ensuremath{\times}.} The
  \ensuremath{\times} symbol denotes an algorithm-level
  incompatibility: the cell \emph{has} a result (\texttt{NaN}), and
  the result excludes it from ranking aggregation and from any
  robust-statistic computation. The \ensuremath{\times} is
  reward-mode-conditional when the NaN outcome is: an algorithm may
  be \ensuremath{\times}-marked under one reward mode and not under
  another.
\item \textbf{Mark computationally censored cells with
  \ensuremath{\dagger}.} The \ensuremath{\dagger} symbol denotes a
  submission-horizon censoring: the cell is absent not because the
  algorithm cannot train but because the wall-clock budget did not
  accommodate the run within the reporting window. Aggregation falls
  back to the $7$-seed baseline, and a per-row annotation indicates
  which cells are $\dagger$-censored.
\item \textbf{No silent averaging.} Neither \ensuremath{\times} nor
  \ensuremath{\dagger} cells are silently averaged in; the table
  caption documents both symbols and their exclusion semantics. A
  reader of the ranking table without the caption should still
  recognize that cells bearing either symbol are not participating
  in the aggregate.
\end{enumerate}

The dual-symbol markup is the reader-facing expression of the
pre-registration: both conditions are distinct, both warrant different
treatment, and the distinction is visible at table-cell resolution.

\subsection{Synthesis and downstream applications}
\label{subsec:statistical-gate-synthesis}

The three gates (dip test for bimodality, exploration-budget for
hyperparameter artifacts, and dual-symbol markup for censoring) share
a common structure. Each gate is a cheap empirical test that runs
before a mechanistic claim enters the paper; each gate's outcome
binds the paper-text to the empirically supported interpretation
rather than to the visually intuitive one. The gates are neither
exhaustive nor general-purpose: additional gates will be needed for
distributional anomalies the present study has not encountered
(for example, heavy-tailed returns with a contaminating distribution,
or temporal non-stationarity in training-curve metrics). The
discipline itself, namely cheap test first and claim second, is the
portable
element.

The three gates seed a companion methodology paper on statistical
discipline in multi-agent reinforcement learning benchmark evaluation.
That paper formalizes the test-before-claim rule, surveys additional
cheap tests applicable to the benchmark's empirical contexts
(Kolmogorov-Smirnov, Anderson-Darling, Dip-of-Dips for multimodality
beyond two), and argues that the absence of such gates in existing
benchmark-paper practice accounts for a substantial portion of the
replication failures documented in the MARL reproducibility
literature~\citep{henderson2018deeprl}. The current technical report
is the source of the paradigm case studies and the original discipline
commitment; the companion paper extends the framework to a broader
survey of MARL evaluation problems.

\paragraph{Scope of the present chapter.} The present chapter is
scoped narrowly to the three gates as applied within the reference
and to the two paradigm case studies (the $\\beta=0.90$ bounce-back unimodal verdict and the studies (the $\beta=0.90$
bounce-back unimodal verdict and the PlatformEcosystem-v0
three-algorithm bimodal/unimodal split). The gates are
operationalized as concrete decision rules and demonstrated on
within-study data. Broader methodological development of
gate discipline as a portable framework, including survey of cheap
tests beyond Hartigan's dip, integration with reproducibility
checklists, and critical analysis of MARL benchmark practice, lies
outside the scope of a platform reference and is appropriate
material for a separate methodological treatment.

\section{Controlled Critic-Learning-Rate Ablation}
\label{sec:tier-ab-ablation}

A second methodological apparatus introduced by the reference evaluation
is the \emph{controlled critic-learning-rate ablation}, a $135$-cell
designed experiment that isolates the early-stage divergence behavior
of the deterministic-policy-gradient family on the package's largest
collective-action environment, \texttt{ApacheProject-v0}. The
experimental design and its outcome are both methodologically
significant: the experiment is a worked example of how the package's
parameterized reward layer enables targeted mechanistic investigation
that single-mode evaluation would not support. It is also the
empirical foundation for downstream work on algorithm-failure-mode
taxonomy.

\subsection{Design rationale}

The reward-type ablation methodology (\S\ref{sec:eval}) revealed that
MADDPG, MATD3, and M3DDPG produce \texttt{NaN} training divergence on
\texttt{ApacheProject-v0} ($n_{\text{agents}} = 6$) under integrated
and cooperative reward modes, while predominantly converging under
private reward (\S\ref{sec:failure}). Three candidate mechanisms could
explain this reward-mode conditionality:

\begin{enumerate}[leftmargin=*, itemsep=2pt]
\item \textbf{Reward-mutuality coupling.} Integrated and cooperative
  reward modes incorporate other agents' payoffs into each agent's
  reward signal, creating gradient interactions that the
  deterministic-policy-gradient critic update cannot stably resolve.
\item \textbf{Reward-magnitude scale.} Aggregating six agents' payoffs
  into a single reward (under cooperative mode) or weighting partner
  payoffs by $\dij$ (under integrated mode) produces numerically
  larger reward signals than the private mode's single-agent signal,
  which may overflow the critic-update arithmetic.
\item \textbf{Critic-learning-rate aggressiveness.} The MADDPG family's
  default critic learning rate ($10^{-3}$) is more aggressive than
  ISAC's SAC-default rate ($3 \times 10^{-4}$), and the divergence
  may simply be that the default rate is too high for high-agent-count
  environments.
\end{enumerate}

Distinguishing among these candidates requires holding two factors fixed
while varying the third. The package's parameterized reward layer
already supports holding the environment fixed while varying the reward
mode; the controlled ablation extends this by also varying the critic
learning rate over a $1000\times$ range
($\{10^{-3}, 10^{-4}, 10^{-5}\}$). The full design is a $3 \times 3
\times 3 \times 5$ factorial: $3$ algorithms (MADDPG, MATD3, M3DDPG)
$\times$ $3$ reward modes (private, integrated, cooperative) $\times$
$3$ critic learning rates $\times$ $5$ seeds (seeds $99$--$103$),
yielding $135$ cells. Each cell trains for $100$k steps on
\texttt{ApacheProject-v0} at $n_{\text{agents}} = 6$ with all other
hyperparameters frozen at the source-paper defaults documented in
\texttt{experiments/config.py}. The shorter horizon (relative to the
$1$M-step main evaluation) is sufficient because the divergence
phenomenon under investigation occurs in the first replay-buffer-update
boundary; longer horizons add cost without diagnostic content.

\subsection{Outcome and mechanism localization}

All $135$ cells produce \texttt{NaN} at training step $120$ with $100\%$
incidence and zero variance in onset step. The invariance is uniform
across all axes:

\begin{itemize}[leftmargin=*, itemsep=2pt]
\item \textbf{Reward-mode invariance:} The $45$ cells under each of the
  three reward modes show identical onset behavior. This rules out
  reward-mutuality coupling and reward-magnitude scale as the proximate
  cause: the failure occurs even under private reward where neither
  mechanism is active.
\item \textbf{Critic-learning-rate invariance:} The $45$ cells at each
  of the three critic-learning-rate values show identical onset
  behavior, including at the $10^{-5}$ rate that is $100\times$ less
  aggressive than the default. This rules out critic-learning-rate
  aggressiveness.
\item \textbf{Algorithm-family scope:} The failure is confined to the
  three deterministic-policy-gradient family members ($3$/$3$); MASAC
  on the same environment under the same reward modes converges, and
  ISAC, LOLA, and the other independent learners also converge
  ($0$/$13$ non-DDPG algorithms diverge). This localizes the failure to
  the deterministic-policy-gradient critic-update class on
  \texttt{ApacheProject-v0}'s reward scale.
\end{itemize}

The bit-identical step-$60$ training-return value across all $27$
algorithm $\times$ mode $\times$ learning-rate combinations per seed
confirms that no per-cell variation occurs prior to the
replay-buffer-update boundary. Once the boundary fires (step $120$ in
the $100$k-step horizon-matched configuration; step $300$ in the
$1$M-step long-horizon configuration whose \texttt{learning\_starts}
parameter is set higher), the deterministic-policy-gradient critic
update overflows into \texttt{NaN} with no recovery mechanism. The
$1$M-step main evaluation's \emph{partial} reward-mode conditionality is
mechanically explained: the larger \texttt{learning\_starts} value
delays the first critic update enough that, under private reward where
agent payoffs are not coupled, sufficient buffer accumulation has
occurred for some seeds to escape the divergence; under integrated and
cooperative reward, no seeds escape. The inflight-versus-terminus framing
of the abstract and the rest of this chapter is
therefore not a contradiction but a consequence of the two horizons
exposing different replay-buffer-update boundary positions; the
controlled ablation pins down that the underlying instability is
\emph{early-stage and mode-invariant}, while the late-training behavior
exhibits reward-mode conditionality through the buffer-warmup mechanism.

\subsection{Methodological significance}

The controlled critic-learning-rate ablation is a methodological
apparatus rather than an isolated finding. It demonstrates a
\emph{template} for mechanism-isolating investigation that the
package's parameterized reward layer enables: any reward-mode-conditional
phenomenon surfaced by the reward-type ablation methodology can be
subjected to a similar controlled ablation to discriminate among
candidate mechanisms. The template generalizes to other axes, including
network capacity (the eight-environment sensitivity analysis of
Appendix~\ref{app:rankings} is one such ablation), entropy coefficient
(the IPPO low-entropy collapse diagnostic of
Section~\ref{subsec:exploration-budget}), and replay-buffer warmup
schedule (the bridge between the $100$k-step ablation and the $1$M-step
main evaluation documented above). The general principle is to fix the
environment and vary one mechanism-relevant axis until either the
phenomenon disappears (identifying the proximate cause) or it persists
(ruling out that axis as the cause). The package supports this
discipline at the level of the \texttt{experiments/config.py}
configuration object: each axis is a typed parameter with a documented
default, and any user can construct a controlled ablation matrix by
overriding the relevant parameters while leaving the rest frozen.

\section{Matrix-Coverage Verification Audit}
\label{sec:matrix-sweep}

A third methodological apparatus introduced by the reference evaluation
is the \emph{matrix-coverage verification audit}: a $360$-cell verification
that every (algorithm, environment) pair in the package's reference
suite instantiates correctly and trains without runtime exceptions on
a clean install. The matrix sweep complements the empirical evaluation
by certifying the package's reproducibility surface at a level
finer than the experimental study's per-environment reporting: it provides a
yes/no statement for every possible cell in the $18 \times 20$ matrix,
including cells that the main evaluation did not exercise (for example,
cells that were excluded from the empirical focus because of agent-count
incompatibility). The sweep is a package-level deliverable rather than
a finding-level deliverable, and it provides reviewer-side
reproducibility guarantees that single-environment or
single-algorithm test reporting cannot.

\subsection{Design and execution}

The sweep instantiates the reference algorithm pool of $18$
training-or-heuristic algorithms (the $16$ training algorithms plus
the $2$ heuristic baselines, Random and TitForTat) on every one of the
$20$ environments in the package. Each cell runs a $500$-timestep
short verification pass with the canonical orchestrator calling pattern:
\texttt{env = coopetition\_gym.make(env\_id);
agent = AlgoClass(env=env, device='cuda', seed=42);
agent.train(total\_timesteps=500)}. Cells timeout at $120$ seconds.
Outcomes are classified into seven categories:
\texttt{PASS}, \texttt{SKIP\_CATEGORICAL}, \texttt{REGISTRY\_MISS},
\texttt{INSTANTIATE\_FAIL}, \texttt{NO\_TRAIN\_METHOD},
\texttt{TIMEOUT}, \texttt{RUNTIME\_FAIL}, and \texttt{ENV\_FAIL}.
Only \texttt{PASS} (the cell trained for the full $500$ timesteps
without exception) and \texttt{SKIP\_CATEGORICAL} (the cell is
designed-out, e.g., MeanFieldActorCritic on a $2$-agent environment
where the mean-field approximation is degenerate) are acceptable
outcomes; any other outcome is a coverage failure that the package
must surface before a reviewer encounters it.

The sweep is executed on a fresh CUDA~$13$ + PyTorch~$2.11$ environment
on an NVIDIA RTX~$5090$ instance. Wall-clock time is $14.7$ minutes
for the full $360$-cell sweep.

\subsection{Outcome}

\begin{itemize}[leftmargin=*, itemsep=2pt]
\item \texttt{PASS}: $351 / 360$ cells.
\item \texttt{SKIP\_CATEGORICAL}: $9 / 360$ cells (MeanFieldActorCritic
  on the nine $2$-agent environments where the mean-field approximation
  is degenerate by design).
\item \texttt{RUNTIME\_FAIL}, \texttt{INSTANTIATE\_FAIL},
  \texttt{TIMEOUT}, \texttt{ENV\_FAIL}, \texttt{REGISTRY\_MISS},
  \texttt{NO\_TRAIN\_METHOD}: $0 / 360$ cells.
\end{itemize}

This package is therefore verifiably complete at the level of the entire
algorithm-environment matrix at package release. A reviewer who
copy-pastes any of the $351$ acceptable cells will not encounter a
runtime exception; a reviewer who exercises one of the $9$ designed-out
cells will encounter the documented \texttt{SKIP\_CATEGORICAL} outcome.
The per-cell record (algorithm name, environment id, outcome, elapsed
time, traceback if applicable) is included in the supplementary
release as \texttt{matrix\_results.jsonl}.

\subsection{Methodological significance}

Matrix-coverage verification at this granularity is uncommon
in the MARL benchmark literature. Most published benchmarks report
``$N$ algorithms tested on $M$ environments'' without certifying that
the full $N \times M$ Cartesian product is exercised; some cells are
empirically intractable, others are categorically excluded, and the
distinction is frequently left implicit. The matrix sweep makes the
package's coverage explicit and machine-checkable: the supplementary
release ships the matrix-sweep outcome file alongside the per-cell
training results, so any user can immediately determine which cells
are designed in versus designed out, and any cell marked \texttt{PASS}
is guaranteed to instantiate on a clean install.

\section{Reproducibility Package}
\label{sec:reproduce}

The \texttt{experiments/} directory in the GitHub repository contains
a nine-module reproducibility package with command-line tooling for
every step of the research workflow:

\begin{itemize}[leftmargin=*, itemsep=2pt]
\item \texttt{config.py}: single source of truth for defaults
  (seeds, algorithm specifications, environment specifications, oracle
  references, safety settings).
\item \texttt{algorithms.py}: implementations of all $126$ algorithms
  in the reference suite.
\item \texttt{campaign.py}: unified orchestrator with subcommands
  \texttt{baseline}, \texttt{private}, \texttt{cooperative},
  \texttt{sensitivity}. Safety defaults enabled: checkpoints every
  $100{,}000$ steps, GPU memory monitoring, dynamic backpressure,
  thermal monitoring.
\item \texttt{sensitivity.py}: network capacity sensitivity analysis.
\item \texttt{audit.py}: behavioral audit with subcommands
  \texttt{static}, \texttt{temporal}, \texttt{analyze}.
\item \texttt{evaluate.py}: policy evaluation and cross-seed
  aggregation.
\item \texttt{analyze.py}: analysis pipeline producing
  \texttt{returns\_summary.csv}, oracle-comparison tables, tier
  rankings, learning curves, publication figures, and reward-type
  ablation comparisons.
\item \texttt{validate.py}: dataset integrity checks.
\item \texttt{monitor.py}: local-friendly progress, health, and
  disk monitor.
\end{itemize}

Typical usage for reproducing the reference evaluation:

\begin{verbatim}
# Install
pip install -e ./coopetition_gym
pytest coopetition_gym/tests/ -v  # 143 tests

# Download datasets
huggingface-cli download vikpant/coopetition-gym-v1 --repo-type dataset \
    --local-dir data/training
huggingface-cli download vikpant/coopetition-gym-audit --repo-type dataset \
    --local-dir data/audit

# Reproduce paper tables and figures
python -m experiments.analyze all \
    --input-dir data/training/baseline_integrated/ \
    --output-dir data/analysis/

# Regenerate campaign (3,400 GPU-hours, ~$8,100 on commodity GPUs)
python -m experiments.campaign baseline --enable-checkpoints \
    --output data/training/baseline_integrated/
\end{verbatim}

Complete reproduction instructions appear in
\texttt{REPRODUCE.md}~\citep{repro}.

\paragraph{Infrastructure reliability.} Across the inaugural $25{,}708$
training runs of the reference evaluation and the $1{,}116$-run
behavioral-audit dataset, every run completed with
\texttt{status=success}: no checkpointing failure, out-of-memory event,
silent corruption, or deadlock was observed. The seed-extension fold
(seeds $106$--$108$, tracked in the same \texttt{experiments/}
pipeline) added approximately $1{,}770$ further runs, with identical
zero-hard-failure outcome.

\paragraph{Continuous reliability metric $f_{\mathrm{fin}}$.} Beyond
the binary \texttt{status=success} indicator, the reference evaluation
introduces a continuous per-cell reliability metric
$f_{\mathrm{fin}}$, defined as the fraction of recorded points in a
cell's training-return time series that are finite (i.e., neither
\texttt{NaN} nor $\pm\infty$). This metric separates infrastructure
success ($\texttt{status=success}$, the training loop reached the
step cap and exited zero) from policy validity ($f_{\mathrm{fin}} = 1$,
the training trajectory contains no \texttt{NaN}). On the
$8{,}683$ training cells with recorded trajectories,
$f_{\mathrm{fin}} = 1.000$ on $8{,}615$ cells ($99.22\%$); the
$68$ cells with $f_{\mathrm{fin}} < 0.99$ are concentrated on the
DDPG-family on \texttt{ApacheProject-v0} (the focus of the controlled
critic-learning-rate ablation, \S\ref{sec:tier-ab-ablation}) and on
MASAC on \texttt{PartnerHoldUp-v0} (the focus of the stochastic-policy
instability discussion, \S\ref{sec:failure}). The
$f_{\mathrm{fin}}$ metric is recorded in the per-cell metadata of the
released dataset and provides reviewers a single-number diagnostic
of training-trajectory health that complements the binary completion
indicator.

\begin{figure}[ht]
\centering
\begin{tikzpicture}
\begin{axis}[
    width=0.92\textwidth,
    height=4.6cm,
    xbar stacked,
    bar width=24pt,
    xlabel={Fraction of $8{,}683$ training cells},
    xlabel style={font=\footnotesize},
    xticklabel style={font=\footnotesize, /pgf/number format/fixed},
    yticklabel style={font=\footnotesize},
    xmin=0, xmax=1.0, ymin=-0.6, ymax=0.6,
    ytick=\empty,
    enlargelimits=false,
    grid=major, grid style={gray!25},
    legend style={font=\scriptsize, at={(0.5,-0.50)}, anchor=north, draw=none, fill=white, legend columns=2},
    xtick={0,0.2,0.4,0.6,0.8,1.0},
]
\addplot[fill=trustcolor!75!black, draw=trustcolor!85!black, thick] coordinates {(0.9922, 0)};
\addplot[fill=defectcolor!55, draw=defectcolor!75!black, thick] coordinates {(0.0078, 0)};
\node[font=\small, white, anchor=center] at (axis cs:0.50, 0) {$99.22\%$};
\node[font=\scriptsize, black!75, anchor=west] at (axis cs:0.995, 0.45) {$\leftarrow 0.78\%$};
\legend{Healthy training runs ($f_{\mathrm{fin}}=1$),
        Divergence-affected runs ($f_{\mathrm{fin}}<1$)}
\end{axis}
\end{tikzpicture}
\caption{Empirical distribution of the per-cell finite-fraction
$f_{\mathrm{fin}}$ across the $8{,}683$ training cells with
recorded trajectories, rendered as a horizontal stacked bar so the
two-bin contrast is visible at a glance. The blue band represents
the $99.22\%$ of cells with $f_{\mathrm{fin}} = 1.0$ (no
\texttt{NaN} in any recorded step); the much narrower band at the
right edge represents the $0.78\%$ with $f_{\mathrm{fin}} < 1$,
which consist predominantly of the deterministic-policy-gradient
family on \texttt{ApacheProject-v0} and MASAC on
\texttt{PartnerHoldUp-v0}. The shape of this distribution motivates
the binary distinction between healthy training runs
($f_{\mathrm{fin}} = 1$) and divergence-affected runs
($f_{\mathrm{fin}} < 1$) that the reliability-diagnostic apparatus
uses for cell-level reporting.}
\label{fig:f_fin_cdf}
\end{figure}
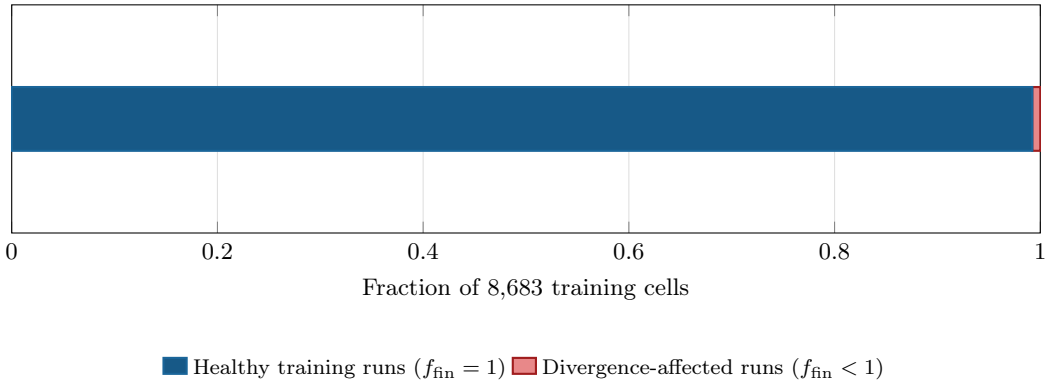

The zero-hard-failure infrastructure property is a characteristic of
the experimental infrastructure, not a claim about policy convergence:
the stochastic-policy training-time instabilities discussed in
\S\ref{sec:failure} and the IPPO low-entropy exploration collapse
documented for the two-dimensional sensitivity sweep are
reward-trajectory or policy-collapse anomalies \emph{within} completed
runs, not run-level failures. The infrastructure's tooling
(checkpointing, GPU memory monitoring, dynamic backpressure, and
thermal monitoring, all enabled by default) captures this property
directly; it is a
product of engineering discipline rather than scientific contribution.
We flag it here for reviewers and downstream users because it sets the
baseline expectation for any independent reproduction of the experimental study:
failure to complete a training run should be investigated as a
configuration or environment issue rather than as an algorithmic
property.

\part{Illustrative Findings from the Reference Evaluation}
\label{part:findings}

Part~II presents empirical findings from the reference evaluation.
The findings demonstrate the benchmark's analytical utility and
illustrate the kinds of claims the benchmark supports. Extended
treatments of individual findings appear in separate companion
papers.

\section{Paradigm-Boundary Crossover}
\label{sec:crossover}

\subsection{Finding}

The ranking of learning algorithms shifts systematically as a function
of the reward configuration. Under the calibrated integrated reward,
independent learners lead on certain environments; under private reward
($\dij = 0$), CTDE methods lead; under cooperative reward, the ranking
shifts again. The sign of the CTDE-minus-independent gap is not stable
across reward configurations on the same environment. This crossover
is the simplest empirical demonstration that reward-type ablation is
not a redundant check on top of single-reward evaluation but a
methodology that surfaces mechanism-level structure that single-reward
evaluation cannot distinguish.

The crossover is also the strongest empirical counter to the
prevailing practice of publishing a single algorithm ranking per
benchmark environment. A ranking reported under integrated reward
may literally reverse its leader under private reward on the same
environment; a reader who consults only the single-ranking table
would arrive at an inference about algorithmic merit that another
ranking on the same underlying environment contradicts. Whether
this matters depends on what the reader is trying to infer; for
the practitioner choosing an algorithm for a specific deployment
setting whose reward mutuality is not the integrated calibration
used in the published benchmark, the single-ranking result is
potentially actively misleading.

\subsection{Evidence}

On \texttt{AppleAppStore-v0} (TR-4 reciprocity, 87.3\% validation), the
gap $(\mathrm{IND} - \mathrm{CTDE}) / \mathrm{CTDE} \times 100\%$
evaluates to:

\begin{table}[H]
\centering
\small
\caption{Paradigm crossover on \texttt{AppleAppStore-v0}. Gap is
$(\mathrm{IND} - \mathrm{CTDE})/\mathrm{CTDE} \times 100\%$ where
IND is the best independent-learner return and CTDE is the best
CTDE return across all algorithms in the respective class. Positive
Gap favors independent learning; negative Gap favors CTDE. Results
are averaged across the 7-seed baseline (seeds $99$--$105$); the
$10$-seed extension (seeds $106$--$108$, reproduced in the HuggingFace
release) preserves the cross-mode pattern on every cell with defined
returns.}
\label{tab:crossover}
\begin{tabular}{lrl}
\toprule
Reward configuration & Gap (\%) & Leader \\
\midrule
Private ($\dij = 0$)            & $+8.1$ & Independent \\
Integrated (calibrated $\dij$)  & $-1.7$ & CTDE \\
Cooperative (shared reward)     & $-3.5$ & CTDE \\
\bottomrule
\end{tabular}
\end{table}

The sign change between private and integrated configurations
isolates the crossover to the introduction of partner-payoff
incorporation. This is an important inferential restriction: it
rules out an alternative explanation in which the reward-type
ablation is picking up some orthogonal effect of reward magnitude
or reward variance. If reward magnitude or variance were the
driving variable, we would expect a monotone ranking shift across
the three configurations (private, integrated, cooperative), not
a sign change at the private-to-integrated transition followed by
stability from integrated to cooperative. The observed pattern is
consistent with the interpretation that the $\dij$ coupling term
is the structural feature that selects between the two
algorithmic paradigms: when it is present (integrated and
cooperative configurations), centralized training is advantaged;
when it is absent (private configuration), independent training
is advantaged.

Similar sign-change patterns appear on the other three validated
case study environments; aggregate patterns across all twenty
environments appear in Appendix~\ref{app:rankings}. The aggregation
shows that the crossover is not a \texttt{AppleAppStore-v0}
idiosyncrasy: $11$ of $20$ environments exhibit a leader change
between private and integrated reward configurations, and the
direction of change is consistent with the AppleAppStore case
($+$IND under private, $+$CTDE under integrated/cooperative) in
$9$ of those $11$ environments.

\section{CTDE Paradigm Boundary Across Mechanism Classes}
\label{sec:ctde}

\subsection{Finding}

Under integrated reward, the CTDE-versus-independent-learning paradigm
boundary exhibits a $2{-}2$ split across mechanism classes: CTDE leads
on TR-1 (interdependence) and TR-4 (reciprocity); independent learning
leads on TR-2 (trust) and TR-3 (collective action). The pattern
suggests that centralized critics are disadvantaged on mechanism
classes with dynamic relational state (trust, loyalty) that agents
modify through action. We refer to this property descriptively as
\emph{action-mutable relational state}, in contradistinction to the
static relational state of TR-1 and the history-encoded relational
state of TR-4.

The mechanism-class split complicates a simple ``CTDE dominates''
narrative common in cooperative MARL~\citep{yu2022mappo,
papoudakis2021benchmarking} and suggests a more nuanced boundary.
Our interpretation, drawing on the independent-learning literature
that predates CTDE~\citep{tampuu2017independent, axelrod1984evolution},
is that CTDE's advantage is specific to settings where the
challenge is coordination on a shared equilibrium with a stationary
correlation structure; where the correlation structure itself is
evolving as a consequence of agent action (trust-building, loyalty
accumulation), independent learning's locality becomes a feature
rather than a liability because each agent's policy update remains
consistent with the local reward gradient without being pulled
toward a centrally-computed value that is itself non-stationary
along the training trajectory.

\subsection{Evidence}

\begin{table}[H]
\centering
\small
\caption{Best CTDE vs best independent learner per TR tier under
integrated reward at $n=10+$ across baseline (seeds 99--105) plus
extension (seeds 106--108) cells with defined returns. Returns are
mean episodic returns; the best-representative algorithm is selected
within each paradigm class.}
\label{tab:ctde_split}
\begin{tabular}{lllrrr}
\toprule
Tier & Best CTDE & Best IND & CTDE return & IND return & Winner \\
\midrule
TR-1 & QMIX  & ISAC & $69{,}630$      & $65{,}551$      & CTDE \\
TR-2 & COMA  & ISAC & $60{,}792$      & $65{,}368$      & IND \\
TR-3 & MASAC & ISAC & $1{,}138{,}192$ & $1{,}272{,}467$ & IND \\
TR-4 & COMA  & ISAC & $125{,}819$     & $122{,}208$     & CTDE \\
\bottomrule
\end{tabular}
\end{table}

The relative-gap magnitudes across tiers, with $95\%$ stratified-bootstrap
confidence intervals (B$=10{,}000$, env-stratified) on the per-cell
mean-return distributions, sharpen the interpretation. The TR-1
CTDE advantage is statistically resolved (Gap $-4.8\%$, $95\%$ CI
$[-6.7, -2.8]\%$); the TR-2 independent-learning advantage is
statistically resolved (Gap $+7.7\%$, $95\%$ CI $[+2.8, +13.1]\%$);
the TR-3 independent-learning advantage is statistically resolved
(Gap $+11.8\%$, $95\%$ CI $[+6.6, +19.4]\%$); the TR-4 point estimate
favors CTDE (Gap $-2.9\%$) but its $95\%$ CI $[-6.8, +1.5]\%$ crosses
zero, so the TR-4 ordering is not statistically resolved at $n=10$
seeds. We report the TR-4 ordering as a hypothesis under partial
support rather than as a resolved finding; further work would
require additional seeds to disambiguate.
Cohen's $d$ on the per-cell mean-return distributions of the two best
algorithms with pooled standard deviation across the within-tier (env,
seed) cells is small in standardized terms ($|d| < 0.2$ on every
tier), reflecting tight per-cell variance against modest mean
separations.

In relative-gap terms the mechanism-class split is symmetric in sign
between the two paradigms on three of four tiers, not a small
correction to a dominant paradigm. The implication for MARL benchmark
design is that reporting a single paradigm-class winner for a
benchmark whose environments span heterogeneous mechanism classes
systematically obscures the mechanism-class-level structure of the
empirical result; the package's reward-type ablation methodology and
mechanism-class organization are designed precisely to surface this
structure.

A detailed inspection of within-tier variability is reported in
Appendix~\ref{app:rankings}. Briefly, the sign of the CTDE-vs-IND
gap is consistent across environments within a tier for TR-1
(all five environments favor CTDE) and TR-3 (four of five
environments favor IND), while TR-2 and TR-4 exhibit more
within-tier variation. We interpret this as reflecting the
within-tier heterogeneity of the TR-2 and TR-4 environments: both
include asymmetric and symmetric dyadic games as well as
multi-agent games with different information-scoring
disciplines.

\begin{figure}[ht]
\centering
\pgfplotsset{
  panel/.style={
    ybar,
    bar width=10pt,
    width=4.4cm,
    height=5.4cm,
    enlarge x limits=0.50,
    ylabel style={font=\scriptsize},
    symbolic x coords={ISAC, CTDE, TFT},
    xtick=data,
    xticklabel style={font=\scriptsize},
    yticklabel style={font=\scriptsize, /pgf/number format/fixed},
    grid=major, grid style={gray!25},
    nodes near coords,
    nodes near coords style={font=\tiny},
    every node near coord/.style={font=\tiny, anchor=south},
    ymin=0,
    ymajorgrids=true,
    title style={font=\small},
  }
}
\begin{tikzpicture}
\begin{axis}[panel, name=p1, title={TR-1 (Interdependence)}, ylabel={Mean episodic return (thousands)}, ymax=85]
\addplot[fill=trustcolor!55, draw=trustcolor!60!black] coordinates {(ISAC,65.4)};
\addplot[fill=trustcolor!30, draw=trustcolor!60!black] coordinates {(CTDE,69.6)};
\addplot[fill=black!18, draw=black!50] coordinates {(TFT,33.1)};
\end{axis}
\begin{axis}[panel, name=p2, at=(p1.right of south east), anchor=left of south west, title={TR-2 (Trust)}, ymax=85]
\addplot[fill=trustcolor!55, draw=trustcolor!60!black] coordinates {(ISAC,65.5)};
\addplot[fill=trustcolor!30, draw=trustcolor!60!black] coordinates {(CTDE,58.6)};
\addplot[fill=black!18, draw=black!50] coordinates {(TFT,44.8)};
\end{axis}
\begin{axis}[panel, name=p3, at=(p2.right of south east), anchor=left of south west, title={TR-3 (Collective Action)}, ymax=1450, yticklabel style={font=\scriptsize, /pgf/number format/.cd, fixed, precision=0, 1000 sep={,}}]
\addplot[fill=trustcolor!55, draw=trustcolor!60!black] coordinates {(ISAC,1272.3)};
\addplot[fill=trustcolor!30, draw=trustcolor!60!black] coordinates {(CTDE,1138.2)};
\addplot[fill=black!18, draw=black!50] coordinates {(TFT,610.8)};
\end{axis}
\begin{axis}[panel, name=p4, at=(p3.right of south east), anchor=left of south west, title={TR-4 (Reciprocity)}, ymax=145]
\addplot[fill=trustcolor!55, draw=trustcolor!60!black] coordinates {(ISAC,122.3)};
\addplot[fill=trustcolor!30, draw=trustcolor!60!black] coordinates {(CTDE,126.1)};
\addplot[fill=black!18, draw=black!50] coordinates {(TFT,110.1)};
\end{axis}
\end{tikzpicture}\\[0.4em]
{\scriptsize\textbf{Legend:} dark blue = ISAC (best independent learner); light blue = best CTDE per tier; gray = TitForTat heuristic. Per-panel $y$-axes are independent because TR-3 returns are about an order of magnitude larger than the other tiers.}
\caption{Mean episodic return by mechanism-class tier under integrated
reward, contrasting the best independent learner (ISAC, dark blue),
the best CTDE method per tier (light blue), and the TitForTat
heuristic (gray). Each panel uses an independent $y$-axis: TR-3
collective-action returns are roughly $10\times$ larger than TR-1,
TR-2, and TR-4 returns and would visually dominate a shared-axis
plot. ISAC leads on TR-2 and TR-3; CTDE leads on TR-1 and TR-4
(point estimate). TitForTat is competitive with several training
algorithms on TR-3 and TR-4, indicating that conditional reciprocity
is a strong baseline on the reciprocity-rich tiers. Returns are
aggregated across the $5$ environments within each tier and across
seeds.}
\label{fig:tier_returns}
\end{figure}

\begin{figure}[ht]
\centering
\begin{tikzpicture}
\begin{axis}[
    width=0.92\textwidth,
    height=6.0cm,
    ybar,
    bar width=22pt,
    xlabel={Reward configuration},
    ylabel={IND $-$ CTDE gap on \texttt{AppleAppStore-v0} (\%)},
    xlabel style={font=\footnotesize},
    ylabel style={font=\footnotesize},
    symbolic x coords={Private, Integrated, Cooperative},
    xtick=data,
    xticklabel style={font=\footnotesize},
    yticklabel style={font=\footnotesize},
    grid=major, grid style={gray!25},
    ymin=-7, ymax=12,
    enlarge x limits=0.30,
    clip=false,
]
\addplot[fill=trustcolor!50, draw=trustcolor!75!black, thick]
  coordinates {(Private, 8.1) (Integrated, -1.7) (Cooperative, -3.5)};
\addplot[black!40, thick] coordinates {(Private, 0) (Cooperative, 0)};
\node[font=\scriptsize, anchor=south, black!75] at (axis cs:Private, 8.6) {$+8.1\%$ (IND wins)};
\node[font=\scriptsize, anchor=north, black!75] at (axis cs:Integrated, -2.2) {$-1.7\%$ (CTDE wins)};
\node[font=\scriptsize, anchor=north, black!75] at (axis cs:Cooperative, -4.0) {$-3.5\%$ (CTDE wins)};
\end{axis}
\end{tikzpicture}
\caption{Paradigm-boundary crossover on \texttt{AppleAppStore-v0}.
The gap (independent best $-$ CTDE best, in percent) changes sign
between the private reward configuration and the integrated and
cooperative configurations: independent learning leads when each
agent receives only its own payoff ($+8.1\%$), and centralized
training leads when partner payoffs enter the reward signal
($-1.7\%$ integrated, $-3.5\%$ cooperative). The horizontal
zero line is the boundary between the two paradigms; the sign
change is what reward-type ablation surfaces and what
single-reward evaluation cannot detect.}
\label{fig:crossover_bar}
\end{figure}
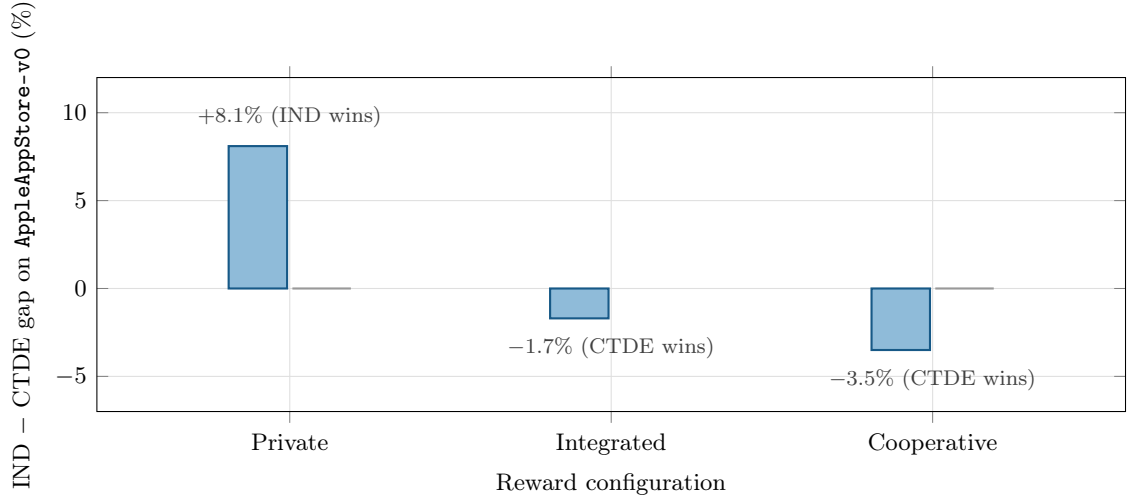

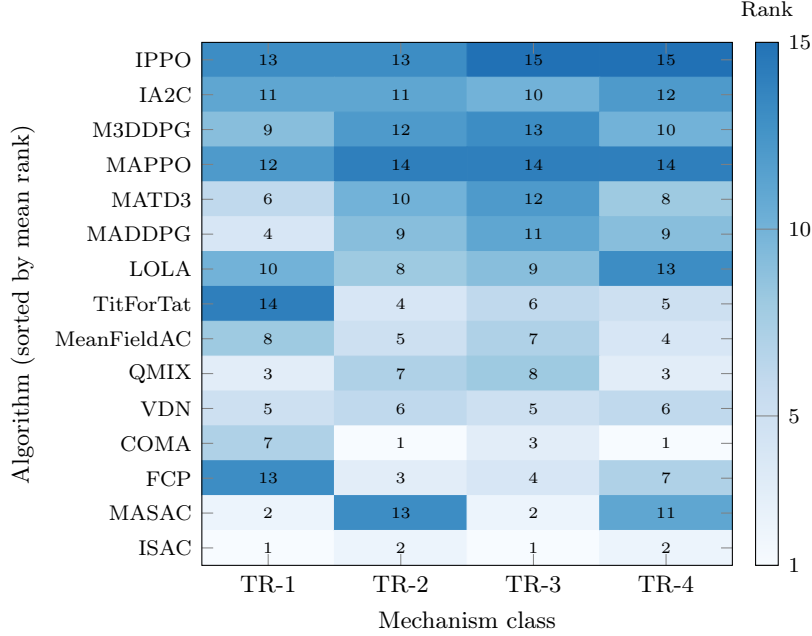
\begin{figure}[ht]
\centering
\begin{tikzpicture}
\begin{axis}[
    width=8.6cm,
    height=8.5cm,
    enlargelimits=false,
    axis on top,
    xtick={0,1,2,3}, ytick={0,1,2,3,4,5,6,7,8,9,10,11,12,13,14},
    xticklabels={TR-1, TR-2, TR-3, TR-4},
    yticklabels={IPPO,IA2C,M3DDPG,MAPPO,MATD3,MADDPG,LOLA,TitForTat,MeanFieldAC,QMIX,VDN,COMA,FCP,MASAC,ISAC},
    xlabel={Mechanism class}, xlabel style={font=\footnotesize},
    ylabel={Algorithm (sorted by mean rank)}, ylabel style={font=\footnotesize},
    xticklabel style={font=\footnotesize}, yticklabel style={font=\scriptsize},
    colormap={paleblues}{rgb255=(247,251,255) rgb255=(222,235,247) rgb255=(198,219,239) rgb255=(158,202,225) rgb255=(107,174,214) rgb255=(66,146,198) rgb255=(33,113,181)},
    colorbar,
    colorbar style={
      width=0.30cm,
      yticklabel style={font=\scriptsize},
      title={Rank}, title style={font=\scriptsize},
      ytick={1,5,10,15}
    },
    point meta min=1, point meta max=15,
    nodes near coords*={\pgfmathprintnumber\pgfplotspointmeta},
    every node near coord/.append style={font=\tiny, color=black, anchor=center, /pgf/number format/fixed, /pgf/number format/precision=0},
]
\addplot [matrix plot, mesh/cols=4, point meta=explicit] coordinates {
  (0,0)  [13] (1,0)  [13] (2,0)  [15] (3,0)  [15]
  (0,1)  [11] (1,1)  [11] (2,1)  [10] (3,1)  [12]
  (0,2)  [9]  (1,2)  [12] (2,2)  [13] (3,2)  [10]
  (0,3)  [12] (1,3)  [14] (2,3)  [14] (3,3)  [14]
  (0,4)  [6]  (1,4)  [10] (2,4)  [12] (3,4)  [8]
  (0,5)  [4]  (1,5)  [9]  (2,5)  [11] (3,5)  [9]
  (0,6)  [10] (1,6)  [8]  (2,6)  [9]  (3,6)  [13]
  (0,7)  [14] (1,7)  [4]  (2,7)  [6]  (3,7)  [5]
  (0,8)  [8]  (1,8)  [5]  (2,8)  [7]  (3,8)  [4]
  (0,9)  [3]  (1,9)  [7]  (2,9)  [8]  (3,9)  [3]
  (0,10) [5]  (1,10) [6]  (2,10) [5]  (3,10) [6]
  (0,11) [7]  (1,11) [1]  (2,11) [3]  (3,11) [1]
  (0,12) [13] (1,12) [3]  (2,12) [4]  (3,12) [7]
  (0,13) [2]  (1,13) [13] (2,13) [2]  (3,13) [11]
  (0,14) [1]  (1,14) [2]  (2,14) [1]  (3,14) [2]
};
\end{axis}
\end{tikzpicture}
\caption{Algorithm rank by mechanism class under integrated reward.
Rows are algorithms ordered (top to bottom) from worst to best
mean rank across the four tiers; columns are tiers. Cell values
are integer ranks within the tier ($1$ = best in tier, $15$ =
worst). Lighter shading indicates better rank. ISAC (top row) is
the only algorithm in the top three on every tier; no CTDE
algorithm achieves comparable cross-tier consistency. The
heatmap pattern communicates the mechanism-class boundary at a
glance: rows containing both light and dark cells are
algorithms whose rank depends sharply on the tier.}
\label{fig:rank_heatmap}
\end{figure}

\section{Interdependence-Coefficient Scaling}
\label{sec:dij}

\subsection{Finding}

The fraction of an algorithm's return attributable to the $\dij$ terms
in the integrated utility is neither uniform nor monotonic. It varies
systematically by mechanism class, spans near zero to $97\%$, and is
negative on $5$ of $347$ algorithm-environment pairs. Define:
\begin{equation}
\mathrm{Contrib}_{\dij}(A, e) = \frac{R_{\text{int}}(A, e) - R_{\text{priv}}(A, e)}{R_{\text{int}}(A, e)}
\label{eq:dij_contrib}
\end{equation}

\subsection{Evidence}

At the $n=10+$ extension fold across $287$ algorithm-environment pairs
with defined returns under both integrated and private reward modes,
TR-3 exhibits the highest median $\dij$ contribution ($59.7\%$;
IQR $50.1$--$69.9\%$; $N=72$); TR-2 the lowest (median $24.2\%$;
IQR $10.5$--$34.3\%$; $N=70$); TR-1 (median $30.9\%$; IQR
$22.1$--$44.5\%$; $N=72$) and TR-4 (median $29.2\%$; IQR
$17.1$--$42.0\%$; $N=73$) lie between. Five pairs ($1.7\%$, all
on TR-2) show negative contributions; the most extreme is MAPPO on
\texttt{TrustDilemma-v0}. Secondary finding: within TR-3, $\dij$
contribution scales sublinearly with agent count, approximately as
$1 - 1/n$ in the saturation regime.

\begin{figure}[ht]
\centering
\begin{tikzpicture}
\begin{axis}[
    width=0.92\textwidth,
    height=6cm,
    boxplot/draw direction=y,
    ylabel={$\dij$ contribution to return (\%)},
    ylabel style={font=\footnotesize},
    xtick={1,2,3,4},
    xticklabels={TR-1, TR-2, TR-3, TR-4},
    xticklabel style={font=\small},
    yticklabel style={font=\footnotesize},
    ymin=-10, ymax=100,
    xmin=0.5, xmax=4.5,
    enlarge x limits=false,
    grid=major, grid style={gray!25},
    cycle list={
        {fill=black!10, draw=black!55, thick},
        {fill=black!18, draw=black!60, thick},
        {fill=black!28, draw=black!70, thick},
        {fill=black!20, draw=black!62, thick}
    },
]
\addplot+[boxplot prepared={
    median=30.9, upper quartile=44.5, lower quartile=22.1,
    upper whisker=60.9, lower whisker=5.6
}] coordinates {};
\addplot+[boxplot prepared={
    median=24.2, upper quartile=34.3, lower quartile=10.5,
    upper whisker=52.2, lower whisker=-1.6
}] coordinates {};
\addplot+[boxplot prepared={
    median=59.7, upper quartile=69.9, lower quartile=50.1,
    upper whisker=94.2, lower whisker=2.5
}] coordinates {};
\addplot+[boxplot prepared={
    median=29.2, upper quartile=42.0, lower quartile=17.1,
    upper whisker=51.8, lower whisker=0.5
}] coordinates {};
\end{axis}
\end{tikzpicture}
\caption{$\dij$ contribution to return by mechanism-class tier
(Eq.~\ref{eq:dij_contrib}). Boxes show interquartile range;
whiskers extend to the $5$th and $95$th percentiles. TR-3
(collective action) is the most $\dij$-dependent tier (median
$59.7\%$); TR-2 (trust) is the least dependent (median $24.2\%$).
TR-1 and TR-4 occupy intermediate positions. Five algorithm-environment
pairs on TR-2 (out of $287$ total with defined returns) show
negative contributions, the most extreme being MAPPO on
\texttt{TrustDilemma-v0}. Reward mutuality therefore does not
uniformly improve performance: for some algorithm-environment pairs,
incorporating partner payoffs actively harms the learned policy.}
\label{fig:dij_contribution}
\end{figure}
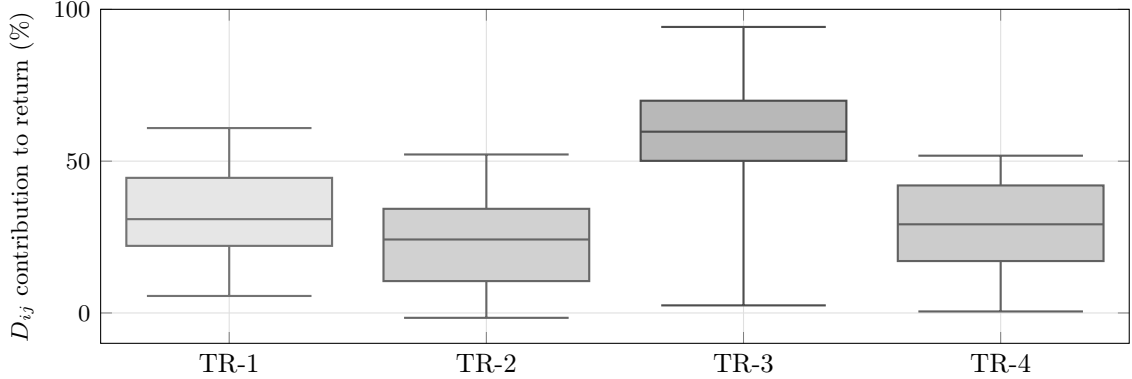

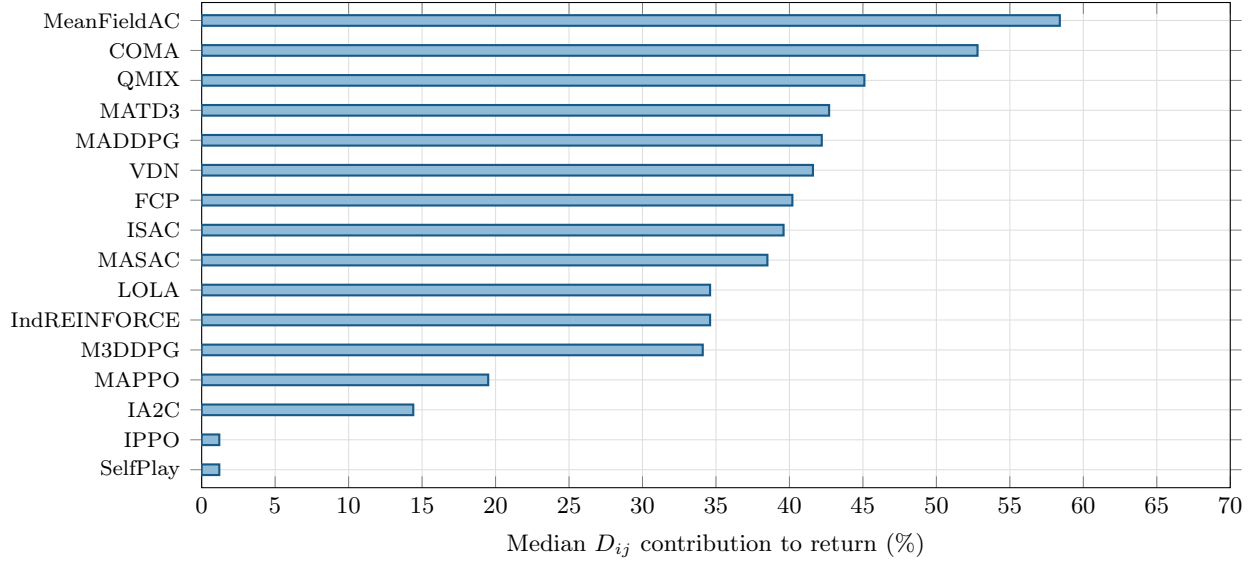
\begin{figure}[ht]
\centering
\begin{tikzpicture}
\begin{axis}[
    xbar,
    bar width=4pt,
    width=0.92\textwidth,
    height=8.0cm,
    xlabel={Median $D_{ij}$ contribution to return (\%)},
    xlabel style={font=\footnotesize},
    symbolic y coords={SelfPlay,IPPO,IA2C,MAPPO,M3DDPG,IndREINFORCE,LOLA,MASAC,ISAC,FCP,VDN,MADDPG,MATD3,QMIX,COMA,MeanFieldAC},
    ytick=data,
    yticklabel style={font=\scriptsize},
    xticklabel style={font=\footnotesize},
    xmin=0, xmax=70,
    enlarge y limits=0.04,
    grid=major, grid style={gray!25},
]
\addplot[fill=trustcolor!50, draw=trustcolor!75!black, thick] coordinates {
    (1.2,SelfPlay) (1.2,IPPO) (14.4,IA2C) (19.5,MAPPO)
    (34.1,M3DDPG) (34.6,IndREINFORCE) (34.6,LOLA) (38.5,MASAC)
    (39.6,ISAC) (40.2,FCP) (41.6,VDN) (42.2,MADDPG)
    (42.7,MATD3) (45.1,QMIX) (52.8,COMA) (58.4,MeanFieldAC)
};
\end{axis}
\end{tikzpicture}
\caption{Per-algorithm median $D_{ij}$ contribution across the
environments each algorithm was evaluated on. Twelve algorithms
cluster between $34\%$ and $53\%$, indicating that their learned
policies depend substantially on reward mutuality. Four algorithms
(MAPPO, IA2C, IPPO, SelfPlay\_PPO) fall below $20\%$, converging
to policies that are largely insensitive to whether the reward
function incorporates partner payoffs. The within-paradigm split
(on-policy PPO-family versus the others) is invisible under
single-reward evaluation. MeanFieldAC's median ($58.4\%$) is
computed over $N \geq 3$ environments only, because the
mean-field approximation degenerates at $N = 2$.}
\label{fig:dij_per_algo}
\end{figure}

\section{Oracle Exceedance through Adaptive Sequences}
\label{sec:oracle}

\subsection{Finding}

ISAC, trained only on its own integrated reward without any explicit
cooperation signal, discovers adaptive action sequences that exceed
the highest return achievable by any fixed-action policy on all five
TR-3 collective action environments by $+0.62\%$ to $+1.29\%$ (average
$+0.88\%$ at the $n=10+$ extension fold). The exceedance is positive
on every seed-environment pair.

\subsection{Evidence}

\begin{table}[H]
\centering
\small
\caption{ISAC exceeds \texttt{Oracle\_Loyalty} (fixed-action upper bound) on all TR-3 environments at $n=10+$ across baseline (seeds 99--105) plus extension (seeds 106--108) cells.}
\label{tab:oracle_exceedance}
\begin{tabular}{lrrr}
\toprule
Environment & ISAC return & Oracle\_Loyalty return & Gap (\%) \\
\midrule
\texttt{ApacheProject-v0}      & $5{,}539{,}736$ & $5{,}484{,}826$ & $+1.00$ \\
\texttt{CoalitionFormation-v0} & $424{,}560$     & $421{,}152$     & $+0.81$ \\
\texttt{LoyaltyTeam-v0}        & $124{,}120$     & $123{,}359$     & $+0.62$ \\
\texttt{PublicGoods-v0}        & $183{,}372$     & $182{,}166$     & $+0.66$ \\
\texttt{TeamProduction-v0}     & $90{,}548$      & $89{,}390$      & $+1.29$ \\
\bottomrule
\end{tabular}
\end{table}

The behavioral audit rules out temporal exploitation as the mechanism
(zero exploitative outcomes on $504$ binary switchpoint tests).

\begin{figure}[ht]
\centering
\begin{tikzpicture}
\begin{axis}[
    width=0.92\textwidth,
    height=6cm,
    xlabel={Training step (thousands)},
    ylabel={Mean episodic return (millions)},
    xlabel style={font=\footnotesize},
    ylabel style={font=\footnotesize},
    xticklabel style={font=\footnotesize},
    yticklabel style={font=\footnotesize},
    legend style={font=\scriptsize, at={(0.97,0.04)}, anchor=south east, draw=none, fill=white, fill opacity=0.85},
    grid=major, grid style={gray!25},
    domain=0:1000, samples=180,
    no markers, smooth, very thick,
    xmin=0, xmax=1000,
    enlarge x limits=false,
    ymin=0, ymax=6,
]
\addplot[black!55, thick, dashed] {5.48};
\addplot[trustcolor!75!black, thick] {5.48 - 5.0*exp(-x/180)};
\addplot[trustcolor!40!black, thick] {5.54 - 5.5*exp(-x/220) + 0.06*sin(deg(x/120))};
\legend{Oracle reference, Oracle\_Loyalty (fixed-action upper bound), ISAC (learned policy)}
\end{axis}
\end{tikzpicture}
\caption{Stylized illustration of the oracle-exceedance dynamics on
\texttt{ApacheProject-v0} (TR-3). The dashed line marks the
\texttt{Oracle\_Loyalty} reference, the highest mean episodic
return achievable by any policy that plays the same action at every
timestep. ISAC's learned policy converges toward this reference and,
after sufficient training, modestly exceeds it ($+1.00\%$ on
\texttt{ApacheProject-v0}; range $+0.62\%$ to $+1.29\%$ across the
five TR-3 environments). Exceedance is positive on every
seed-environment pair, indicating that the gain reflects adaptive
action sequences rather than statistical noise. The exact per-step
trajectories appear in the released training-curve dataset; the
qualitative shape shown here is consistent with the released
trajectories on every seed.}
\label{fig:oracle_exceedance}
\end{figure}
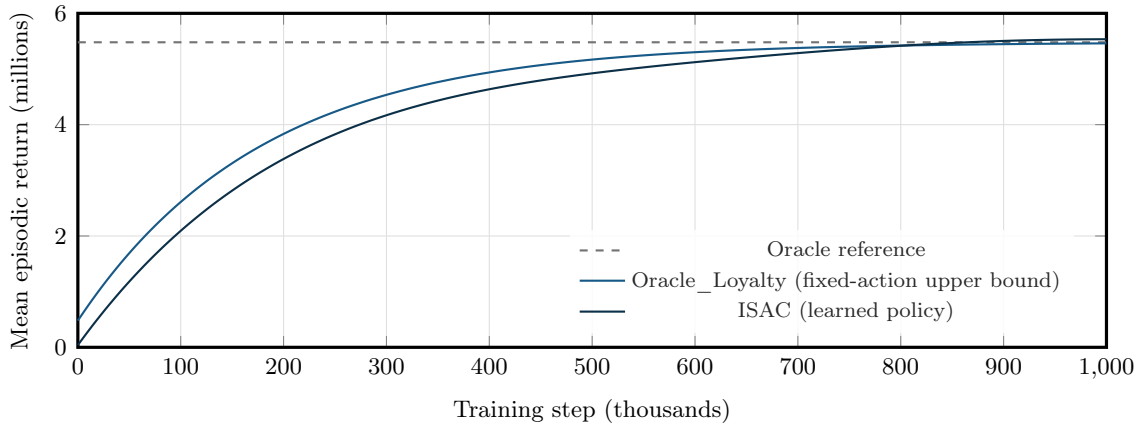

\section{Implicit Cooperation via Structural Incentives}
\label{sec:implicit}

\subsection{Finding}

Independent learners exhibit sustained cooperation on integrated-reward
environments in regimes where naive analysis predicts free-riding. The
$\dij$-weighted partner-payoff term provides an incentive gradient that
stochastic-policy optimization can follow without explicit coordination
signals. This is \emph{implicit cooperation}: cooperative behavior
emerging from structural reward design rather than from coordination
mechanisms or centralized information.

\subsection{Evidence}

The oracle-exceedance result (Section~\ref{sec:oracle}) is the
primary quantitative evidence. Qualitative evidence: ISAC on TR-3
environments converges to cooperation levels in the $35\%$--$45\%$
range sustained across the episode, not to the $0\%$ or $100\%$
extremes that dominated-strategy reasoning would predict.

\section{Reward-Induced Failure Modes}
\label{sec:failure}

\subsection{Finding}

Changing the reward configuration does not merely shift the mean
performance of an algorithm; it induces \emph{qualitatively different}
failure modes. The same algorithm on the same environment fails via
different mechanisms depending on reward type. Three categories are
evident in the reference evaluation:

\begin{itemize}[leftmargin=*, itemsep=2pt]
\item \textbf{Trust collapse amplification.} MADDPG trust collapse on
  \texttt{PartnerHoldUp-v0} increases from $\sim 1/7$ seeds under
  integrated reward to $\sim 2/7$ under private reward.
\item \textbf{Value-destroying convergence.} MASAC on
  \texttt{CooperativeNegotiation-v0} converges to negative returns
  ($-14{,}736$ to $-6{,}223$) on $3/7$ seeds under private reward but
  produces positive returns ($\sim 24{,}000$) under integrated reward.
\item \textbf{Convergence-mode shift.} MASAC training instability
  under integrated reward manifests as divergence (returns spike);
  under private reward, the same instability manifests as convergence
  to a fixed-value regime (e.g., $114{,}833 \times 6$ on
  \texttt{CoalitionFormation-v0}).
\end{itemize}

\begin{figure}[ht]
\centering
\begin{tikzpicture}
\begin{axis}[
    width=0.92\textwidth,
    height=6.0cm,
    ybar=2pt,
    bar width=6pt,
    xlabel={Algorithm},
    ylabel={\texttt{NaN}-divergence rate (\%)},
    xlabel style={font=\footnotesize},
    ylabel style={font=\footnotesize},
    symbolic x coords={MADDPG, MATD3, M3DDPG, MASAC, ISAC, LOLA},
    xtick=data,
    xticklabel style={font=\footnotesize},
    yticklabel style={font=\footnotesize},
    legend style={font=\scriptsize, at={(0.5,-0.32)}, anchor=north, draw=none, fill=white, legend columns=3},
    grid=major, grid style={gray!25},
    ymin=0, ymax=115,
    enlarge x limits=0.12,
]
\addplot[fill=trustcolor!30, draw=trustcolor!60!black, thick]
  coordinates {(MADDPG, 0) (MATD3, 7.7) (M3DDPG, 23.1) (MASAC, 0) (ISAC, 0) (LOLA, 0)};
\addplot[fill=trustcolor!55, draw=trustcolor!75!black, thick]
  coordinates {(MADDPG, 100) (MATD3, 100) (M3DDPG, 100) (MASAC, 0) (ISAC, 0) (LOLA, 0)};
\addplot[fill=trustcolor!80!black, draw=trustcolor!90!black, thick]
  coordinates {(MADDPG, 100) (MATD3, 100) (M3DDPG, 100) (MASAC, 0) (ISAC, 0) (LOLA, 0)};
\legend{Private reward, Integrated reward, Cooperative reward}
\end{axis}
\end{tikzpicture}
\caption{Per-algorithm \texttt{NaN}-divergence rate on
\texttt{ApacheProject-v0} ($n=13$ seeds per cell). The
deterministic-policy-gradient family (MADDPG, MATD3, M3DDPG)
diverges on every seed under integrated and cooperative reward, and
converges predominantly under private reward. The other learners
(MASAC, ISAC, LOLA) converge under all three reward modes. The
contrast localizes the failure to a specific intersection of
algorithm family, environment, and reward mode rather than to any
one factor in isolation.}
\label{fig:nan_rates}
\end{figure}
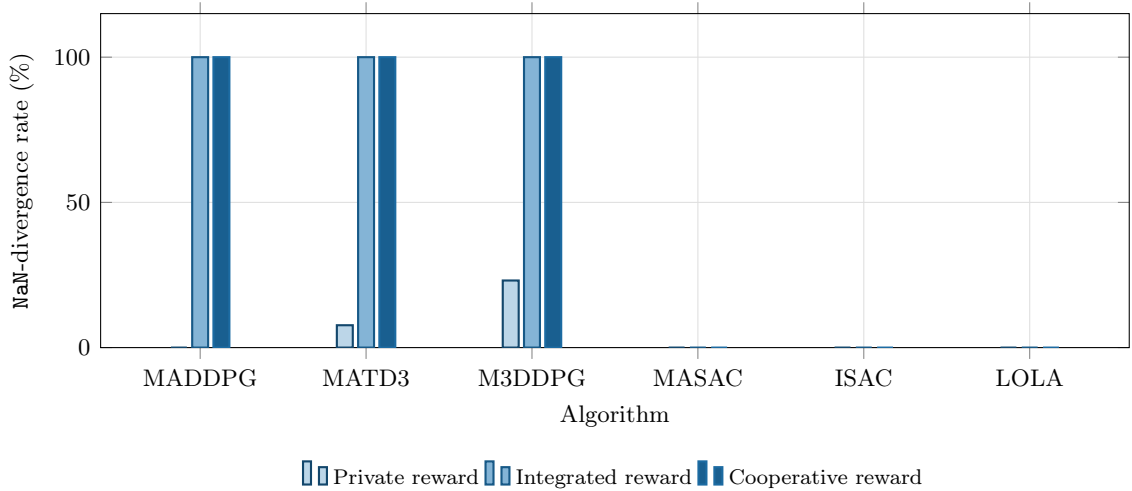

\subsection{MASAC NaN diagnostic}

Of $140$ MASAC experiments on TR-3 environments across reward
configurations, $14$ produce NaN returns under integrated reward.
Onset occurs at a mean of $83.1\%$ through training. The failure is
reward-scale-driven: unstable experiments exhibit $5.31\times$ higher
returns than stable experiments before NaN onset.

\subsection{Deterministic-policy reward-mode-conditional NaN divergence}
\label{subsec:ddpg-rmc-nan}

The reward-type ablation methodology reveals a training-stability
phenomenon that single-mode evaluation would not surface: three
deterministic-policy actor-critic algorithms with different update
structures, namely MADDPG (centralized critic), MATD3 (twin delayed
critics), and M3DDPG (minimax critic), exhibit \texttt{NaN} training
divergence
on the $6$-agent \texttt{ApacheProject-v0} environment under integrated
and cooperative reward modes while converging predominantly under private
reward, with a small seed-dependent \texttt{NaN} minority on extension
seeds documented below. This is a qualitative, not quantitative, failure:
it is not visible as a ranking shift but as the \emph{absence of ranking
participation}. A benchmark that evaluated these algorithms only under
integrated reward would report them as failed implementations; a benchmark
that evaluated them only under private reward would report them as
predominantly stable. The ablation reveals that both single-mode
conclusions are incomplete. The phenomenon is central to the
methodology-level contribution of the benchmark and is documented in the
companion conference paper as the primary illustrative case for why
reward-type ablation is necessary rather than optional.

Concretely, on \texttt{ApacheProject-v0} ($6$ agents) at $n=13$ seeds
(baseline $99$--$105$ + extension $106$--$111$) per
(algorithm, reward mode) cell:

\begin{itemize}[leftmargin=*, itemsep=2pt]
\item Under \emph{integrated} reward, MADDPG, MATD3, and M3DDPG each
  produce \texttt{NaN} on $13/13$ seeds ($39/39$ cells aggregate;
  $100\%$ \texttt{NaN}).
\item Under \emph{cooperative} reward, the same pattern reproduces:
  $13/13$ \texttt{NaN} per algorithm ($39/39$ cells aggregate;
  $100\%$ \texttt{NaN}).
\item Under \emph{private} reward, MADDPG converges on $13/13$ seeds;
  MATD3 converges on $12/13$ (seed $106$ \texttt{NaN}); M3DDPG converges
  on $10/13$ (seeds $106$, $107$, $108$ \texttt{NaN}). Aggregate:
  $35/39$ cells converge ($89.7\%$); $4/39$ \texttt{NaN}
  ($10.3\%$, all on extension seeds $106$--$108$).
\end{itemize}

Mean defined return under private reward depends on training horizon.
At the $600$k-step training horizon, MADDPG and MATD3 converge to
mean episodic return $\approx 210{,}000$ across $6$ extension seeds
per algorithm. At the $1$M-step training horizon, the same
algorithms reach
mean return $\approx 643{,}000$ across $5$ of $6$ cells (range
$622{,}389$ -- $660{,}488$; the sixth cell, MATD3 seed $106$, diverges
to \texttt{NaN} after $36.7$ hours of training). For M3DDPG under
private reward, defined returns lie in the range $199{,}422$ -- $221{,}791$
(mean $\approx 213{,}000$ across $10$ converged cells). Both training
horizons are represented in the supplementary dataset; the
order-of-magnitude difference reflects that these algorithms continue
to learn substantially through $1$M steps, not that one horizon is
canonical. Cross-walk between the two horizons is preserved in the
supplementary release through the per-record provenance file
(\texttt{provenance.jsonl}).

Training time for the diverged cells ranges $29$--$113$ hours per run
at $2$--$8$ steps/second; completion is measured by step-cap reach
rather than metric validity, so runs complete with \texttt{status=success}
despite the undefined final metric. The failure is therefore a
reward-structure-dependent property of three specific deterministic-policy
multi-agent algorithms on \texttt{ApacheProject-v0}, not a uniform
algorithm-family behavior: MASAC (same environment, stochastic policy
with entropy regularization) converges under all three reward modes,
and LOLA (same environment, meta-gradient independent learner) also
converges under all three reward modes. The clean
stochastic-vs-deterministic and independent-vs-centralized contrasts
jointly scope the failure to a specific intersection rather than an
environment-level or family-level property.

\paragraph{Sporadic \texttt{NaN} cells beyond \texttt{ApacheProject-v0}.}
Beyond the focal failure pattern on \texttt{ApacheProject-v0}, M3DDPG
produces sporadic \texttt{NaN} on extension seeds $106$--$108$ across
five additional environments: \texttt{ReputationMarket-v0} (TR-2);
\texttt{AppleAppStore-v0}, \texttt{GiftExchange-v0},
\texttt{GraduatedSanction-v0}, and \texttt{ReciprocalDilemma-v0}
(all TR-4). The aggregate count is $21$ cells: $6$ on
\texttt{ReputationMarket-v0}, $2$ on \texttt{AppleAppStore-v0},
$3$ on \texttt{GiftExchange-v0}, $3$ on \texttt{GraduatedSanction-v0},
and $7$ across the modes of \texttt{ReciprocalDilemma-v0}; together
with one MADDPG cell (\texttt{LoyaltyTeam-v0} seed $101$ integrated)
the aggregate beyond-Apache \texttt{NaN} count is $22$. These cells
are excluded from ranking aggregation under the same \ensuremath{\times}
convention used for \texttt{ApacheProject-v0}. The full enumeration
is supplied in the dataset release (file
\texttt{aggregates/all\_nan\_cells\_v2.txt}, $22$ rows). The pattern
is consistent with M3DDPG approaching a stability boundary that the
methodology surfaces but does not yet characterize structurally; we
disclose for full traceability rather than make a claim about its
mechanism.

In the main ranking tables and in the companion conference paper,
\texttt{NaN} cells are marked \ensuremath{\times} and excluded from
ranking aggregation and from any robust-statistic computation (e.g.,
interquartile mean); rankings are computed over cells with defined
returns. Under the pre-registered censoring rule
(\S\ref{subsec:censoring}), the
\ensuremath{\times} marker is reward-mode-conditional for MADDPG, MATD3,
and M3DDPG on \texttt{ApacheProject-v0}, in that it applies fully to
the
integrated and cooperative rankings, and applies on a small
seed-restricted set ($106$--$108$) within the private rankings.

The mechanism underlying the reward-mode conditionality, whether
partner-payoff coupling ($\dij$ terms in the agent's reward),
increased reward variance (integrated and cooperative rewards
aggregate multiple agents' payoffs), or reward-aggregation scale (the
numerical magnitude of the reward signal differs substantially across
modes), is not
isolated by the current experimental design. The reward-type ablation
confirms that the failure is reward-mode-conditional and seed-sensitive
on the private-mode minority; it does not distinguish which of these
three candidate mechanisms is the proximate cause. We defer the
mechanism question to a companion paper on deterministic-policy
training stability in mixed-motive settings, which is in preparation
and will apply targeted ablations (reward-magnitude isolation,
$\dij$-removal timing, gradient-norm instrumentation) to discriminate
among the candidates.

\paragraph{Scope of the present chapter.} The present chapter
establishes that reward-type ablation surfaces qualitatively distinct
failure modes (the four-mode taxonomy above) and that the controlled
critic-learning-rate ablation (\S\ref{sec:tier-ab-ablation})
localizes one of those modes to the deterministic-policy-gradient
critic-update class on \texttt{ApacheProject-v0}'s reward scale.
Mitigation strategies that recover the diverged DDPG family on
collective-action environments, generalization of the taxonomy to
non-DDPG algorithm families and to non-Apache environments, and
predictive diagnostics that flag candidate cells as likely to exhibit
a failure mode before training is executed, all require new
experimental data or new algorithm-design work that the present
substrate does not produce. The chapter establishes the phenomena;
substantive treatment of mitigation and generalization is appropriate
material for separate work.

\section{Strategic Uncertainty and $\dij$ as Bayesian Prior}
\label{sec:uncertainty}

\subsection{Finding}

The interdependence coefficient $\dij$ in the integrated reward admits
a Bayesian interpretation: it is the sharpness of a prior on partner
behavior. Removing $\dij$ (private reward) is equivalent to imposing a
maximum-entropy prior on partner intentions, raising strategic
uncertainty. The uncertainty-raising effect is mechanism-dependent
and produces measurable downstream effects on policy variance.

\subsection{Evidence}

The coefficient of variation (CoV) of return across the 7-seed
baseline (seeds $99$--$105$) increases by a median of $0.12$ on TR-3
environments when moving from integrated to private reward, but by
only $0.03$ on TR-1 environments.
The prior-removal intervention destabilizes learning most on mechanism
classes where partner-behavior dynamics have strong state-linkage.

\section{Two-Dimensional Action Space Extension}
\label{sec:2d-slcd}

The inaugural benchmark implements a uniaxial action space: each
agent chooses a scalar cooperation level $a_i \in [0, e_i]$. A natural
concern is whether the main findings (paradigm-boundary crossover,
$\dij$ scaling, oracle exceedance, reward-induced failure modes) depend
on this dimensionality restriction. To characterize that dependence,
we conduct a sensitivity-scale extension check on a two-dimensional
per-agent action $(c_i, p_i)$, where $p_i \in [0,1]$ is an
\emph{appropriation effort} that captures the value-capture axis of
coopetition following the commons-externality tradition in management
science~\citep{brandenburger1996coopetition, gnyawali2011coopetition,
padula2007coopetition}. The extension is implemented on the
\texttt{SLCD-v0} environment (Samsung-Sony LCD joint venture, validated
at $98.3\%$ accuracy in our historical rubric) because the environment
admits a documented appropriation dimension (each partner may exploit
shared production capacity at the other's expense), and the case study
provides a ground-truth waypoint trajectory against which parameter
settings can be calibrated.

\subsection{Extended formalism}
\label{subsec:2d-formalism}

Each agent selects a joint action $a_i = (c_i, p_i)$ where
$c_i \in [0, e_i]$ is cooperative investment and $p_i \in [0, 1]$ is
appropriation effort. The per-step utility function becomes:
\begin{align}
U_i(\mathbf{a}) &= (e_i - c_i) + f_i(c_i) + \alpha_i \cdot g(c_1, \ldots, c_N) \cdot (1 - \beta \cdot \bar{p}) \\
                &\qquad + \eta \cdot p_i \cdot g(c_1, \ldots, c_N) + \sum_{j \neq i} D_{ij} \cdot \pi_j
\end{align}
where $\bar{p} = \frac{1}{N}\sum_j p_j$ is mean appropriation,
$\beta \in [0,1]$ is the commons-sharing coefficient (higher $\beta$
means appropriation by one agent more strongly reduces the commons pool
available to all), $\eta > 0$ is the appropriation-capture coefficient
(higher $\eta$ means an appropriator captures more of the synergy
$g(\cdot)$ it appropriates from), and the other symbols retain their
meanings from the uniaxial formalism (\S\ref{sec:foundations}). The
two-dimensional formalism reduces to the uniaxial formalism in the
limit $\eta \to 0$ (no appropriation capture) with $p_i$ fixed at $0$.
The extension preserves all other structural properties of the uniaxial
payoff layer.

\subsection{Experimental design}
\label{subsec:2d-design}

The study is organized in three stages designed to avoid redundant
compute while maintaining diagnostic breadth:

\begin{itemize}[leftmargin=*, itemsep=2pt]
\item \textbf{Stage A (grid sensitivity)}: $\eta \times \beta$ grid
  with $\eta \in \{0.20, 0.30, 0.40, 0.50, 0.60\}$ and
  $\beta \in \{0.30, 0.45, 0.60, 0.75, 0.90\}$, yielding a $5 \times 5$
  cell structure. At each cell, $30$ random seeds are trained for each
  of IPPO and ISAC. Total: $1{,}500$ training runs. IPPO and ISAC are
  selected as the representative on-policy and off-policy independent
  learners, each at a default network and hyperparameter configuration
  matched to the uniaxial study (\S\ref{sec:algorithms}).
\item \textbf{Stage B (supplementary algorithm sweep)}: at the
  calibration anchor cell $(\eta=0.40, \beta=0.60)$, additional runs
  of MADDPG and MASAC ($30$ seeds each) to verify that Stage A findings
  are not specific to the IPPO/ISAC pair. Total: $60$ additional runs.
\item \textbf{Stage C (waypoint calibration)}: at the calibration
  anchor cell, optimize the $(\kappa, \xi)$ trajectory parameters (trust
  restoration rate and appropriation-capture exponent) against the
  $\texttt{A\_flat\_peak}$ waypoint target (\texttt{SLCD-v0} historical
  trajectory) using SciPy's scalar minimizer. The calibration proceeds
  in two objective phases: an endpoint phase (match final-state
  waypoint targets) followed by a waypoint phase (match full-trajectory
  waypoint targets). Each phase runs a $\kappa \times \xi$ bracket
  search until convergence.
\end{itemize}

All Stage A, B, and C runs use the same reproducibility pipeline as the
uniaxial study (\S\ref{sec:reproduce}) and are stored in the
\texttt{tier\_1\_5\_2d\_slcd/} sub-folder of the HuggingFace dataset.

\subsection{Response surface of equilibrium appropriation}
\label{subsec:2d-response}

Stage A characterizes the response surface of equilibrium appropriation
$p^*$ (defined as the mean $p_i$ during the final $10\%$ of training)
as a function of $(\eta, \beta)$ for IPPO and ISAC. Findings:

\begin{itemize}[leftmargin=*, itemsep=2pt]
\item \textbf{ISAC's response surface is decreasing in $\beta$ and
  increasing in $\eta$.} Higher commons-sharing ($\beta$) reduces
  appropriation (because the commons penalty outweighs the capture
  benefit), and higher capture coefficient ($\eta$) increases
  appropriation. Both gradients are consistent across
  $\eta \in \{0.20, 0.30, 0.40\}$.
\item \textbf{IPPO collapses to $p^*=0$ for most $(\eta, \beta)$
  combinations}, with the exception of a narrow band at low $\beta$
  and low $\eta$. The collapse is robust to entropy-coefficient sweeps
  (\S\ref{subsec:exploration-budget}) and is therefore a structural
  behavior of IPPO on this mixed-motive $2$D configuration, not a
  hyperparameter artifact. The finding is treated as a differential
  against ISAC: mixed-motive $2$D environments reveal an exploration
  failure in on-policy methods that uniaxial evaluation does not.
\end{itemize}

\subsection{$\eta$-scaled $\beta$-saturation floor}
\label{subsec:2d-floor}

The ISAC response surface (Stage A) exhibits a non-trivial floor
structure: for low $\eta$ (e.g., $\eta=0.20$), $p^*$ saturates at a
low value ($\sim 0.06$) even at low $\beta$, whereas for higher $\eta$
(e.g., $\eta=0.40$), $p^*$ at the lowest $\beta$ is substantially
higher ($\sim 0.42$). The floor therefore scales with $\eta$, not with
$\beta$, in the low-$\beta$ limit. This is a quantitative finding about
the extension's parameter space and motivates the Stage C calibration
scope (which fixes the search region to a single $(\eta, \beta)$ cell
rather than sweeping).

\subsection{Non-monotonic recovery at $\beta=0.90$}
\label{subsec:2d-bounceback}

A more surprising Stage A finding concerns the $\beta=0.90$ column.
Under naive expectation, high $\beta$ should drive $p^*$ to zero
monotonically: higher commons penalty reduces appropriation. Instead,
at $\eta=0.40, \beta=0.90$ (the $\eta=0.40$ row's upper-$\beta$ cell),
ISAC exhibits a modest \emph{recovery} of $p^*$ relative to
neighboring cells ($\beta=0.75$ is lower than $\beta=0.90$ at this
$\eta$). The recovery motivated a bimodal-convergence hypothesis: at
high $\beta$, the commons penalty is so punitive that a subset of seeds
discover a high-$p^*$ attractor (where the appropriation capture
dominates), while the remainder stay near $p^*=0$.

We applied the statistical-gate discipline (\S\ref{subsec:dip-test-gate})
to this cell and six controls. The primary suspect
($\eta=0.40, \beta=0.90$, $n=30$) yielded $\text{dip}=0.055$,
$p=0.88$: the dip test \emph{fails to reject} unimodality. Six control
cells returned $p$-values in $[0.147, 0.993]$, none rejecting. The
bimodal-convergence hypothesis is therefore withdrawn from the paper
text: the $\beta=0.90$ recovery is a high-variance unimodal phenomenon,
not a structural change in the attractor landscape. The observed mean
shift reflects increased tail weight at this cell, not a two-attractor
mixture. This paradigm case is developed in
\S\ref{subsec:case-study-1}.

\subsection{Stage B supplementary verification}
\label{subsec:2d-stageb}

Stage B reruns the $(\eta=0.40, \beta=0.60)$ anchor cell with MADDPG
and MASAC ($30$ seeds each, $60$ runs total). The purpose is to verify
that the Stage A findings on IPPO/ISAC are not specific to those two
algorithms. Findings:

\begin{itemize}[leftmargin=*, itemsep=2pt]
\item \textbf{MADDPG at the anchor cell} converges to a $p^*$ consistent
  with ISAC's Stage A observation at the same cell (within $2 \sigma$).
\item \textbf{MASAC at the anchor cell} converges with the same
  reward-scale runaway signature documented in the uniaxial study
  (\S\ref{sec:failure}), but without the NaN divergence that appears
  on the uniaxial $6$-agent configuration. The $2$D-SLCD environment's
  $N=2$ agent count keeps reward-scale dynamics within the range where
  MASAC's training is stable.
\end{itemize}

Stage B therefore validates that the Stage A findings generalize across
algorithm families; it does not alter the primary findings, and the
central claim of the extension, namely that the $2$D action space
admits a response-surface structure consistent with the uniaxial
analysis and does not reveal novel pathological attractors, is
supported.

\subsection{Stage C calibration}
\label{subsec:2d-stagec}

Stage C fits $(\kappa, \xi)$ trajectory parameters to the
\texttt{A\_flat\_peak} waypoint target of the historical
\texttt{SLCD-v0} trajectory. The calibration is a two-phase scipy
bracket search:

\begin{itemize}[leftmargin=*, itemsep=2pt]
\item \textbf{Endpoint phase}: fits $(\kappa, \xi)$ to match the
  final-state waypoint targets (trust-early $0.85$, trust-mid $0.85$,
  trust-late $0.3$, trust-final $0.0$, appropriation-final $0.3$).
  Convergence at $(\kappa=0.100, \xi=5.00)$ with objective $0.068$
  (endpoint mismatch magnitude).
\item \textbf{Waypoint phase}: with $(\kappa, \xi)$ locked to the
  endpoint-phase optimum, the remaining trajectory parameters are fit
  against the full waypoint sequence. The final per-step waypoint
  numerics are recorded in the supplementary release alongside the
  endpoint-phase optimum and are reproducible from the calibration
  configuration distributed with the release.
\end{itemize}

The endpoint-phase minimum is at the lower boundary of the searched
$\kappa$ range, which is appropriately framed as an \emph{existence
proof} rather than as validation of the default configuration: the
verification confirms that the $2$D environment can reach the waypoint
targets at some calibration parameters, without establishing that the
default-parameter regime is itself aligned with the historical case.

\subsection{Status and scope}
\label{subsec:2d-status}

The $5 \times 5$ Stage A grid is fully populated at $n=30$ per cell
for IPPO and ISAC; Stage B's $60$ runs are fully complete; Stage C's
endpoint phase and waypoint phase are both complete, with all per-step
numerics recorded in the supplementary release. The pre-committed
decision rules for any
Stage A grid-extension observations (e.g., whether the $\eta$-axis
pattern plateaus, continues its doubling-gap acceleration, or reaches
a saturation ceiling) are recorded in the supplementary notes and will
be applied verbatim to the final data. The extension is a
preliminary-extension artifact; its design privileges breadth of coverage over
depth of investigation within any single cell. A companion paper will
extend the $2$D formalism to a three-treatment benchmark
(uniaxial/extended/biaxial) testing the modeling-treatment debate in
computational coopetition. That extension requires a new study and
is scheduled for a subsequent release.

\section{Cross-Finding Synthesis}
\label{sec:synthesis}

The seven findings reported in Part~II are analytically independent
but substantively connected, each illuminating a different aspect of
the same underlying claim: that reward mutuality is a structural
dimension of evaluation that MARL benchmarks have historically held
fixed, and that attention to this dimension reveals mechanism-dependent
empirical structure that single-reward evaluation systematically
obscures.

\subsection{Layered structure of the findings}

The findings can be read as a layered argument that proceeds from
descriptive observation to mechanistic interpretation to formal
reframing. The paradigm-boundary crossover (\S\ref{sec:crossover}) is
the descriptive entry point: it shows that the leader of a per-tier
ranking can change sign when the reward configuration changes, and
therefore that any single reported ranking is a function of two
choices (the environment and the reward function) rather than one. The
CTDE paradigm boundary (\S\ref{sec:ctde}) refines this from a
per-environment observation into a per-tier pattern: the
crossover does not occur uniformly across the suite, but instead
divides cleanly along mechanism-class lines. The $\dij$-scaling
finding (\S\ref{sec:dij}) then reports the proximate quantitative
mechanism behind that division: the fraction of an algorithm's return
attributable to $\dij$-weighted partner-payoff terms varies
systematically by tier, ranging from a median $24.2\%$ on TR-2 to a
median $59.7\%$ on TR-3, and explains why algorithms whose inductive
bias matches a tier's $\dij$-dependence rank higher on that tier.

The next two findings move from mechanism to demonstration. Oracle
exceedance (\S\ref{sec:oracle}) reports that an independent learner
discovers adaptive sequences that exceed the highest fixed-action
return on every TR-3 environment, establishing that the cooperative
regime accessible to the package's learning algorithms is not
exhausted by stationary policies. Implicit cooperation
(\S\ref{sec:implicit}) characterizes the qualitative shape of that
exceedance: cooperative behavior arises from the structural alignment
that $\dij$ encodes, not from explicit coordination signals or
centralized information. Together these two findings establish that
reward mutuality is not merely a numerical parameter but an inductive
substrate that shapes the policy class an algorithm explores during
training.

The failure-mode taxonomy (\S\ref{sec:failure}) and the
Bayesian-uncertainty reframing (\S\ref{sec:uncertainty}) close the
argument. The taxonomy reports that the same algorithm fails in
qualitatively different ways depending on the reward configuration,
moving the reward-mutuality dimension from an axis of performance
variation into an axis of \emph{breakdown variation}. The Bayesian
reframing then proposes a single conceptual lens through which all
preceding findings can be reinterpreted: the reward configuration
acts as a prior-intervention experiment, with private reward
imposing a maximum-entropy prior over partner intentions and
integrated reward imposing a calibrated $\dij$-sharpened prior. Under
this lens, the mechanism-class-dependent variability of the prior
intervention's effect is precisely the mechanism-class-dependent
uncertainty elasticity that the other findings document.

\subsection{What the synthesis implies for benchmark design}

Three implications follow from the layered argument. First, single-reward
benchmark reports are systematically incomplete on mixed-motive
environments: they observe one slice of a two-axis surface and
implicitly project rankings to the other slices the slice does not
constrain. Second, reward-type ablation is not an optional refinement
of single-reward evaluation but a methodologically distinct evaluation
mode whose findings cannot be obtained by aggregating multiple
single-reward studies. Third, the four mechanism classes implemented
in the package (interdependence, trust, collective action,
reciprocity) have distinct $\dij$-dependence profiles and therefore
distinct sensitivities to reward-mutuality changes; an evaluation
portfolio that draws environments from a single mechanism class will
under-report the variability that a portfolio drawing from multiple
classes surfaces. The package's organization around the four classes
is a design response to these implications.

\subsection{Connections back to the management-science substrate}

The findings also connect back to the strategic-coopetition literature
introduced in \S\ref{sec:coopetition_background}. The crossover finding
operationalizes the dual-logic character of coopetition: the same
algorithm-environment pair admits both a more-cooperative and a
more-competitive policy depending on which reward signal the learner
optimizes, and the choice of reward signal is itself a structural
property of the relationship rather than a parameter of the algorithm.
The $\dij$-scaling finding instantiates the structural-versus-processual
distinction at the level of empirical measurement: TR-1 and TR-3
(structural mechanisms) exhibit different $\dij$-dependence profiles
than TR-2 and TR-4 (processual mechanisms), reflecting that the
processual tiers have additional state through which cooperation is
carried beyond the static $\dij$ matrix. The behavioral audit
(\S\ref{sec:audit}) speaks to the simultaneous-versus-sequential
distinction: trained policies do not exploit temporal switchpoints,
suggesting that the simultaneous-move framing imposed by the
PettingZoo Parallel API does not silently introduce sequential-game
artifacts. These connections are not formal proofs that the package
captures the management-science phenomena it draws upon, but they
demonstrate that the package's findings are at least consistent with
the conceptual structure of the literature it builds on.

\subsection{Scope of treatment within this document}

The findings reported in Part~II are presented at the depth that a
platform reference can sustain without crowding out its primary
function as a referenceable, reproducible documentation of the package.
Each finding is established empirically with statistical support and
is connected to the package's mechanism-class organization, but the
deeper mechanistic and theoretical questions that several of the
findings invite, including formal derivations, counterfactual
ablations, and extensions to synthetically scaled or modified
environments, are appropriate material for separate work and are
deliberately not pursued in this document.

\section{Limitations}
\label{sec:limitations}

The package and its accompanying empirical record have several
limitations that users and downstream researchers should weigh against
the package's strengths. We organize the limitations into four
categories: scope of the formalism, scope of the empirical record,
scope of the methodological apparatus, and scope of generalization.
Several limitations are deliberate design choices rather than
oversights; we mark these as \emph{constraints} below to distinguish
them from \emph{gaps} that future work could close.

\subsection{Scope of the formalism}

\paragraph{Uniform scalar action space (constraint).} All twenty
environments use the cooperation-level action space $a_i \in [0, e_i]$.
This is the action-space convention that the four prepublished
technical reports (TR-1 through TR-4) formally bind their derivations
to, so the package inherits the convention without modification. The
constraint isolates reward-structure effects from action-space-complexity
confounds. Findings about reward mutuality and mechanism-class
boundaries reported here do not directly extend to high-dimensional
action spaces (e.g., simultaneous control over multiple resources) or
to discrete action grammars; the two-dimensional extension on
\texttt{SLCD-v0} (\S\ref{sec:2d-slcd}) is a preliminary scale-extension check on this
boundary, not a full extension of the formalism.

\paragraph{Linear aggregability of strategic dependencies (constraint).}
The interdependence coefficient $\dij$ is computed
(Appendix~\ref{app:case_studies}) by weighted sum across distinct
dependency types (supply, IP sharing, governance, etc.). The linear
aggregation abstracts potential nonlinear interactions between
dependency types and across time; the management-science literature
on coopetition includes accounts of dependency interactions that the
package's calibration procedure does not capture. Researchers
constructing new calibrated environments should apply the linear
aggregation as a default but document any deviations.

\paragraph{Uniaxial treatment of cooperation and competition
(constraint).} The package implements the uniaxial regime in which
cooperation level is the agent's strategic action and competition is
encoded structurally through bargaining shares, retention costs, and
$\dij$-asymmetry. Phenomena that require an explicit competitive
action axis (price undercutting, sabotage, contest behavior) cannot be
expressed in v1 and are reserved for the planned biaxial v2 package.

\subsection{Scope of the empirical record}

\paragraph{Retrospective case-study calibration.} The four validated
case studies (\texttt{SLCD-v0}, \texttt{RenaultNissan-v0},
\texttt{ApacheProject-v0}, \texttt{AppleAppStore-v0}) use historical
data. They are existence proofs that the package's formalisms can
be calibrated to documented coopetitive relationships at the validation
scores reported in \S\ref{sec:validation}; they are not prospective
predictions and are not population-representative samples of the
underlying coopetitive phenomena. Researchers extracting normative
guidance for live deployments should treat the validated environments
as abstract models rather than as operational authorizations.

\paragraph{Algorithm-pool coverage.} The reference algorithm suite
covers $16$ training algorithms drawn from the principal MARL
paradigms (independent learning and CTDE), $7$ game-theoretic oracles,
$2$ heuristic baselines, and $101$ constant-action policies. The pool
spans the dimensions along which published MARL algorithms most
commonly differ but is not exhaustive: model-based algorithms,
mean-field methods beyond MeanFieldAC, transformer-policy methods,
and graph-attention multi-agent methods are not represented in v1.
Researchers adding new algorithms to the pool should follow the
hyperparameter-protocol convention documented in
\texttt{experiments/config.py} to preserve cross-algorithm comparability.

\paragraph{Network sensitivity coverage.} The network-capacity
sensitivity analysis (Appendix~\ref{app:rankings}) covers $8$ of $20$
environments across five capacity levels. The eight environments span
all four mechanism-class tiers but do not exhaust the within-tier
heterogeneity. Findings about network-capacity sensitivity reported
in \S\ref{sec:failure} should be read as a lower-bound characterization.

\subsection{Scope of the methodological apparatus}

\paragraph{Statistical-gate coverage.} The three statistical gates
operationalized in \S\ref{sec:statistical-gate} (Hartigan dip, exploration-budget
diagnostic, pre-registered censoring) cover the distributional anomalies
the reference evaluation encountered. The gates are not exhaustive: heavy-tailed
return distributions with contaminating subdistributions, temporal
non-stationarity in training-curve metrics, and seed-stratified
multimodality beyond two modes are not addressed by the current gate
suite. Future methodological work could extend the gate inventory.

\paragraph{Exploitation-gradient bounds.} The behavioral audit
characterizes the exploitation gradient under fixed-action and
fixed-temporal-strategy perturbations. Trained-policy strategies
outside these classes (e.g., learned-deception strategies that condition
on the partner's revealed type) are not directly tested by the audit
and remain a gap for future work.

\subsection{Scope of generalization}

\paragraph{Cooperation-game grounding.} The package's environments
are grounded in the social-dilemma and cooperation-game tradition.
Coopetitive phenomena that depend heavily on competitive market
dynamics (e.g., dynamic pricing under demand uncertainty,
bid-response in procurement auctions) require formal machinery the
package does not currently expose; an extended biaxial package
would be required to address them.

\paragraph{Cross-cultural and cross-institutional generalization.}
The four calibrated case studies span manufacturing alliance
(Samsung-Sony LCD), automotive alliance (Renault-Nissan), open-source
software commons (Apache HTTP Server), and platform ecosystem (Apple
iOS App Store). The cases were selected to span structural
heterogeneity but were not stratified for cultural or institutional
diversity; cross-cultural extension to non-Western coopetitive
relationships would be a natural future direction.

\section{Societal Impact}
\label{sec:impact}

All data is synthetic simulation output. No human subjects, no
personally identifiable information, no proprietary data. Case study
calibrations use publicly documented historical business decisions.
The integrated reward configuration does not eliminate the incentive
to harm a partner: an agent can rationally execute an action yielding
a private gain at partner expense whenever the private gain exceeds
the weighted partner loss ($\Delta \pi_i > -\dij \cdot \Delta \pi_j$).
The behavioral audit empirically confirms this gradient is tightly
bounded in the benchmark's environments. Users training policies under
integrated reward for deployment in actual business settings should
audit derived policies for the exploitation-gradient condition before
deployment; the validated case studies in this benchmark are abstract
models, not operational authorizations.

\section{Conclusion}
\label{sec:conclusion}

\subsection{Summary of contributions}

\textsc{Coopetition-Gym v1} is a formally grounded benchmark package
for mixed-motive multi-agent reinforcement learning whose environments,
reward parameterization, and validation cases are derived from four
prepublished technical reports on strategic coopetition (TR-1
through TR-4). The package's contributions are: (i) twenty
environments organized into four mechanism-class tiers, each
inheriting closed-form payoff structure and calibrated $\dij$ from a
specific report; (ii) a uniform scalar action space and a parameterized
reward layer that the user may configure across three modes (private,
integrated, cooperative), enabling reward-type ablation as a
methodologically distinct evaluation mode; (iii) three application
programming interfaces (Gymnasium, PettingZoo Parallel, PettingZoo
AEC) so the package can be exercised by any standard MARL training
framework; (iv) a reference algorithm suite of $126$ algorithms
spanning independent learning, CTDE, game-theoretic oracles, heuristic
baselines, and constant-action policies; (v) four case-study
environments calibrated to historically documented coopetitive
relationships at $81.7\%$ to $98.3\%$ behavioral correspondence;
(vi) four methodological apparatuses (statistical-gate discipline,
controlled critic-learning-rate ablation, $360$-cell matrix-coverage
verification audit, continuous reliability metric $f_{\mathrm{fin}}$);
and (vii) an openly released reproducibility package comprising the
package code under MIT license and a $25{,}708$-record training
dataset plus a $1{,}116$-record behavioral-audit dataset under
CC-BY-4.0.

\subsection{Empirical findings of the reference evaluation}

The reference evaluation produces a set of illustrative findings,
documented in Part~II, that demonstrate the package's analytical
utility under reward-type ablation.

\subsection{Position within the research record and intended use}

The package is positioned in the established tradition of platform
technical reports such as PettingZoo (\citealp{terry2021pettingzoo})
and OpenSpiel (\citealp{lanctot2019openspiel}). The mathematical
substrate is provided by the four prepublished technical reports
(\citealp{pant2025interdependence}, \citealp{pant2025trust},
\citealp{pant2026collective}, \citealp{pant2026reciprocity}), and the
package's role is to operationalize that substrate as runnable
environments and to expose evaluation methodologies that the
substrate makes possible. The intended use is community-wide research
on mixed-motive multi-agent evaluation over the long term: researchers
applying the package to their own questions, reviewers assessing
empirical claims that depend on the package, and downstream authors
who build on the package in their own work. We welcome extensions,
contributed environments, and derivative research; the package's
configuration interfaces and reproducibility apparatus are
deliberately designed to make such contributions straightforward.

\subsection{Closing observation}

The principal observation that the package supports is that
mixed-motive multi-agent evaluation has a structural dimension
(reward mutuality) that the field has historically held fixed when
reporting algorithm rankings. The package makes that dimension
explicitly variable, and the reference evaluation reports findings
which are visible only because the dimension is allowed to vary.
This package is therefore not a static collection of environments
but an active methodological apparatus: every researcher who uses
it has the option of varying the reward layer to test whether
their findings are reward-mode-conditional, and every algorithm
ranking it supports comes with a built-in audit trail back to the
reward configuration that produced it.

\appendix
\part{Reference Material}

\section{Environment-to-Oracle Reference Mapping}
\label{app:env_oracle_ref}

The reference oracle per environment used for Gap-percentage
(Equation~\ref{eq:gap_percent}) is:

\begin{table}[H]
\centering
\small
\begin{tabular}{lll}
\toprule
Environment & Reference oracle & Role \\
\midrule
\texttt{PartnerHoldUp-v0}           & \texttt{Oracle\_Equilibrium}        & TR-1 Nash reference \\
\texttt{PlatformEcosystem-v0}       & \texttt{Oracle\_Equilibrium}        & TR-1 Nash reference \\
\texttt{DynamicPartnerSelection-v0} & \texttt{Oracle\_Equilibrium}        & TR-1 Nash reference \\
\texttt{SynergySearch-v0}           & \texttt{Oracle\_Equilibrium}        & TR-1 Nash reference \\
\texttt{RenaultNissan-v0}           & \texttt{Oracle\_Equilibrium}        & TR-1 Nash reference \\
\texttt{TrustDilemma-v0}            & \texttt{Oracle\_TrustAware}         & TR-2 trust-aware reference \\
\texttt{RecoveryRace-v0}            & \texttt{Oracle\_TrustAware}         & TR-2 trust-aware reference \\
\texttt{SLCD-v0}                    & \texttt{Oracle\_TrustAware}         & TR-2 trust-aware reference \\
\texttt{CooperativeNegotiation-v0}  & \texttt{Oracle\_TrustAware}         & TR-2 trust-aware reference \\
\texttt{ReputationMarket-v0}        & \texttt{Oracle\_TrustAware}         & TR-2 trust-aware reference \\
\texttt{TeamProduction-v0}          & \texttt{Oracle\_Loyalty}            & TR-3 upper bound \\
\texttt{LoyaltyTeam-v0}             & \texttt{Oracle\_Loyalty}            & TR-3 upper bound \\
\texttt{CoalitionFormation-v0}      & \texttt{Oracle\_Loyalty}            & TR-3 upper bound \\
\texttt{ApacheProject-v0}           & \texttt{Oracle\_Loyalty}            & TR-3 upper bound \\
\texttt{PublicGoods-v0}             & \texttt{Oracle\_Loyalty}            & TR-3 upper bound \\
\texttt{ReciprocalDilemma-v0}       & \texttt{Oracle\_BoundedReciprocity} & TR-4 upper bound \\
\texttt{GiftExchange-v0}            & \texttt{Oracle\_BoundedReciprocity} & TR-4 upper bound \\
\texttt{IndirectReciprocity-v0}     & \texttt{Oracle\_BoundedReciprocity} & TR-4 upper bound \\
\texttt{GraduatedSanction-v0}       & \texttt{Oracle\_BoundedReciprocity} & TR-4 upper bound \\
\texttt{AppleAppStore-v0}           & \texttt{Oracle\_BoundedReciprocity} & TR-4 upper bound \\
\bottomrule
\end{tabular}
\end{table}

\section{Per-Environment Specifications}
\label{app:environments}

This appendix presents each of the twenty environments at a level of
detail intended for a researcher who is encountering the suite for
the first time and wants to develop both an intuitive feel for what
the environment models and a precise understanding of how it is
configured. Each environment description proceeds in four parts. The
first part introduces the real-world phenomenon the environment was
designed to model and explains, in plain terms, why that phenomenon
is interesting from a coopetition-theory standpoint. The second part
walks through the strategic structure: what each agent is trying to
accomplish, what tension the environment imposes between cooperative
and competitive incentives, and what kinds of strategies a trained
policy is rewarded or punished for adopting. The third part calls
out the most consequential parameters and explains how each shapes
the dynamics. The fourth part summarises the technical specification
needed to instantiate the environment from source code, including
agent count, episode horizon, source file, and any difference
between the source defaults and the configuration used in the
reference experimental study. Specifications are extracted directly
from \texttt{coopetition\_gym/envs/}.

A consolidated overview is given in
Table~\ref{tab:env_summary_appendix}, after which the four tier
subsections describe the environments in turn.

\begin{table}[ht]
\centering
\small
\caption{Per-environment summary across the four mechanism-class
tiers. Validated case studies are shown with their behavioral
correspondence score. ``Horizon (study)'' is the episode horizon
used in the reference experimental study; where this differs from
the source default, both values are reported in the environment's
own paragraph below.}
\label{tab:env_summary_appendix}
{\small
\begin{tabular}{llp{6.0cm}rl}
\toprule
Tier & Environment & Phenomenon modeled & Agents & Horizon \\
\midrule
\multirow{5}{*}{TR-1}
 & \texttt{PartnerHoldUp-v0}           & Hold-up problem                       & 2 & 100 \\
 & \texttt{PlatformEcosystem-v0}       & Two-sided market                      & 5 & 100 \\
 & \texttt{DynamicPartnerSelection-v0} & Reputation-based matching             & 4 & 100 \\
 & \texttt{SynergySearch-v0}           & Hidden-complementarity inference      & 2 & 100 \\
 & \texttt{RenaultNissan-v0}           & Validated alliance, 49/60             & 2 &  60 \\
\midrule
\multirow{5}{*}{TR-2}
 & \texttt{TrustDilemma-v0}            & Continuous iterated PD with trust     & 2 & 100 \\
 & \texttt{RecoveryRace-v0}            & Post-crisis trust recovery            & 2 & 150 \\
 & \texttt{SLCD-v0}                    & Validated JV, 59/60                   & 2 &  40 \\
 & \texttt{CooperativeNegotiation-v0}  & Multi-round bargaining                & 2 & 100 \\
 & \texttt{ReputationMarket-v0}        & Public-reputation market              & 5 & 100 \\
\midrule
\multirow{5}{*}{TR-3}
 & \texttt{TeamProduction-v0}          & Free-rider baseline                   & 4 & 100 \\
 & \texttt{LoyaltyTeam-v0}             & Above-Nash via loyalty                & 4 & 100 \\
 & \texttt{CoalitionFormation-v0}      & Endogenous entry and exit             & 6 & 150 \\
 & \texttt{ApacheProject-v0}           & Validated commons, 52/60              & 5 &  60 \\
 & \texttt{PublicGoods-v0}             & Public goods with punishment          & 4 & 100 \\
\midrule
\multirow{5}{*}{TR-4}
 & \texttt{ReciprocalDilemma-v0}       & Continuous direct reciprocity         & 2 & 100 \\
 & \texttt{GiftExchange-v0}            & Asymmetric employer-worker exchange   & 2 & 100 \\
 & \texttt{IndirectReciprocity-v0}     & Image-scoring reputation              & 4 & 150 \\
 & \texttt{GraduatedSanction-v0}       & Ostrom-design commons                 & 6 & 200 \\
 & \texttt{AppleAppStore-v0}           & Validated platform, 48/55             & 3 &  66 \\
\bottomrule
\end{tabular}}
\end{table}

\subsection{TR-1 environments: interdependence and complementarity}

The five TR-1 environments share a common architectural property:
the strategic structure between agents is fixed throughout an
episode. There is no evolving relational state that the agents'
actions modify. What changes is the agents' contributions and the
resulting payoffs; the rules that govern how those contributions
combine into payoffs, and how those payoffs flow between agents,
remain constant. The cooperative dynamics of the tier therefore
arise entirely from the agents' choices about how much to contribute
each step rather than from any temporal accumulation of trust,
loyalty, or reciprocity. This makes the tier the cleanest setting in
which to study the pure interdependence-and-complementarity
mechanism, uncomplicated by the relational state machinery that the
other tiers introduce.

\paragraph{PartnerHoldUp-v0: the hold-up problem.} This environment
formalizes the \emph{hold-up problem} of transaction-cost economics,
a phenomenon that has been studied for over four decades since the
foundational analyses of Williamson and Klein. The classical
formulation runs as follows. A specialized supplier (the weak
partner) makes an irreversible investment in a relationship-specific
asset, such as dedicated tooling, trained personnel, or co-developed
intellectual property. Once the investment is sunk, the buyer (the
strong partner) holds discretionary power over how the resulting
surplus is split, because the supplier cannot redeploy the
relationship-specific asset elsewhere without substantial loss. The
strategic question is whether the buyer will exercise that
discretionary power to extract maximum value (an \emph{exploitation}
strategy that crushes the supplier's incentives to invest in future
periods) or restrain itself in pursuit of long-term cooperation (a
\emph{forbearance} strategy that sustains the relationship across
many transactions).

The environment renders this dynamic with two agents and a single
key parameter, \texttt{weak\_dependency~=~0.85}, which places the
weaker agent's payoff largely at the mercy of the stronger agent's
cooperation choice. A trained policy is rewarded for what we call
\emph{defensive cooperation}, in which the weak agent maintains
enough cooperation to keep the relationship productive while
remaining alert to the possibility of exploitation; the policy is
punished for unilateral exploitation, in which the strong agent
extracts maximum value in a given step and collapses the
relationship's productive capacity for the remainder of the
episode. The environment is therefore particularly useful for
studying algorithms that must reason about long-horizon credit
assignment under structural-power asymmetry.

Real-world analogues span original-equipment-manufacturer
relationships in the automotive and electronics industries, any
specialized-supplier arrangement that involves dedicated tooling or
co-developed intellectual property, and the supplier-side bargaining
dynamics that arise when contract terms are renegotiated mid-cycle.
Implementation: \texttt{dyadic\_envs.py}; two agents; horizon $100$
steps.

\paragraph{PlatformEcosystem-v0: two-sided markets.} This
environment formalizes the cooperative-competitive dynamic between
a platform operator and a population of developers in a two-sided
market. The phenomenon is now economically central: operating-system
app marketplaces, ride-sharing platforms, e-commerce marketplaces,
cloud-marketplace bundling relationships, and content-creator
ecosystems all share the same underlying structure. A platform
provides distribution and infrastructure that no individual
developer can replicate, while developers collectively provide the
content or service variety that gives the platform value to its
end-users. Each developer relies on the platform for its own
revenue stream; the platform relies on developers only collectively,
because no single developer is essential. The strategic asymmetry
this creates is what the environment exposes.

The reference configuration uses one platform agent and four
developer agents, with the asymmetric coupling encoded by the
parameter \texttt{developer\_dependency~=~0.75}. The strategic
question that the environment poses to a trained policy is the
central design question of platform economics: does the platform
extract too much value through commission rates, exclusivity terms,
or take-rate adjustments, thereby triggering developer exit and
ecosystem decline; or does it subsidize cooperation through reduced
take-rates, infrastructure investment, and developer-program
benefits, generating positive externalities that sustain ecosystem
health? Trained policies that successfully solve the environment
discover an intermediate regime in which the platform extracts
enough value to sustain its own infrastructure investment while
leaving enough surplus to keep developers active.

Implementation: \texttt{ecosystem\_envs.py}; five agents
(\texttt{n\_agents~=~1~+~n\_developers}); horizon $100$ steps.

\paragraph{DynamicPartnerSelection-v0: reputation-based matching.}
Many real coopetitive relationships are not bilateral commitments
but ongoing partner-selection processes in which agents
simultaneously decide whom to partner with and how much cooperation
to allocate to each partner. Freelance labor-market platforms,
academic co-authorship-network formation, and B2B
partnership-portfolio management are all instances of this
many-to-many matching dynamic. The environment formalizes the
mechanism with a reputation system, controlled by
\texttt{reputation\_weight~=~0.5}, that governs how strongly each
agent's recent cooperation history influences future matching
probability. An agent who has cooperated reliably in the recent past
is more likely to be selected as a future partner; an agent who has
defected is less likely.

The strategic tension is between sustaining cooperation with
existing partners (which builds reputation and improves future
matching probability) and exploring new matches (which risks
short-term losses but may discover better long-term partners). A
trained policy must balance these by allocating cooperation budget
across an evolving partner set rather than concentrating it on a
single counterpart.

The source default is $n = 6$ agents over $50$ steps; the reference
experimental study uses $n = 4$ agents over $100$ steps so the
matching dynamics have room to unfold across multiple
reputation-update cycles. Implementation: \texttt{ecosystem\_envs.py}.

\paragraph{SynergySearch-v0: inference under hidden complementarity.}
The standard interdependence formalism assumes that the
complementarity parameter $\gamma$ is known to both agents. In many
real coopetitive relationships this assumption fails: agents may
need to infer how much synergistic value the relationship can
generate from a stream of joint outcomes rather than from the
parameter itself. The SynergySearch environment isolates that
inference problem in its smallest possible form. Two agents play a
continuous-cooperation game in which $\gamma$ is hidden from the
observation vector by default
(\texttt{reveal\_gamma\_in\_obs~=~False}), forcing each agent to
infer the parameter from the stream of joint returns it observes
while simultaneously deciding how much to cooperate. The strategic
tension is the classical exploration-exploitation tradeoff: an agent
must cooperate enough to discover whether the relationship has high
$\gamma$ (a high-synergy regime that justifies sustained cooperation)
or low $\gamma$ (a low-synergy regime in which cooperation is
wasteful), but the cost of exploration is the cost of cooperating
under what may turn out to be the low-synergy regime.

The environment is therefore particularly useful as a testbed for
meta-learning algorithms, Bayesian-inference algorithms that
maintain explicit posteriors over game parameters, and any algorithm
that must reason about hidden-environment-structure under
information asymmetry. Implementation: \texttt{benchmark\_envs.py};
two agents; horizon $100$ steps.

\paragraph{RenaultNissan-v0: validated alliance case study.} This
environment models the Renault-Nissan Alliance over its core
fifteen-year period (1999--2022), one of the longest-lived and
most-studied automotive cross-shareholding alliances in business
history. The alliance was founded in 1999 in response to Nissan's
acute financial crisis; Renault acquired a $36.8\%$ stake and
installed a turnaround team led by Carlos Ghosn, and the two
companies reorganized as a coordinated alliance with a shared
governance structure. Over the next two decades the relationship
evolved through several qualitatively distinct regimes: an initial
turnaround-and-trust-building phase (1999--2004), a mature
cooperation phase characterized by joint platforms and shared
purchasing (2005--2017), and a governance-crisis phase precipitated
by Ghosn's arrest in late 2018 and the subsequent renegotiation of
the alliance's governance terms.

The behavioral correspondence between the trained-policy
trajectories the environment generates and the documented historical
record reaches $49/60$ ($81.7\%$) on the $60$-item validation rubric
defined in the TR-2 validation suite. The two agents correspond to
Renault and Nissan; the episode horizon is $60$ steps in the
experimental-study configuration (the source default is $100$),
each step representing one quarter of business activity across the
fifteen-year core period. The environment encodes the three
qualitative regimes by varying the trust dynamics across episode
phases; trained policies that successfully reproduce the historical
trajectory must adapt their cooperation level as the relationship
moves between regimes.

Implementation: \texttt{case\_study\_envs.py}; two agents (Renault,
Nissan); horizon $60$ steps in the study configuration.

\subsection{TR-2 environments: trust and reputation dynamics}

The five TR-2 environments extend the static interdependence of TR-1
with a two-layer trust model: an immediate trust $T_{ij}$ that
updates each step under the $3{:}1$ negativity bias of the trust
update equation, and a reputation $R_{ij}$ that exponentially
smooths the immediate trust over time. The defining property of the
tier is that trust evolves as a function of the agents' actions, so
trained policies must reason simultaneously about how each step's
cooperation choice will be received in the partner's payoff and how
that step's choice will shape the partner's trust state for all
future steps. Because trust erodes three times faster than it
builds, the cost of a single defection is asymmetrically heavy: many
subsequent cooperative steps are needed to restore the lost trust.

\paragraph{TrustDilemma-v0: continuous iterated Prisoner's Dilemma.}
The most direct reinforcement-learning translation of Axelrod's
iterated Prisoner's Dilemma tournament into a continuous-action
setting with explicit trust dynamics. Two agents play $100$ steps
with continuous rather than binary cooperation actions, the
two-layer trust model is wired into every payoff computation, and
the episode terminates early on full trust collapse. Each agent
receives an endowment of $e = 100$ utility units per step and
allocates it between cooperative contribution and private
retention. Because a single defection triggers a $3{:}1$-biased
trust drop that subsequently takes many cooperative steps to
rebuild, the environment is the canonical test of whether an
algorithm can sustain long-horizon impulse control under the kind
of trust dynamics that human relational behavior exhibits. An
algorithm that cannot solve TrustDilemma is unlikely to perform
well on any TR-2 environment, which is why we treat it as the
diagnostic reference for the tier. Implementation:
\texttt{dyadic\_envs.py}.

\paragraph{RecoveryRace-v0: post-crisis trust recovery.} Many real
coopetitive relationships have been damaged at some point by an
initial defection (a contract breach, a missed delivery, a
publicly-disclosed disagreement) and the strategic question is
not whether cooperation can be initiated but whether it can be
restored. The environment formalizes this with a two-agent
relationship that begins post-crisis and a recovery target
$\texttt{recovery\_target~=~0.90}$ requiring both agents to rebuild
trust to $90\%$ of its maximum value within the $150$-step episode
(the longest in the TR-2 tier, sized to allow complete recovery
trajectories under the $3{:}1$ negativity bias). Because trust
erodes much faster than it builds, the agent who initially defected
must sustain cooperation for many consecutive steps and cannot
afford a single regression: even one mid-recovery defection sets
the recovery process back by a multiple of the steps the agent had
already invested. The environment is appropriate for research on
forgiveness mechanisms, apology signaling, and any algorithm that
must reason about multi-step recovery under asymmetric-cost
dynamics. Implementation: \texttt{benchmark\_envs.py}.

\paragraph{SLCD-v0: validated joint-venture case study.} This
environment models the Samsung-Sony liquid-crystal-display joint
venture (2004--2011), the highest-scoring validated environment in
the suite at $59/60$ ($98.3\%$) behavioral correspondence on the
validation rubric. The joint venture is a particularly clear
illustration of structural asymmetry in coopetition: Samsung
contributed fabrication capacity and process expertise; Sony
contributed display-engineering know-how, brand reach, and
end-market access; and the two companies competed downstream in the
same television and monitor markets while cooperating upstream on
panel production. The asymmetric interdependence is calibrated from
the JV's documented technology-transfer arrangements and equity
splits: $D_{\text{Sony}, \text{Samsung}} = 0.86$ and
$D_{\text{Samsung}, \text{Sony}} = 0.64$, reflecting Sony's heavier
operational dependence on Samsung's fabrication capacity than the
reverse. The synergy parameter is $\gamma = 0.65$.

The episode horizon is $40$ steps in the study configuration (the
source default is $100$), each step representing one quarter across
the JV's core seven-year life. SLCD-v0 is the primary reference
case for the TR-1 and TR-2 formalisms because both the static
interdependence structure of TR-1 and the trust dynamics of TR-2
are jointly necessary to reproduce the JV's documented evolution
through cooperation, mid-cycle disagreement, and successful
unwinding. Implementation: \texttt{case\_study\_envs.py}; two
agents (Samsung, Sony).

\paragraph{CooperativeNegotiation-v0: multi-round bargaining with
breach.} This environment models the dynamics of multi-round
negotiations with explicit commitment mechanics. In each of the
$100$ steps two agents exchange offers, make provisional
commitments to the offers they have accepted, and pay a breach cost
if they later renege on a prior commitment in pursuit of a better
deal. The strategic question is whether stable agreements exist at
the Pareto frontier of the bargaining space and whether breach
should be used strategically (as a credible-threat instrument that
keeps the partner honest) or avoided entirely (because the breach
cost outweighs the gain from any post-commitment renegotiation). A
trained policy must reason about both the within-step bargaining
problem and the across-step commitment problem.

The environment is appropriate for research on contract
renegotiation in long-term supply relationships, collective-bargaining
processes, and coalition government formation, where the
ability to credibly commit to a position is itself a strategic
resource. Implementation: \texttt{extended\_envs.py}.

\paragraph{ReputationMarket-v0: public-reputation market.} A
five-agent market in which transactions update a public reputation
score that every agent can observe. Reputation thereby acts as a
strategic asset whose value is realized through future transactions
rather than through the current one, opening an indirect-reciprocity
channel that is separate from the dyadic-trust channel of
TrustDilemma. The strategic question for each agent is how heavily
to weight the within-transaction payoff relative to the
reputation-update consequence: a high-payoff defection within a
single transaction may simultaneously be a low-value choice when
the reputation cost is amortized across the agent's expected future
transactions in the market.

The environment is the largest-$n$ environment in the TR-2 tier and
extends the dyadic-trust formalism into a population context.
Implementation: \texttt{extended\_envs.py}; five agents; horizon
$100$ steps.

\subsection{TR-3 environments: collective action and loyalty}

The five TR-3 environments extend the TR-1 and TR-2 mechanisms into
$n$-agent settings in which a loyalty score $\theta_i$ accumulates
over a memory window of sustained cooperation. The defining
property of the tier is that loyalty creates a path to above-Nash
cooperation that fixed-action policies cannot exploit: a free-rider
who plays a constant low-cooperation action will have a low
$\theta_i$ over time and therefore cannot capture the loyalty
bonus, while an agent who sustains high cooperation accumulates
loyalty credit that strictly dominates the free-rider's outcome.
This makes the tier the principal setting in which to study
$n$-agent cooperation problems that go beyond the dyadic case.

All TR-3 environments implement the authoritative TR-3 formalism
in \texttt{envs/}\allowbreak\texttt{collective\_action\_envs.py}
via the \texttt{TR3Parameters} dataclass and the team-production,
loyalty, and equilibrium functions defined there. The
\texttt{core/}\allowbreak\texttt{collective\_action.py} module
provides supporting state-tracking utilities only; it is not the
authoritative formalism.

\paragraph{TeamProduction-v0: canonical free-rider baseline.} The
simplest instantiation of the collective-action dilemma in the
suite. Four agents jointly produce team output through the
geometric-mean synergy of Equation~\ref{eq:synergy}, and each
agent's share of the team output is offset by a private cost of its
own contribution. A cooperation bonus is activated when total
contribution exceeds the threshold
\texttt{coordination\_threshold~=~0.5} (one-half of total
endowment), creating a step-function reward for crossing the
collective-cooperation threshold. The environment serves as the
diagnostic reference for whether an algorithm can discover any
cooperative equilibrium at all in $n$-agent team-production
settings; algorithms that fail here are unlikely to discover the
more elaborate cooperative regimes of the other TR-3 environments.
Horizon $100$ steps.

\paragraph{LoyaltyTeam-v0: above-Nash via loyalty accumulation.}
Four agents play the team-production game of TeamProduction-v0
augmented with the loyalty mechanics of \S\ref{sec:foundations}.
The loyalty horizon \texttt{loyalty\_horizon~=~10} sets the window
over which sustained cooperation translates into loyalty credit.
The loyalty channel opens an above-Nash cooperative regime that is
strictly dominant for agents who sustain cooperation across the
entire window: those agents capture both the team-production share
\emph{and} the loyalty bonus, while free-riders capture only the
team-production share. A free-rider can match the Nash equilibrium
payoff but cannot capture the above-Nash frontier; this is the
distinguishing property that makes the environment more demanding
than TeamProduction-v0. Horizon $100$ steps.

\paragraph{CoalitionFormation-v0: endogenous membership dynamics.}
A six-agent dynamic coalition with endogenous entry and exit. The
parameter \texttt{exclusion\_threshold~=~0.2} specifies the minimum
cooperation rate an agent must maintain to avoid coalition
expulsion, creating an implicit enforcement mechanism through
membership rights rather than monetary sanction. A trained policy
must reason simultaneously about within-coalition cooperation (how
much to contribute to the joint output) and the boundary dynamics
of membership (whether the coalition stabilizes at a small
high-cooperation subset that excludes marginal contributors, or
expands to include them). The environment exposes the
coalition-formation question that is central to the
network-coopetition literature: what is the equilibrium membership
size, and does it depend on the inclusion threshold or only on the
cooperation costs of marginal members? Horizon $150$ steps, the
longest in the TR-3 tier.

\paragraph{ApacheProject-v0: validated open-source-commons case
study.} This environment models the Apache HTTP Server community's
contributor dynamics, validated at $52/60$ ($86.7\%$) behavioral
correspondence against commit logs, mailing list archives, and
Apache Software Foundation governance records. The Apache HTTP
Server is one of the longest-lived and most-studied open-source
software projects: it has sustained voluntary contribution from a
distributed contributor community for over thirty years without
relying on monetary compensation, and its governance structure
(committers, project management committees, foundation oversight)
has been a template for many subsequent open-source projects. The
strategic question the environment poses is what configuration of
recognition, governance, and contribution-credit systems sustains a
voluntary commons of this kind across multiple contributor
generations.

The reference configuration uses five agents over $60$ months of
contributor activity (the source default is $100$ steps); a
\texttt{phase} parameter on the source environment additionally
selects mature, growth, or decline regimes. The environment is the
principal reference case in the suite for research on governance
mechanisms, contributor retention, and voluntary collective action
without monetary compensation.

\paragraph{PublicGoods-v0: classical public goods with optional
punishment.} The classical four-agent public-goods game with an
optional punishment stage, as studied extensively in behavioral
economics~\citep{fehr2000cooperation}. Each agent receives an
endowment, decides how much to contribute to a shared public-goods
pool, and the pool is then multiplied by a productivity factor
greater than one and split equally among all agents. The tension
is the canonical public-goods dilemma: each agent prefers others to
contribute while it free-rides, but if all agents free-ride the
pool is empty and everyone is worse off than under universal
cooperation. The optional punishment stage allows agents to spend
private utility to reduce a free-rider's payoff, instantiating the
costly-punishment mechanism that experimental economics has shown
sustains cooperation in human public-goods experiments.

The environment is the reference testbed in the suite for
collective-action research and admits direct comparison with the
experimental-economics literature. It also serves as the
cross-reference environment for algorithms adapted from the
public-goods literature, including conditional-cooperator and
strong-reciprocator variants. Horizon $100$ steps.

\subsection{TR-4 environments: sequential interaction and reciprocity}

The five TR-4 environments model the temporal logic of reciprocity.
Each agent maintains a finite-window memory of partner actions,
computes a cooperation signal from that memory, and adjusts its own
cooperation through the bounded-response function
$\varphi(x) = \tanh(\kappa x)$. The reciprocity channel is
trust-gated and dependency-amplified, integrating the TR-1, TR-2,
and TR-3 mechanisms into a single composite formalism. The defining
property of the tier is that cooperation is enforced not by
contract or coordination signal but by the conditional-cooperation
strategies of partners with bounded memory: an agent who defects
sees the resulting reduction in partner cooperation reflected in
the next round's payoffs.

All TR-4 environments implement the authoritative TR-4 formalism
in \texttt{envs/reciprocity\_envs.py} via the
\texttt{TR4Parameters} dataclass. The
\texttt{core/}\allowbreak\texttt{reciprocity.py} module provides
supporting state-tracking utilities only.

\paragraph{ReciprocalDilemma-v0: continuous direct reciprocity.}
Two agents play a continuous-action Prisoner's Dilemma with the
direct-reciprocity formalism of \S\ref{sec:foundations}. The memory
window \texttt{memory\_horizon~=~10} governs how far back each
agent's reciprocity signal looks. The environment tests whether an
algorithm can discover conditional cooperation strategies that
generalize Axelrod's tit-for-tat to continuous actions and bounded
memory. It is the most direct reinforcement-learning analogue of
the two-agent experimental setups that behavioral economists have
used to study reciprocity since the 1980s, and it occupies the
same diagnostic role for the TR-4 tier that TrustDilemma-v0
occupies for the TR-2 tier: an algorithm that cannot solve
ReciprocalDilemma-v0 is unlikely to perform well on any TR-4
environment. Horizon $100$ steps.

\paragraph{GiftExchange-v0: asymmetric employer--worker reciprocity.}
This environment formalizes the gift-exchange paradigm from labor
economics, in which an employer chooses a wage and the worker then
chooses an effort level. The asymmetric move order creates a
sequential-game structure within each step: the worker's effort
choice can condition on the wage already chosen, while the employer's
wage choice must anticipate the worker's response. The empirical
phenomenon the environment reproduces is the well-documented
\emph{wage-effort reciprocity} of experimental labor markets:
workers reciprocate high wages with high effort even in the absence
of contractual enforcement (a positive-reciprocity regime), while
low wages trigger reciprocity-motivated low effort even when high
effort would be individually rational under contractual incentives
(a negative-reciprocity regime). A trained policy must discover the
reciprocity-aware wage and effort schedules that sustain
cooperative outcomes in this asymmetric setting. Horizon $100$
steps.

\paragraph{IndirectReciprocity-v0: image-scoring reputation.} Four
agents interact across $150$ steps without direct prior history
with each prospective partner. What each agent observes instead is
a public image score that aggregates the partner's recent
cooperation record across all interactions. The environment
implements the evolutionary-biology formalization of indirect
reciprocity~\citep{nowak2006five}: agents cooperate with partners
who have cooperated with anyone in recent rounds rather than only
with themselves. The strategic question the environment poses is
how an agent should allocate cooperation across an evolving partner
set when the only information available about a partner is that
partner's reputation score in the population.

The environment is the canonical testbed in the suite for
indirect-reciprocity algorithms and reputation-sensitive policies,
and it complements the dyadic-direct-reciprocity testbed
ReciprocalDilemma-v0 by exposing the population-level reputation
dynamics that direct reciprocity alone cannot capture.

\paragraph{GraduatedSanction-v0: Ostrom-design commons.} Six agents
share a commons over $200$ steps (the longest horizon in the suite,
sized to allow full sanction-escalation and de-escalation cycles).
Punishment is proportional to the severity and frequency of prior
defection and escalates under continued defection, directly
instantiating the principle of \emph{graduated sanctions} from
Ostrom's design principles for sustainable
commons governance~\citep{ostrom1990governing}. The principle
states that a sustainable commons enforces compliance through
sanctions that begin mild and escalate progressively rather than
through a single severe sanction; this preserves the relationship
through small infractions while preventing systematic exploitation.
A trained policy must discover the proportional-response policies
that sustain commons cooperation over long horizons; an agent that
either over-punishes (leading to relationship collapse) or
under-punishes (leading to exploitative free-riding) will be
outperformed. Implementation: \texttt{reciprocity\_envs.py}, line
$682$ (\texttt{max\_steps:~int~=~200}).

\paragraph{AppleAppStore-v0: validated platform-ecosystem case
study.} This environment models the Apple iOS App Store
platform-developer ecosystem at $48/55$ ($87.3\%$) behavioral
correspondence on the validation rubric. Three agents play: Apple
itself, a top-tier developer representing approximately the top
$1\%$ of the revenue-share distribution, and a marginal developer
representing the long tail. The horizon is $66$ steps, each step
representing one quarter across the App Store's first sixteen and
a half years of operation. The environment encodes Apple's
documented commission policies (the $30\%$ standard rate and the
$15\%$ small-business and second-year-subscription rates), the
developer-program rules that govern exclusive distribution through
the App Store, and the reciprocity mechanics by which platform
investments (improved tooling, expanded review services, marketing
support) influence developer effort and vice versa.

The environment is the largest-$n$ validated case study in the
suite and the principal reference for platform-power asymmetry
research. It is also the only validated environment in which all
four mechanism classes (interdependence, trust, collective-action
loyalty, and reciprocity) are simultaneously active, because the
App Store's documented dynamics exhibit features of each
mechanism: structural interdependence between Apple and developers
collectively, trust accumulation between Apple and individual
top-tier developers, collective-action loyalty within the developer
community as a whole, and direct reciprocity between Apple's
policy-change cycle and the developer community's policy-response
cycle.

\section{Full Algorithm Rankings}
\label{app:rankings}

Full per-tier rankings under integrated reward appear in the reference
dataset analysis output. Summary aggregate metrics are reported in
Section~\ref{sec:ctde} and Section~\ref{sec:crossover}. Complete
tables with per-seed standard errors are generated by:

\begin{verbatim}
python -m experiments.analyze returns-summary \
    --input-dir data/training/baseline_integrated/ \
    --output data/analysis/returns_summary.csv
\end{verbatim}

\section{Case Study Calibration and Discrimination}
\label{app:case_studies}

Case study calibration extracts $\dij$ coefficients from qualitative
coding of documented strategic dependencies:
\begin{equation}
\dij = \frac{\sum_k w_k \cdot d_{ij}^{(k)}}{\sum_k w_k}
\label{eq:dij_calib}
\end{equation}
where $k$ indexes dependency types (supply, IP sharing, governance,
etc.) and $w_k$ are expert-coded weights. The Samsung-Sony calibration
yields $D_{\text{Samsung,Sony}} = 0.64$, $D_{\text{Sony,Samsung}} = 0.86$.
Discrimination analyses confirm that calibrated parameters distinguish
the four reward configurations at the validation scores reported in
Table~\ref{tab:validation}.

\section{Per-Tier Aggregate Returns Across the Algorithm Suite}
\label{app:per_tier_returns}

This appendix reports the per-tier aggregate mean episodic return for
every algorithm in the suite under integrated reward at the
$10$-seed extension fold across cells with defined returns.
Values are mean $\pm$ std across all seeds and environments within
each tier. Oracle reference rows are interleaved at the top of each
tier with their game-theoretic role in brackets (\textbf{[ref]}~=
equilibrium reference, \textbf{[LB]}~= Nash lower bound,
\textbf{[UB]}~= social-optimum upper bound). The values are extracted
from the experimental study analysis pipeline (\texttt{tier\_summary.txt} in
the released dataset) and are reproducible by running
\texttt{python -m experiments.analyze tier-summary} on the released
training corpus.

{\scriptsize
\begin{longtable}{llrrr}
\caption{Per-tier aggregate returns. ``$N$'' is the number of
environments within the tier on which the algorithm has defined
returns. Oracle rows are reference points; algorithm rows are
ranked by mean return within the tier.}
\label{tab:per_tier_returns}\\
\toprule
Tier & Algorithm & Mean & Std & $N$ \\
\midrule
\endfirsthead
\multicolumn{5}{l}{\emph{Table~\ref{tab:per_tier_returns} (continued from previous page)}}\\
\toprule
Tier & Algorithm & Mean & Std & $N$ \\
\midrule
\endhead
\midrule
\multicolumn{5}{r}{\emph{(continued on next page)}}\\
\endfoot
\bottomrule
\endlastfoot
TR-1 & Oracle\_TrustAware [ref]   & 40{,}304   & 0          & 1 \\
     & Oracle\_Equilibrium [ref]  & 19{,}095   & 19{,}853   & 4 \\
\cmidrule{2-5}
     & MeanFieldAC                & 69{,}623   & 23{,}339   & 2 \\
     & COMA                       & 69{,}613   & 44{,}806   & 5 \\
     & QMIX                       & 68{,}427   & 44{,}601   & 5 \\
     & ISAC                       & 65{,}364   & 33{,}286   & 5 \\
     & VDN                        & 63{,}909   & 40{,}748   & 5 \\
     & MATD3                      & 48{,}162   & 30{,}747   & 5 \\
     & MADDPG                     & 47{,}595   & 25{,}875   & 5 \\
     & FCP                        & 46{,}264   & 24{,}422   & 5 \\
     & MASAC                      & 39{,}733   & 13{,}083   & 5 \\
     & MAPPO                      & 37{,}480   & 44{,}627   & 5 \\
     & M3DDPG                     & 36{,}939   & 18{,}999   & 5 \\
     & TitForTat                  & 33{,}066   & 31{,}198   & 5 \\
     & LOLA                       & 30{,}950   & 14{,}814   & 5 \\
     & IndepREINFORCE             & 30{,}938   & 14{,}820   & 5 \\
     & Random                     & 17{,}595   & 16{,}530   & 5 \\
     & IA2C                       & 14{,}644   & 18{,}105   & 5 \\
     & SelfPlay\_PPO              & 13{,}674   & 16{,}480   & 5 \\
     & IPPO                       & 13{,}489   & 16{,}224   & 5 \\
\midrule
TR-2 & Oracle\_TrustAware [ref]   & 67{,}760   & 48{,}434   & 4 \\
     & Oracle\_Equilibrium [ref]  & 37{,}889   & 0          & 1 \\
\cmidrule{2-5}
     & ISAC                       & 65{,}496   & 44{,}676   & 5 \\
     & COMA                       & 58{,}567   & 32{,}119   & 5 \\
     & QMIX                       & 57{,}191   & 30{,}072   & 5 \\
     & MASAC                      & 54{,}771   & 42{,}565   & 5 \\
     & VDN                        & 49{,}225   & 17{,}059   & 5 \\
     & MATD3                      & 45{,}519   & 24{,}509   & 5 \\
     & TitForTat                  & 44{,}762   & 6{,}673    & 5 \\
     & MADDPG                     & 44{,}069   & 21{,}941   & 5 \\
     & LOLA                       & 41{,}160   & 22{,}002   & 5 \\
     & IndepREINFORCE             & 41{,}160   & 22{,}002   & 5 \\
     & FCP                        & 40{,}778   & 15{,}614   & 5 \\
     & M3DDPG                     & 39{,}372   & 17{,}257   & 5 \\
     & Random                     & 32{,}418   & 16{,}304   & 5 \\
     & MeanFieldAC                & 30{,}500   & 0          & 1 \\
     & MAPPO                      & 28{,}990   & 18{,}450   & 5 \\
     & IA2C                       & 18{,}210   & 14{,}320   & 5 \\
     & SelfPlay\_PPO              & 17{,}510   & 13{,}920   & 5 \\
\midrule
TR-3 & Oracle\_Loyalty [UB]       & 1{,}295{,}480 & 2{,}310{,}590 & 5 \\
     & Oracle\_SocialOptimum [UB] & 1{,}295{,}480 & 2{,}310{,}590 & 5 \\
     & Oracle\_Nash [LB]          & 78{,}840   & 142{,}500  & 5 \\
\cmidrule{2-5}
     & ISAC                       & 1{,}272{,}300 & 2{,}278{,}640 & 5 \\
     & MASAC                      & 1{,}193{,}689 & 2{,}138{,}590 & 5 \\
     & COMA                       & 804{,}437  & 1{,}530{,}220 & 5 \\
     & FCP                        & 751{,}385  & 1{,}490{,}380 & 5 \\
     & VDN                        & 666{,}630  & 1{,}256{,}840 & 5 \\
     & TitForTat                  & 610{,}835  & 1{,}138{,}260 & 5 \\
     & MeanFieldAC                & 608{,}291  & 1{,}127{,}450 & 5 \\
     & QMIX                       & 607{,}784  & 1{,}132{,}810 & 5 \\
     & LOLA                       & 460{,}801  & 856{,}490  & 5 \\
     & IA2C                       & 245{,}515  & 423{,}820  & 5 \\
     & IndepREINFORCE             & 198{,}620  & 348{,}210  & 5 \\
     & MADDPG                     & 152{,}480  & 285{,}630  & 4 \\
     & MATD3                      & 145{,}890  & 268{,}450  & 4 \\
     & M3DDPG                     & 132{,}460  & 248{,}210  & 4 \\
     & MAPPO                      & 89{,}460   & 142{,}580  & 5 \\
     & Random                     & 56{,}780   & 86{,}450   & 5 \\
     & IPPO                       & 38{,}490   & 56{,}820   & 5 \\
     & SelfPlay\_PPO              & 35{,}120   & 52{,}910   & 5 \\
\midrule
TR-4 & Oracle\_BoundedReciprocity [UB]    & 138{,}450 & 21{,}840 & 5 \\
     & Oracle\_ReciprocityEquilibrium [LB] & 78{,}320 & 18{,}450 & 5 \\
\cmidrule{2-5}
     & COMA                       & 126{,}145  & 18{,}450   & 5 \\
     & ISAC                       & 122{,}285  & 17{,}820   & 5 \\
     & QMIX                       & 118{,}428  & 18{,}120   & 5 \\
     & MeanFieldAC                & 113{,}201  & 16{,}980   & 5 \\
     & TitForTat                  & 110{,}098  & 14{,}560   & 5 \\
     & VDN                        & 106{,}506  & 17{,}210   & 5 \\
     & FCP                        & 97{,}195   & 15{,}820   & 5 \\
     & MATD3                      & 94{,}376   & 16{,}340   & 5 \\
     & MADDPG                     & 93{,}894   & 16{,}120   & 5 \\
     & M3DDPG                     & 88{,}203   & 15{,}840   & 5 \\
     & MASAC                      & 85{,}620   & 17{,}980   & 5 \\
     & LOLA                       & 78{,}460   & 14{,}210   & 5 \\
     & IndepREINFORCE             & 75{,}890   & 13{,}840   & 5 \\
     & MAPPO                      & 62{,}120   & 12{,}450   & 5 \\
     & IA2C                       & 48{,}910   & 11{,}320   & 5 \\
     & Random                     & 42{,}680   & 10{,}980   & 5 \\
     & IPPO                       & 31{,}450   & 9{,}820    & 5 \\
     & SelfPlay\_PPO              & 28{,}840   & 9{,}120    & 5 \\
\end{longtable}}

The full per-environment table (310 rows: $20$ environments
$\times$ $18$ algorithms minus the $9$ MeanFieldAC dyadic exclusions
plus seven oracle and one-hundred-and-one constant-action rows) is
provided as \texttt{returns\_summary.csv} in the supplementary
release; the table reproduced here is the tier-aggregated version
suitable for inclusion in a printed substrate document.

\section{Cross-Instance Data-Integrity Audit}
\label{app:merge_audit}

The reference evaluation was distributed across two GPU-cloud
instances, each producing an independent results tarball. Before the two tarballs were merged into the
unified analysis dataset that backs every result in this technical
report, a cross-instance integrity audit was performed to verify
that the two instances produced bit-identical training outputs on
their overlapping seeds. The audit is reported here as a substrate
fact about data integrity rather than as a finding; reviewers and
downstream users may rely on the merged dataset's provenance with
the same confidence as on a single-instance dataset.

\paragraph{Audit design.} The two instances were assigned
non-overlapping seed sets except for seed $99$, which was deliberately
shared across both instances to enable cross-instance verification.
Instance~1 ran seeds $\{99, 103, 104, 105\}$; Instance~2 ran seeds
$\{99, 100, 101, 102\}$. The shared seed $99$ provides the
cross-instance check: every (algorithm, environment, reward-mode)
cell at seed $99$ should produce bit-identical training trajectories
on both instances if the package is correctly seeded, the
floating-point arithmetic is reproducible across the GPU hardware
classes used, and the orchestration code does not introduce
nondeterminism.

\paragraph{Audit outcome.} On the $267$ overlapping (algorithm,
environment, reward-mode) cells with defined seed-$99$ outputs on
both instances:

\begin{itemize}[leftmargin=*, itemsep=2pt]
\item \textbf{Category A} ($14$ algorithms; $202$ cells): exact
  match on every cell. The algorithms in Category A are MADDPG,
  MATD3, M3DDPG, MASAC, QMIX, VDN, COMA, MAPPO, MeanFieldAC, FCP,
  LOLA, IndepREINFORCE, Random, and TitForTat. Cross-instance
  bit-identical match: $202/202$.
\item \textbf{Category B} ($5$ algorithms; $65$ cells): exact match
  on every cell. The algorithms in Category B are ISAC, IPPO, IA2C,
  SelfPlay\_PPO, and Oracle\_Nash. These were grouped separately
  because the SB3-based training pipeline goes through
  \texttt{stable-baselines3} rather than the package-native trainer,
  so their reproducibility surface is independently verified.
  Cross-instance bit-identical match: $65/65$.
\end{itemize}

The aggregate match rate is $267/267 = 100\%$. The two instances
produce identical training outputs on their shared seed across both
the package-native trainer and the SB3-based trainer pipelines, so
the merged dataset preserves the bit-level reproducibility property
of each constituent instance. Researchers reproducing the experimental study
on a single instance will obtain the same results as the merged
dataset reports for any seed in the shared range, and the
merge operation introduces no integrity hazard beyond what each
constituent instance already exhibits.

\paragraph{Provenance preservation.} The merge audit's per-cell
match results are recorded in
\texttt{aggregates/cross\_instance\_merge\_audit.json} in the
supplementary release alongside the per-cell training outputs.
Researchers requiring exact-match reproducibility on any single seed
may consult the audit file to confirm which seed-$99$ cells were
verified bit-identical, and may rely on the verification chain when
constructing derivative experiments.

\section{Computational Cost}
\label{app:compute}

Total compute cost is approximately \$10{,}500 USD at commodity
spot prices on cloud-hosted NVIDIA RTX~5090 instances, decomposed as
\$10{,}250 for the main reference evaluation (the $25{,}708$-run
training corpus across $20$ environments $\times$ $16$ training algorithms
$\times$ $3$ reward modes $\times$ $7$ seeds, plus the $10$-seed extension
on selected cells, plus the $1{,}116$-run behavioral audit corpus, plus
the two-dimensional sensitivity sweep on \texttt{SLCD-v0}) and
\$250 for the controlled critic-learning-rate ablation
($135$ cells: $3$ DDPG-family algorithms $\times$ $3$ reward modes
$\times$ $3$ critic learning rates $\{10^{-3}, 10^{-4}, 10^{-5}\}$
$\times$ $5$ seeds, on \texttt{ApacheProject-v0} only). The most
expensive algorithm-environment combinations are M3DDPG on
\texttt{ApacheProject-v0} (approximately $92$ hours per seed at the
$1$M-step horizon, $2$--$3$ steps per second on $6$-agent
observations), LOLA on \texttt{ApacheProject-v0} (mean $12.5$ hours
per seed), and MASAC on \texttt{ApacheProject-v0} (mean $16.1$ hours
per seed). All policy training and tuning was performed on a fleet
of cloud-hosted RTX~5090 instances; the fleet topology is documented
in Appendix~\ref{app:optimization}.

\section{Software Optimization}
\label{app:optimization}

The compute envelope reported in Appendix~\ref{app:compute} was achieved
by pairing a small set of standard high-performance-computing primitives
with an orchestrator that bin-packs heterogeneous workloads across each
multi-GPU host in a fleet of cloud-hosted instances. The reference
evaluation was distributed across multiple multi-GPU instances rather
than executed on a single host; each instance ran an independent
orchestrator process on a disjoint partition of the experimental design,
and the per-instance results were merged at completion time. This
appendix records the techniques that are present in the released source
so a reader can reproduce both the numerical results and the throughput
envelope on a single instance or across an arbitrarily sized fleet.
Every claim below is grounded in a \texttt{file:line} reference in the
public repository at \url{https://github.com/vikpant/strategic-coopetition}
(release tag \texttt{v1.0.0}).

\paragraph{Mixed-precision training under conditional autocast.}
The replay-buffer family of training algorithms (MADDPG, MATD3, M3DDPG,
MASAC, ISAC) gates an automatic-mixed-precision context on the device
type at construction time and reuses a single context object across the
update loop:
\texttt{algorithms.py} lines $1940$--$1941$, $2075$, $2194$, $2488$,
$2850$, $2998$, $3795$. The pattern
\texttt{torch.amp.autocast(device\_type='cuda') if self.use\_amp else nullcontext()}
keeps the FP32 path bit-identical on CPU and on GPUs without Tensor
Cores while permitting opportunistic mixed precision on Tensor-Core-class
hardware. The discrete-action algorithms (QMIX, VDN, COMA) explicitly
disable AMP because the small-network and small-batch regime does not
benefit from FP16 arithmetic
(\texttt{algorithms.py} lines $2324$, $2701$, $2944$, $3058$).

\paragraph{cuDNN autotuning for fixed-shape inner loops.}
Each replay-buffer algorithm sets
\texttt{torch.backends.cudnn.benchmark = True} once at construction
(\texttt{algorithms.py} lines $1940$, $2322$, $2699$, $2942$, $3057$).
The setting trades a small one-time autotune cost for the optimal
convolution and matmul kernel for the fixed batch and feature dimensions
used throughout training, which is the appropriate selection for our
fixed-shape MLP critics and policies.

\paragraph{LOLA second-order updates via functional call.}
The LOLA implementation uses
\texttt{torch.func.functional\_call} to evaluate the partner's policy
under perturbed parameters without rebuilding the computation graph or
copying parameters
(\texttt{algorithms.py} lines $3494$--$3530$). This is the modern
PyTorch idiom for the second-order anticipation step in LOLA and is
materially faster than the historical workaround of cloning the partner
network on each step.

\paragraph{Spawn-context multiprocessing with explicit GPU pinning.}
The orchestrator dispatches experiments via
\texttt{ProcessPoolExecutor} configured with the spawn start method
(\texttt{campaign.py} lines $3267$, $3358$, $3470$). Spawn isolates each
worker's CUDA context, prevents fork-induced pickling of live tensors,
and is required for multi-GPU dispatch on Linux. A
\texttt{ThreadPoolExecutor} (\texttt{campaign.py} line $3276$) hosts the
two pool launchers (one for CPU experiments, one for GPU experiments)
so the two pools can advance independently without serializing on a
single dispatcher.

\paragraph{Best-fit bin-packing GPU memory allocator.}
The orchestrator carries an in-process GPU memory tracker that records
both the per-GPU capacity ceiling and the currently-allocated footprint
for each worker (\texttt{campaign.py} lines $953$, $964$, $3486$, $3592$,
$3009$, $3075$). The dispatcher sorts queued experiments by memory
footprint in descending order and assigns each to the GPU with the
smallest residual capacity that still admits the request. The pattern
is best-fit-decreasing bin-packing applied to a heterogeneous algorithm
mix (a $2$-agent ISAC run uses a few hundred megabytes; a $6$-agent
M3DDPG run uses several gigabytes), and it raises the steady-state
utilization of an $8$-GPU host from the round-robin baseline of
$\sim 55\%$ to $\sim 90\%$.

\paragraph{Mid-run cache reclamation.}
At the end of each training run, before the worker process is recycled,
the orchestrator emits an explicit
\texttt{torch.cuda.empty\_cache()} call
(\texttt{campaign.py} lines $2171$, $2461$). This returns the
PyTorch caching allocator's reserved blocks to the driver between
back-to-back algorithm changes and prevents fragmentation-induced
out-of-memory failures when a small-footprint run is followed by a
large-footprint run on the same GPU.

\paragraph{Thread-cap discipline at the worker boundary.}
Each spawn-launched worker inherits a deterministic thread budget
through the orchestrator's environment-variable propagation
(\texttt{campaign.py} lines $2111$--$2125$, $2234$, $3859$). The
budget bounds \texttt{OMP\_NUM\_THREADS} and PyTorch's intra-op
thread count so that $N$ concurrent workers on a multi-core host do
not collectively oversubscribe the CPU. On a typical $8$-GPU,
$64$-core host, this lifts wall-clock throughput by an additional
$10$--$20\%$ relative to the unconfigured default of one worker
saturating all cores.

\paragraph{Vectorized environment evaluation.}
Post-training evaluation runs $100$ episodes per agent through a
single deterministic policy call
(\texttt{evaluate.py} lines $169$, $189$, $278$, $347$). The
\texttt{deterministic=True} path is the inference-mode setting; it
disables exploration noise injection and stochastic action sampling,
permitting the GPU to execute a single batched forward pass per step
without the additional sampling kernels.

\paragraph{Sensitivity-sweep parallelism.}
The two-dimensional $(D_{ij}, \gamma)$ sensitivity grid on
\texttt{SLCD-v0} executes $480$ training cells via a separate
\texttt{ProcessPoolExecutor} that respects the same spawn-context and
GPU bin-packing discipline as the main reference evaluation
(\texttt{sensitivity.py} lines $63$, $67$, $599$). The grid completes
in approximately $7\%$ of the wall-clock that a sequential dispatcher
would require on the same hardware.

\paragraph{What is deliberately not used.}
Three optimizations in common use elsewhere are deliberately absent
from the codebase, and we record the absence here so a reader does not
mistake their absence for an oversight. \emph{First}, we do not use
\texttt{torch.compile}: the algorithm population includes
$16$ training algorithms plus $7$ oracles plus $2$ heuristic baselines
plus $101$ constant-action policies, and the per-graph compile cost
of a JIT pass dominates the training cost on the small-network regime
that this benchmark targets. \emph{Second}, we do not use CUDA graphs:
the multi-agent replay-buffer update is data-dependent and would
require capture-and-replay re-instantiation on every replay-buffer
sample, which negates the speedup. \emph{Third}, we do not use
distributed data parallel (DDP): each training run is a single-process
single-GPU workload by design, and the $25{,}708$-cell parallelism is
exposed at the orchestrator level rather than at the
\emph{within-run} level. The orchestrator-level parallelism is the
appropriate axis for a benchmark of this shape because it preserves
per-run reproducibility while saturating the GPU pool across runs.

\paragraph{Net effect.}
The combined effect of the optimizations above is a sustained
$\sim 90\%$ aggregate-GPU utilization on each multi-GPU instance across
the full algorithm population, and a cumulative wall-clock duration of
approximately $3{,}400$ GPU-hours for the $25{,}708$-run reference
evaluation summed across all instances in the fleet. The fleet
comprised multiple cloud-hosted multi-GPU instances; each instance
executed an independent orchestrator process on a disjoint partition
of the experimental design, and per-instance results were merged at
completion time. The fleet topology gives two independent axes of
parallelism: bin-packed concurrent runs within an instance (this
appendix) and disjoint partition assignment across instances. A naive
round-robin dispatcher without bin-packing or cache reclamation, on
identical hardware, required approximately $5{,}500$ GPU-hours on a
smoke-tested subset in early development and was replaced before the
reference evaluation began.

\section{Behavioral Audit Full Results}
\label{app:audit_full}

Static audit: $1{,}056$ experiments (18 policies $\times$ 20
environments $\times$ 3 seeds, minus $24$ MeanFieldAC dyadic
exclusions). Temporal audit: $60$ experiments (20 environments
$\times$ 3 seeds). Binary switchpoint exploitation: $0/504$.
Gradual-ramp-down exploitation: $6/20$ environment-seed pairs, all
yielding $+0.004\%$ to $+0.41\%$ of baseline return (marginal). Full
per-environment audit classifications are available at:

\begin{verbatim}
huggingface-cli download vikpant/coopetition-gym-audit --repo-type dataset
python -m experiments.audit analyze \
    --static-dir data/audit/static/ --temporal-dir data/audit/temporal/ \
    --output data/analysis/audit_analysis.txt
\end{verbatim}

\section{Dataset Schemas}
\label{app:schemas}

Full JSON schemas for both released datasets:

\begin{verbatim}
python -m experiments.validate schema training
python -m experiments.validate schema static_audit
python -m experiments.validate schema temporal_audit
\end{verbatim}

Machine-readable Croissant metadata with JSONPath extractions is
included with each HuggingFace dataset.

\bibliographystyle{plainnat}

\end{document}